%% file: GCMagicc_arxiv.tex
\PassOptionsToPackage{compatibility=false}{caption}
\PassOptionsToPackage{countmax}{subfloat}
\PassOptionsToPackage{hypertexnames=false}{hyperref}
\makeatletter
\IfFileExists{caption-standard.sto}{\def\caption@documentclass{standard}}{}  
\makeatother
\documentclass[gmd, manuscript]{copernicus}
\usepackage{longtable}
\usepackage{booktabs}
\makeatletter
\renewcommand\AB@authnote[1]{\hbox{\raisebox{0.8ex}{\fontsize{7}{7}\selectfont #1}}}
\renewcommand\AB@affilnote[1]{\textsuperscript{\normalfont #1}\nobreakspace}
\makeatother

\providecommand{\introductionname}{Introduction}
\providecommand{\conclusionname}{Conclusions}

\begin{document}
\nolinenumbers  

\title{GCMagicc v1: a fast generative emulator for multivariate climate-impact ensembles}

\Author[1,3]{Nicolai}{Meinshausen}
\Author[1,2][malte.meinshausen@unimelb.edu.au, malte.meinshausen@climate-resource.com]{Malte}{Meinshausen}
\Author[2]{Jared}{Lewis}
\Author[1,2,4]{Zebedee}{Nicholls}
\Author[4,5]{Sarah}{Schöngart}
\Author[2]{Alister}{Self}
\Author[3]{Xinwei}{Shen}
\Author[2]{Karla}{Spiller}
\Author[1]{Elisabeth}{Vogel}

\affil[1]{School of Geography, Earth and Atmospheric Sciences, The University of Melbourne, Swanston Street, Parkville, Victoria 3010, Australia}
\affil[2]{Climate Resource, Victoria Rd 105, Fitzroy, Victoria 3065, Australia}
\affil[3]{Department of Statistics, University of Washington, Padelford Hall, Seattle, Washington 98195, USA}
\affil[4]{Energy, Climate, and Environment Program, International Institute for Applied Systems Analysis, Laxenburg, Austria}
\affil[5]{Institute for Atmospheric and Climate Science, ETH Zürich, Zurich, Switzerland}

\runningtitle{GCMagicc v1: a fast generative emulator}
\runningauthor{N. Meinshausen et al.}
\correspondence{Nicolai Meinshausen (meinshausen@gmail.com) and Malte Meinshausen (malte.meinshausen@unimelb.edu.au)}

\received{}
\pubdiscuss{}
\revised{}
\accepted{}
\published{}

\firstpage{1}

\maketitle

\begin{abstract}
Projecting the impacts of climate change requires large ensembles of climate variables that match historical observations, align with the warming ranges assessed by the IPCC, and can efficiently run new future emissions scenarios, including the newest generation of climate model scenarios (CMIP7) and pathways consistent with countries' Paris Agreement pledges. Generating such ensembles at the scale needed for impact studies is normally computationally prohibitive. We close this gap with GCMagicc, a hybrid model that pairs a simple physical climate model with machine learning to generate ensembles of 10 climate variables at the resolution of full-scale Earth system models, without relying on GPU resources or retraining for new scenarios. Being trained on 32 CMIP6 Earth system models and observational/reanalysis data, GCMagicc complements rather than replaces Earth system models. We apply it to a range of future pathways: the canonical set of SSP scenarios in the latest Intergovernmental Panel on Climate Change (IPCC) report (1.2--6.1$^{\circ}$C warming, min-max range across scenarios of 5-95 percentile ranges), current policies (2.3--4.0$^{\circ}$C); national pledges under the Paris Agreement (1.5--3.3$^{\circ}$C) and the CMIP7 range from the so-called 'VL' to 'H' scenarios (1.2--4.2$^{\circ}$C), releasing a large public dataset. As an illustration of the usefulness of GCMagicc, we perform an attribution analysis on the severe 2025 Iranian drought using GCMagicc ensembles (and three CMIP6 ESMs large ensembles as comparisons) with and without anthropogenic forcings. Our GCMagicc results suggest a strong anthropogenic signal, with a median probability of drought conditions at least as severe as the observed ones of 29\% (median) with anthropogenic forcing and zero probability (median) under natural-forcing only simulations. In the future, drought conditions are projected to materially worsen, amplifying the potential for severe agricultural and food security impacts and geopolitical conflicts that use water scarcity as a weapon. GCMagicc data is available at \url{https://gcmagicc.org}.
\end{abstract}


\introduction\label{main}


Translating up-to-date scenarios into regional climate hazard projections faces multiple hurdles as Earth System Model runs are often not observationally aligned, have too few ensemble members, take months and are best-estimate ensembles of opportunity rather than aligned with assessed uncertainty distributions. For example, data to project or attribute the climate-induced impacts often relies on older scenario generations, such as the Representative Concentration Pathways (RCPs) \citep{VanVuuren2011RCPOverview,Meinshausen2011RCPConcentrations} or so-called 'Shared Socio-economic Pathways' (SSPs) \citep{ONeill2016ScenarioMIP,Meinshausen2020SSPConcentrations}, not taking into account the latest current policies or Paris Agreement pledges, the so-called 'Nationally Determined Contributions' (NDCs).

Newly available scenarios for CMIP7 reflect the new realities, ranging from high-overshoot 1.5\,$^{\circ}$C scenarios to upper-end scenarios that sit just above current-policy projections \citep{VanVuuren2026ScenarioMIPCMIP7}. While this constrained scenario space better reflects plausible emission futures, it creates another issue: the high-end SSP5-8.5 or SSP3-7.0 scenarios often served as proxies for high-impact, low-likelihood outcomes for lower-emission scenarios, and without perturbed-physics ensembles with Earth System Models (ESMs), a narrowed scenario range alone does not provide a basis for tail-risk assessment. Climate risk assessment therefore needs to reflect not only best-estimate climate projections for plausible emission futures and the ESM-specific outliers of those models' best-estimate projections \citep{Bevacqua2026}, but also the broader geophysical uncertainty that an appropriate perturbed-physics ensemble approach through the whole cause-effect chain (e.g. incl. carbon cycle uncertainties) and observational constraints would theoretically show.

The question arises of whether and how new machine-learning and emulation approaches can fill this gap. Weather forecasting is being transformed by observation-driven methods that now rival or surpass physics-based forecasting models \citep{lam_2023_graphcast, bi_2023_pangu, kochkov_neural_2024, price_2024_gencast}. Future climate is not just a replication of the past, though. That is why machine-learning approaches for climate projections will most likely continue to have to rely on process based ESMs in their training. Those machine learning approaches in the climate realm therefore provide a complementary extension, rather than a replacement of ESMs - which is a key difference between the climate and weather realms. ESMs will likely continue to provide the best avenue for scientific knowledge to inform future projections, for example, by investigating non-linear feedback processes, and bio-geo-chemical process interactions.

The main theoretical benefits of a combination of CMIP6-like ESM output, ML approaches and/or emulators is the ability to project climate change a) for new scenarios, b) to support climate impact attribution, i.e. running hypothetical forcing scenarios like natural-only forcings, c) via large ensembles that allow considering the return frequencies of rare events, d) while maintaining consistency across variables to evaluate compound risks in large ensembles \citep[e.g. ][]{bevacqua2023advancing}, e) while being aligned with recent observations, for use in impact models and f) in a way that allows to disentangle ESM model uncertainty and natural variability.
As an example, neither CMIP6-like ESM output nor - to our knowledge - any existing ML emulators jointly satisfy the seven capabilities that the multivariate drought-attribution and projection use case requires (Table~\ref{tab:emulator-comparison-main}). (1) \textit{Variable completeness}: at least the seven covariate variables needed for Penman--Monteith reference evapotranspiration \citep{Allen1998FAO56} and the precipitation-minus-PET water balance, yet many CMIP6 archived runs lack one or more of these, and many ML emulators are restricted to temperature and precipitation. (2) \textit{Free-running operation}: factual and counterfactual climates remain dynamically distinguishable only if the model evolves freely rather than under prescribed sea-surface temperatures or ocean state. (3) \textit{Decomposed forcing} inputs (such as GHG, aerosol, solar, volcanic, ozone) - either via separate input emissions, concentrations or forcings - are essential for constructing natural-only and single-forcing counterfactuals, yet CMIP6 single-forcing experiments exist for only a handful of ESMs. (4) \textit{Stochastic large ensembles}: 100 or more realisations per scenario are key for tail-risk assessment, but the available CMIP6 large ensembles are restricted to a handful of ESMs and a small set of scenarios. (5) \textit{Temporally coherent multi-decadal output}: cumulative indices such as the 48-month Standardised Precipitation Evapotranspiration Index (SPEI-48) require stable low-frequency variability over many decades, which short-window or independently sampled ML emulators do not generally demonstrate. (6) \textit{Present-day observational alignment}: ESM output typically diverges from recent observational data, and bias-correction techniques such as ISIMIP3BASD \citep{lange_biascorrection, lange_2021_isimip3basd} or NEX-GDDP-CMIP6 \citep{thrasher2022nasa} introduce their own limitations in a high-dimensional projection space \citep{maraun_2017_bias, vogel2023evaluation, lehner_etal2023_biasadj}. (7) \textit{Applicability to new scenarios} without re-training: ESM output is often not yet available for the most decision-relevant scenario space (CMIP7, NDC-aligned, current-policy) due to the lag from scenario generation to ESM runs, and many ML emulators are scenario-bound.

These joint capability gaps come on top of structural and parameter uncertainties: ESM selections used at the regional scale face the `hot model' problem \citep{Hausfather2022HotModel} or other unrepresentativeness (e.g.\ being ensembles of best guesses rather than systematic explorations of geophysical uncertainty), which can bias local and regional impact assessments. And while the 'hot model' problem might be temporary given recent studies pointing towards higher warming \citep{Gyuleva2026TCR}, consistently reflecting the assessed knowledge from multiple lines of evidence at the global scale down to regional impact studies will remain a challenge for decades to come.

Multiple machine-learning (ML) approaches to emulate climate have emerged in recent years \citep{deburgh-dayMachineLearningNumerical2023, bodnar_2025_aurora, tebaldi_emulators_2025}, which in principle can overcome the constraints of parameterised approaches given the high dimensionality of the problem. Some ML approaches focus on temporal or spatial downscaling \citep{bassetti_diffesm_2024}. Many hybrid or `pure' ML models act as atmospheric models conditioned on boundary conditions such as sea-surface temperatures \citep{kochkov_neural_2024, wattmeyer_2025}, with recent advances allowing `free-running' projections by coupling the ocean (e.g.\ SamudrACE for the ACE family \citealp{Duncan_2025}) or sufficing with CO$_2$ as forcing predictor \citep{guan_lucie3d}. Foundational weather and climate ML models such as ClimaX \citep{nguyen_climax_2023} and Aurora \citep{bodnar_2025_aurora} are pre-trained on combinations of reanalysis and CMIP6 datasets and predict a broad set of atmospheric state variables. Recent score-based diffusion work on an equal-area HEALPix mesh learns the joint monthly distribution of near-surface temperature, precipitation, relative humidity, and wind speed, efficiently generates stochastic ensembles, and reproduces cross-variable statistics, distribution tails, and responses in held-out SSP2-4.5 and SSP3-7.0 experiments \citep{bouabid2026score}. Its published configuration is trained separately for each parent ESM and draws independent monthly states conditioned on global-mean surface temperature and calendar month rather than temporally coherent trajectories. 

GCMagicc instead introduces explicit temporal dependence through lagged fields at the coarser hierarchy levels and generates continuous monthly rollouts, while providing all seven Penman--Monteith inputs among its ten output variables, training across CMIP6 models and ERA5, and accepting decomposed forcing predictors; this structural difference does not remove the low-frequency and spectral limitations documented in Section~\ref{limitations}. Most closely related to the machine-learning model we present here (GCMagicc), the flow-matching generative emulator ArchesClimate~\citep{clyne_2026_archesclimate} emulates a single Earth system model (IPSL-CM6A-LR) as a free-running coupled system and, notably, transfers to unseen scenarios without retraining, though it targets only that one ESM rather than a multi-model ensemble, with small ensembles and no observational anchoring (Table~\ref{tab:emulator-comparison-main}).

Emulator capabilities are heterogeneous across the published literature more generally \citep{tebaldi_emulators_2025,nath_2025_mercury}; the FASTMIP pilot experiment provides a coordinated intercomparison of spatial emulators against a common protocol \citep{fastmip_protocol}. 

As an illustration of why all seven capabilities matter simultaneously, we apply GCMagicc to the late-2025 Iranian drought state in Section~\ref{drought-attrib}. The 2026 European drought and its likely impact on agricultural yields \citep{JRC147969} is another example of a recent event that would benefit from a similar attribution analysis, but is outside the scope of this study.

\begin{table}[t]
\centering
\footnotesize
\setlength{\tabcolsep}{4pt}
\caption{Capability coverage for the multivariate drought-attribution use case across representative climate emulators and ESM-based workflows. The key-reference column gives representative primary citation(s). Columns: 'Penman-M vars': $\ge 7$ Penman--Monteith variables; 'Free-run': free-running with no prescribed sea-surface temperature or ocean state, i.e., being driven by emission scenarios; 'Dec.ERF': decomposed effective radiative forcing inputs; 'Stoch.': stochastic large-ensemble formulation; 'Multi-dec.': temporally coherent multi-decadal output sufficient for SPEI-48; 'Obs.': present-day observational alignment; 'New scen.': applicability to new scenarios without re-training. \ensuremath{\checkmark}\ supported; \ensuremath{\checkmark^{1}}\ when used in combination with MAGICC (as in this study); \ensuremath{\checkmark^{2}}\ imposed by construction; \ensuremath{\circ}\ partial or conditional support; \ensuremath{\times}\ not supported. For GCMagicc, 'Obs.' is marked as imposed because alignment is imposed for global-mean \textit{tas} by construction while the regional and multivariate alignment is learned and evaluated separately (Section~\ref{sec:Obs-aligned}); 'Free-run' is marked 'supported' because the free-running claim holds for the coupled MAGICC+GCMagicc package, with the conditioning on MAGICC global temperature and heat uptake. The variants GCMagicc-PM and GCMagicc-XS are run based on forcing input only; and 'Multi-dec.' is marked partial because the amplitude and frequency spectra of the low-frequency modes are not yet at parity with the range across CMIP6 ESMs or observations (Section~\ref{limitations}). See Sect.~\ref{sec:supp-emulator-comparison} in the Supplement for detailed per-cell footnotes and citations.}
\label{tab:emulator-comparison-main}
\begin{tabular}{@{}>{\raggedright\arraybackslash}p{0.17\textwidth}>{\raggedright\arraybackslash}p{0.15\textwidth}ccccccc@{}}
\tophline
Emulator/ML model & Key reference & Penman-M\,vars & Free-run & Dec.ERF & Stoch. & Multi-dec. & Obs. & New scen. \\
\middlehline
CMIP6 + ISIMIP3BASD & \citep{Eyring2016CMIP6Design,lange_biascorrection} & \ensuremath{\checkmark} & \ensuremath{\checkmark} & \ensuremath{\circ} & \ensuremath{\circ} & \ensuremath{\checkmark} & \ensuremath{\circ} & \ensuremath{\circ} \\
MESMER family & \citep{beusch_mesmer_2020,nath_mesmer_m_2022,quilcaille_mesmer_x_2022} & \ensuremath{\times} & \ensuremath{\checkmark} & \ensuremath{\circ} & \ensuremath{\checkmark} & \ensuremath{\checkmark} & \ensuremath{\circ} & \ensuremath{\checkmark} \\
MESMER-MAGICC & \citep{beusch_magicc_mesmer_2022} & \ensuremath{\times} & \ensuremath{\checkmark} & \ensuremath{\circ} & \ensuremath{\checkmark} & \ensuremath{\checkmark} & \ensuremath{\circ} & \ensuremath{\checkmark} \\
STITCHES & \citep{tebaldi_stitches_2022} & \ensuremath{\checkmark} & \ensuremath{\checkmark} & \ensuremath{\times} & \ensuremath{\times} & \ensuremath{\circ} & \ensuremath{\times} & \ensuremath{\circ} \\
ClimaX & \citep{nguyen_climax_2023} & \ensuremath{\circ} & \ensuremath{\circ} & \ensuremath{\circ} & \ensuremath{\times} & \ensuremath{\circ} & \ensuremath{\circ} & \ensuremath{\circ} \\
Aurora & \citep{bodnar_2025_aurora} & \ensuremath{\circ} & \ensuremath{\times} & \ensuremath{\times} & \ensuremath{\times} & \ensuremath{\times} & \ensuremath{\checkmark} & \ensuremath{\times} \\
cBottle & \citep{brenowitz_cbottle_2025} & \ensuremath{\circ} & \ensuremath{\circ} & \ensuremath{\times} & \ensuremath{\checkmark} & \ensuremath{\circ} & \ensuremath{\checkmark} & \ensuremath{\times} \\
NeuralGCM & \citep{kochkov_neural_2024} & \ensuremath{\circ} & \ensuremath{\circ} & \ensuremath{\times} & \ensuremath{\circ} & \ensuremath{\circ} & \ensuremath{\checkmark} & \ensuremath{\circ} \\
ACE2 & \citep{wattmeyer_2025} & \ensuremath{\checkmark} & \ensuremath{\circ} & \ensuremath{\times} & \ensuremath{\times} & \ensuremath{\checkmark} & \ensuremath{\checkmark} & \ensuremath{\circ} \\
SamudrACE & \citep{Duncan_2025} & \ensuremath{\checkmark} & \ensuremath{\checkmark} & \ensuremath{\times} & \ensuremath{\times} & \ensuremath{\checkmark} & \ensuremath{\times} & \ensuremath{\times} \\
LUCIE-3D & \citep{guan_lucie3d} & \ensuremath{\circ} & \ensuremath{\checkmark} & \ensuremath{\times} & \ensuremath{\circ} & \ensuremath{\checkmark} & \ensuremath{\checkmark} & \ensuremath{\circ} \\
DiffESM & \citep{bassetti_diffesm_2024} & \ensuremath{\times} & \ensuremath{\times} & \ensuremath{\times} & \ensuremath{\checkmark} & \ensuremath{\circ} & \ensuremath{\circ} & \ensuremath{\circ} \\
Score-based ESM emulator & \citep{bouabid2026score} & \ensuremath{\circ} & \ensuremath{\times} & \ensuremath{\times} & \ensuremath{\checkmark} & \ensuremath{\times} & \ensuremath{\times} & \ensuremath{\checkmark} \\
ArchesClimate-SSP & \citep{clyne_2026_archesclimate} & \ensuremath{\circ} & \ensuremath{\checkmark} & \ensuremath{\circ} & \ensuremath{\circ} & \ensuremath{\checkmark} & \ensuremath{\times} & \ensuremath{\checkmark} \\
\middlehline
\textbf{GCMagicc (this work)} & This work & \ensuremath{\checkmark} & \ensuremath{\checkmark^{1}} & \ensuremath{\checkmark} & \ensuremath{\checkmark} & \ensuremath{\circ} & \ensuremath{\checkmark^{2}} & \ensuremath{\checkmark} \\
\bottomhline
\end{tabular}
\end{table}

We provide projections for CMIP6 SSPs, NDC-related pathways reflecting NDC and long-term target ambitions (Section~\ref{sec:variablematches}), and the final CMIP7 ScenarioMIP pathways H, HL, M, ML, L, VL, and LN \citep{VanVuuren2026ScenarioMIPCMIP7}. An interactive data supplementary is available at gcmagicc.org, providing 5.6 TB data for further research, including monthly, seasonal maps and statistical comparisons for all 10 variables.

\section{GCMagicc model overview}\label{ml}

GCMagicc is a generative multivariate climate emulator trained on 32 CMIP6 models and ERA5 reanalysis. GCMagicc is focused on monthly timesteps and 10 impact-relevant geophysical variables (\textit{tas}, \textit{pr}, \textit{tasmax}, \textit{tasmin}, \textit{hurs}, \textit{huss}, \textit{psl}, \textit{rsds}, \textit{sfcWind}, \textit{ts}). 

To capture cross-model and assessed warming uncertainty beyond what a single model framework can deliver, we couple GCMagicc with the simple climate model MAGICC \citep{meinshausen_magicc_2011, Nicholls2022}, which provides probabilistic global-mean temperature, global-mean heat uptake, and decomposed forcing predictors (Sect.~\ref{sec:supp-magicc-setup} in the Supplement). Trained on 32 CMIP6 models and ERA5 reanalysis, GCMagicc is evaluated against ERA5 historical conditions and held-out CMIP6 output (Section~\ref{val}); its global-mean response distribution is aligned with the IPCC AR6 assessed warming distribution by predictor construction. Residual offsets of up to 0.46\,$^{\circ}$C at the tails of the assessed IPCC range for some scenarios are still present as the match between the MAGICC predictor distribution and the IPCC assessed warming ranges is imperfect, even though MAGICC is one of the emulators matching the IPCC assessment's multiple lines of evidence the closest(Cross Chapter Box 7.1 in \citealt{forster_2021_ar6_wgi_ch7} and Section~\ref{sec:Obs-aligned}).

The model is hierarchically structured on a HEALPix grid \citep{gorski2005healpix}, with the coarsest layer solved first using the full predictor set and subsequent finer layers using appropriate subsets of predictors and temporal lags (Figure \ref{fig:gcmagicc-model-architecture} and Sect.~\ref{sec:supp-gcmagicc-predictors} in the Supplement). The HEALPix resolution parameter \textit{nside} controls output grid resolution: \textit{nside}~$=$~64 yields an approximate $1^{\circ}\times 1^{\circ}$ grid, while \textit{nside}~$=$~256 yields approximately $0.25^{\circ}\times 0.25^{\circ}$ (without the enhanced gridpoint density at higher latitudes). GCMagicc applies without additional training to scenarios whose forcing trajectories fall within or near the CMIP6 training envelope; the reduced-predictor variant GCMagicc-XS, which holds \textit{ssp585} out of training, provides the high-end probe of that envelope (Sect.~\ref{sec:supp-extrapolation-XS} in the Supplement), while behaviour under forcing compositions outside it, such as strongly aerosol-perturbed or solar-radiation-modification pathways, is untested here.

\begin{figure*}[p]
\centering
\includegraphics[width=\textwidth,height=0.82\textheight,keepaspectratio]{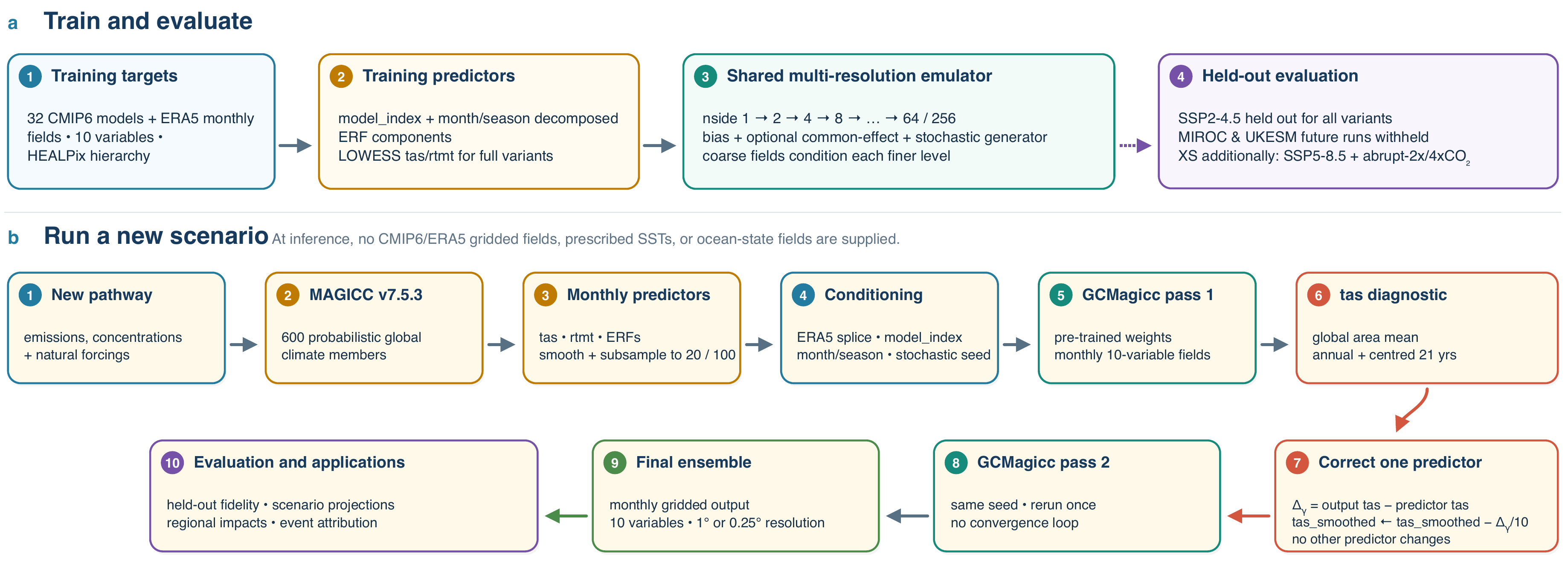}
\caption{\textbf{GCMagicc v1 workflow for training and inference.} \textbf{(a)} Training uses monthly fields from 32 CMIP6 models and ERA5 together with model, seasonal, forcing and -- for the full-predictor variants GCMagicc and GCMagicc-CE -- smoothed global-temperature and heat-uptake predictors. All four variants share the hierarchical HEALPix generator but differ in predictors, training design and evaluation purpose; dashed arrows denote held-out evaluation rather than training input. \textbf{(b)} For a new pathway, the default GCMagicc chain requires scenario emissions or concentrations and natural forcings, from which MAGICC generates a probabilistic global predictor ensemble. No gridded ESM or reanalysis fields, prescribed sea-surface temperatures, or ocean-state fields are supplied at inference. GCMagicc and GCMagicc-CE use exactly two passes: the first diagnoses annual global-area-weighted \textit{tas} bias after centred 21-year smoothing; only smoothed \textit{tas} is adjusted in normalized predictor units before one same-seed rerun. GCMagicc-PM and GCMagicc-XS omit the two smoothed global predictors and use one emulator pass. Final output comprises monthly stochastic fields for \textit{tas}, \textit{pr}, \textit{tasmax}, \textit{tasmin}, \textit{hurs}, \textit{huss}, \textit{psl}, \textit{rsds}, \textit{sfcWind}, and \textit{ts}. Predictor definitions and variant configurations are detailed in Sect.~\ref{sec:supp-gcmagicc-predictors} in the Supplement.}
\label{fig:training-inference-workflow}
\end{figure*}

Predictors are aerosol, solar, volcanic, ozone, GHG and total effective radiative forcings, alongside lowpass-filtered global-mean surface air temperature and global-mean top-of-atmosphere heat uptake. Predictors are produced by the simple climate model MAGICC \citep{meinshausen_magicc_2011} usually in its IPCC AR6 calibration (Cross-Chapter Box 7.1 of \citealp{forster_2021_ar6_wgi_ch7}; \citealp{Nicholls2022}), if not stated otherwise. For a general overview of the GCMagicc training and inference workflow, see Fig.~\ref{fig:training-inference-workflow}.

\begin{figure*}[p]
\centering
\includegraphics[width=\textwidth,height=0.82\textheight,keepaspectratio]{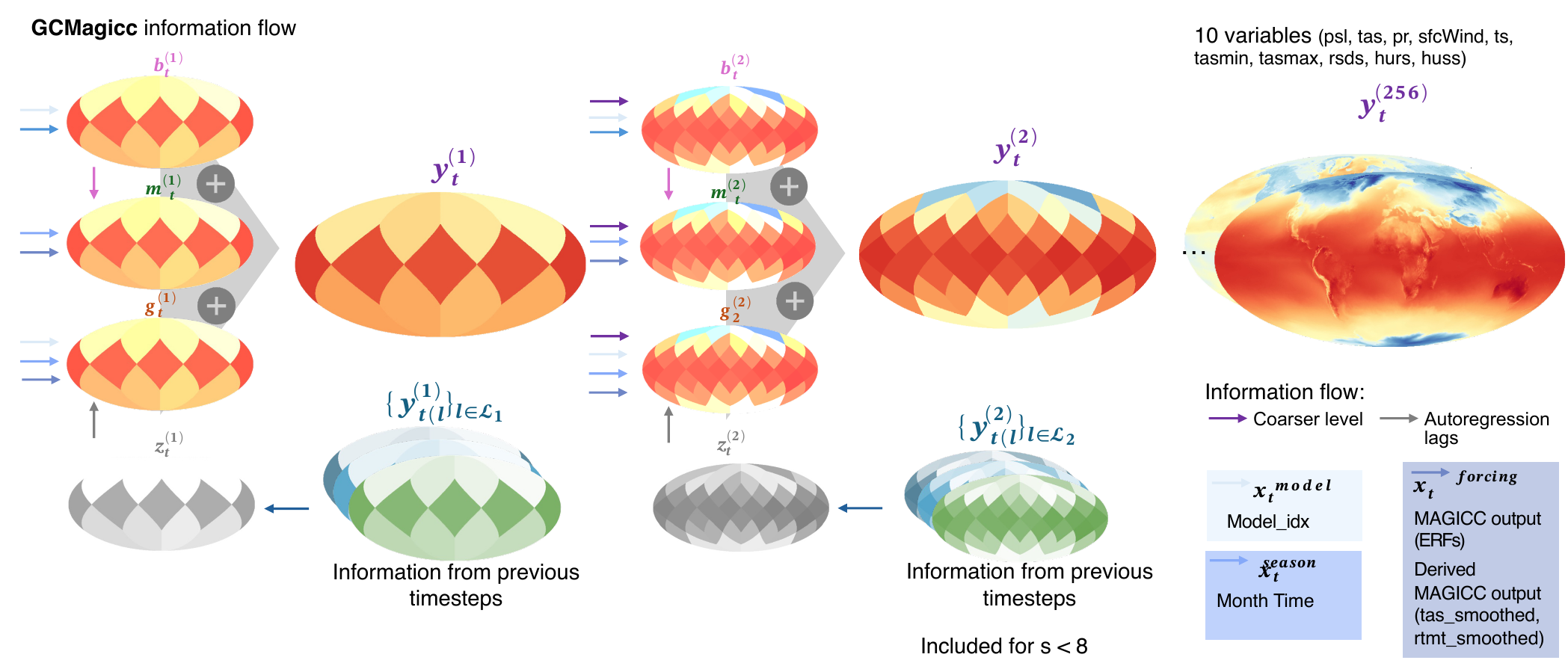}
\caption{\textbf{Hierarchical information flow in GCMagicc.} At each HEALPix resolution $s$, the normalized ten-variable field $\textit{y}^{(s)}_t$ combines a model- and season-dependent bias term $\textit{b}^{(s)}_t$, an optional forcing-driven common-effect term $\textit{e}^{(s)}_t$, and a stochastic generator $\textit{g}^{(s)}_t$ driven by latent input $\textit{z}^{(s)}_t$. Finer levels condition on the contemporaneous coarser field, while temporal dependence at $s\leq8$ is introduced through lagged fields. The model is conditioned on model identity, seasonal covariates, and forcing predictors, including decomposed effective radiative forcings and, for the full-predictor variants, MAGICC-derived smoothed global temperature and heat uptake. The hierarchy proceeds from $s=1$ to $s=256$ and jointly generates $\textit{psl}$, $\textit{tas}$, $\textit{pr}$, $\textit{sfcWind}$, $\textit{ts}$, $\textit{tasmin}$, $\textit{tasmax}$, $\textit{rsds}$, $\textit{hurs}$, and $\textit{huss}$.}
\label{fig:gcmagicc-model-architecture}
\end{figure*}

\section{Methods}

\subsection{Data}

\textbf{Data sources and experiments.}
We use gridded monthly output from CMIP6 Earth system models and ERA5 reanalysis \citep{Eyring2016CMIP6Design,hersbach2020era5}, cited as data at the model/MIP granularity in Table~\ref{tab:cmip6-dois} in the Supplement and in \citet{era5_dataset} respectively. The CMIP6 runs cover idealized $\mathrm{CO_2}$ experiments, historical runs (all-forcing, GHG forcing and natural-only-forcing), and ScenarioMIP projections \citep{ONeill2016ScenarioMIP,gillett2016damip}. Specifically, the experiments used are \textit{1pctCO2}, \textit{abrupt-2xCO2}, \textit{abrupt-4xCO2}, \textit{hist-aer}, \textit{hist-GHG}, \textit{hist-nat}, \textit{historical}, \textit{ssp119}, \textit{ssp126}, \textit{ssp370}, \textit{ssp434}, \textit{ssp460}, \textit{ssp534-over}, and \textit{ssp585}. For ERA5, historical monthly surface data from 1940 to 2024 are used \citep{hersbach2020era5}. Table~\ref{tab:runs} in the Supplement lists all model-scenario combinations of 32 CMIP6 ESMs and the number of ensemble members that were used.

\textbf{Spatial grids and spherical discretization.}
All CMIP6 fields are represented on a regular $1^{\circ}\times 1^{\circ}$ latitude-longitude grid with $360\times 180=64{,}800$ global grid cells. ERA5 fields are additionally used on their native $0.25^{\circ}\times 0.25^{\circ}$ grid \citep{hersbach2020era5}, which for global coverage including the poles corresponds to $1440\times 721=1{,}038{,}240$ grid points. Latitude-longitude grids are mapped to a HEALPix discretization of the sphere with RING indexing \citep{gorski2005healpix}. The default resolution is \textit{nside}$\,=64$, implying $N_{\text{pix}}=12\,\textit{nside}^2=49{,}152$ equal-area pixels; the configuration \textit{nside}$\,=256$ is supported for higher-resolution experiments, for which no evaluation is presented in this paper and for which ERA5 is the only training source at that resolution, implying $N_{\text{pix}}=12\,\textit{nside}^2=786{,}432$ \citep{gorski2005healpix}.
HEALPix provides equal-area pixels, hierarchical resolution changes, and neighbourhood operations without latitude-dependent cell areas, which makes it suitable for the model's multi-resolution factorization.

\textbf{Target variables.}
We use model output for the F=10 impact-relevant variables that the emulator predicts: Near-Surface Air Temperature (\textit{tas}), Precipitation (\textit{pr}), Daily Maximum Near-Surface Air Temperature (\textit{tasmax}), Daily Minimum Near-Surface Air Temperature (\textit{tasmin}), Near-Surface Relative Humidity (\textit{hurs}), Near-Surface Specific Humidity (\textit{huss}), Sea Level Pressure (\textit{psl}), Surface Downwelling Shortwave Radiation (\textit{rsds}), Near-Surface Wind Speed (\textit{sfcWind}), Surface Temperature (\textit{ts}); see Table~\ref{tab:supp-affine} in the Supplement.

\textbf{Affine normalization and inverse transform.}
To bring all target variables to roughly zero mean and unit variance, each variable $\textit{y}^{(v)}$ is affinely transformed before training, $\tilde{\textit{y}}^{(v)} = a_v\,\textit{y}^{(v)} + b_v$, and inverted by $\textit{y}^{(v)} = (\tilde{\textit{y}}^{(v)} - b_v)/a_v$. The coefficients are listed in Table~\ref{tab:supp-affine} in the Supplement.

\subsection{GCMagicc Predictors and Model Architecture}\label{sec:predictors-architecture}

\textbf{GCMagicc versions.}
We describe four versions of GCMagicc, denoted GCMagicc, GCMagicc-CE (common-effect), GCMagicc-PM (perfect-model), and GCMagicc-XS (extrapolation-scenario). They share the multi-resolution HEALPix architecture and training framework described below, and differ in (i) their predictor set, (ii) whether a common-effect stage is included, (iii) the lag sets used at the coarsest resolutions, and (iv) which scenarios are held out from training. GCMagicc and GCMagicc-CE use smoothed global-mean indicators derived from MAGICC as predictors and are trained jointly on many CMIP6 models and scenarios to produce calibrated large-ensemble projections; GCMagicc-CE additionally includes a common-effect stage at coarse resolutions. GCMagicc-PM and GCMagicc-XS share a reduced predictor set that omits the MAGICC-derived smoothed global indicators (\textit{tas\_smoothed} and \textit{rtmt\_smoothed}) but retains the ERF components, model index and seasonal descriptors otherwise used by GCMagicc. GCMagicc-PM is retrained per CMIP6 model on that model's historical run alone and supports the perfect-model uncertainty experiment (Sect.~\ref{sec:supp-perfect-model} in the Supplement) and the leave-one-family-out cross-validation. GCMagicc-XS is trained so that it can be a high-end-of-warming scenario-extrapolation test. The differences are summarized in Table~\ref{tab:gcmagicc-variants}.

\textbf{Forcing and conditioning variables.}
The emulator is conditioned on global descriptors and forcing covariates \textit{model\_index}, \textit{month}, \textit{time}, \textit{tas\_smoothed}, \textit{rtmt\_smoothed}, \textit{stratO3\_ERF}, \textit{sol\_ERF}, \textit{other\_ERF}, \textit{volc\_ERF}, \textit{nat\_ERF}, \textit{totalO3\_ERF}, \textit{GHG\_ERF}, \textit{aer\_ERF}, and \textit{CO2\_ERF}. Here \textit{model\_index} is a categorical identifier of the driving CMIP6 model, so the emulator can learn systematic biases specific to each model. The calendar month \textit{month} is encoded as an integer $m\in\{1,\dots,12\}$, and \textit{time} denotes a sinusoidal seasonal embedding given by $\sin(2\pi m/12)$ and $\cos(2\pi m/12)$. GCMagicc-PM and GCMagicc-XS omit \textit{tas\_smoothed} and \textit{rtmt\_smoothed} and condition only on the remaining predictors above (i.e.\ the ERF components and the time and model descriptors); GCMagicc and GCMagicc-CE use the full predictor set. The remaining predictors are decomposed effective radiative forcing (ERF) components, that is, the net radiative forcing including rapid adjustments \citep{smith2020erf}.

\textbf{Multi-resolution factorization.}
Let $\mathcal{S}=\{1,2,4,8,16,32,64,128,256\}$ denote the HEALPix resolution levels with $P_s=12s^2$ pixels at level $s$, and let $\textit{y}^{(s)}_t\in\mathbb{R}^{F\times P_s}$ be the normalized multivariate field at month $t$ with $F=10$ channels. Let the predictor vector be decomposed as
\begin{equation}
\textit{x}_t=\big(\textit{x}^{\textit{model}}_t,\ \textit{x}^{\textit{season}}_t,\ \textit{x}^{\textit{forcing}}_t\big),
\label{eq:x_decomp}
\end{equation}
where $\textit{x}^{\textit{model}}_t$ contains the model-index information, $\textit{x}^{\textit{season}}_t$ contains seasonal covariates, and $\textit{x}^{\textit{forcing}}_t$ contains the forcing predictors, including smoothed global indicators (in GCMagicc and GCMagicc-CE only) and ERF components.

\textbf{Broad outline (shared structure across variants and dependence versions).}
At each resolution, the model is a superposition of three components: a bias mean term, an optional common-effect mean term, and a stochastic generator. For the coarsest resolution $s=1$, where there is no lower-resolution input, we write
\begin{equation}
\widehat{\textit{y}}^{(1)}_t ={} \textit{b}_1(\textit{x}^{\textit{model}}_t,\textit{x}^{\textit{season}}_t) \; + \; \textit{e}_1(\textit{x}^{\textit{season}}_t,\textit{x}^{\textit{forcing}}_t,\textit{b}_1(\textit{x}^{\textit{model}}_t,\textit{x}^{\textit{season}}_t)) 
 \; + \; \textit{g}_1(\textit{x}_t,\textit{z}^{(1)}_t),
\label{eq:outline_s1}
\end{equation}
and for higher resolutions $s>1$, conditioning on $\textit{y}^{(s/2)}_t$, we write
\begin{equation}
\widehat{\textit{y}}^{(s)}_t ={} \textit{b}_s(\textit{y}^{(s/2)}_t,\textit{x}^{\textit{model}}_t,\textit{x}^{\textit{season}}_t) \;  + \; \textit{e}_s(\textit{x}^{\textit{season}}_t,\textit{x}^{\textit{forcing}}_t,\textit{b}_s(\textit{y}^{(s/2)}_t,\textit{x}^{\textit{model}}_t,\textit{x}^{\textit{season}}_t)) \; + \; \textit{g}_s(\textit{y}^{(s/2)}_t,\textit{x}_t,\textit{z}^{(s)}_t).
\label{eq:outline_shigh}
\end{equation}
Temporal dependence enters only through the latent input $\textit{z}_t$, specified as
\begin{equation}
\textit{z}^{(s)}_t =
\left\{
\begin{array}{ll}
\varepsilon_t, \qquad \varepsilon_t\sim p(\varepsilon), & \text{(i.i.d.\ sampling)},\\[4pt]
\textit{r}_s\!\Big(\{\widehat{\textit{y}}^{(s)}_{t-\ell}\}_{\ell\in\mathcal{L}_s},\eta_t\Big), \qquad \eta_t\sim p(\eta), & \text{(autoregressive)},
\end{array}
\right.
\label{eq:outline_z_cases}
\end{equation}
ensuring that $\textit{r}_s$ has marginally the same distribution as $\varepsilon_t$. Temporal dependence is enabled only at the coarse resolutions $s\le 8$, with $\mathcal{L}_s=\varnothing$ for $s\ge 16$. Unless noted otherwise in Table~\ref{tab:gcmagicc-variants}, the lag sets are $\mathcal{L}_1=\mathcal{L}_2=\mathcal{L}^{\star}$, $\mathcal{L}_4=\{1,\dots,5\}$, and $\mathcal{L}_8=\{1\}$, where $\mathcal{L}^{\star}$ is defined in the table footnote. The common-effect term $\textit{e}_s$ is active in GCMagicc-CE at $s\in\{1,2,4,8\}$ and is omitted at every resolution in GCMagicc, GCMagicc-PM and GCMagicc-XS, as well as at $s\ge 16$ in all versions.

\begin{table}[ht]
\centering
\caption{Configurations distinguishing the four versions of GCMagicc. All four share the multi-resolution HEALPix architecture and training framework; at HEALPix levels $s\ge 16$ they are identical.}\label{tab:gcmagicc-variants}
\small
\renewcommand{\arraystretch}{1.2}
\begin{tabularx}{\columnwidth}{@{}>{\raggedright\arraybackslash}p{0.12\columnwidth}>{\raggedright\arraybackslash}p{0.20\columnwidth}>{\raggedright\arraybackslash}p{0.12\columnwidth}>{\raggedright\arraybackslash}X>{\raggedright\arraybackslash}p{0.16\columnwidth}@{}}
\tophline
Variant & Predictors & Common-effect $\textit{e}_s$ active at & Training data and held-out scenarios & Primary use \\
\middlehline
GCMagicc & full & (none) & 32 CMIP6 models + ERA5, joint multi-scenario; ERA5, UKESM and MIROC contribute historical data only; \textit{ssp245} held out as in-distribution test & default projections \\
GCMagicc-CE & full & $s\in\{1,2,4,8\}$ & same as GCMagicc & sensitivity of evaluation diagnostics to modelling-architecture assumptions \\
GCMagicc-PM & no MAGICC-derived indicators: model index, seasonality and forcing predictors only & (none) & retrained independently on each CMIP6 model's historical run; per-model instances plus one instance per model family (13 families; Table~\ref{tab:supp-emergent-families} in the Supplement) & perfect-model uncertainty; LOFO cross-validation \\
GCMagicc-XS & no MAGICC-derived indicators: model index, seasonality and forcing predictors only & (none) & 32 CMIP6 models + ERA5, joint multi-scenario; \textit{abrupt-4xCO2}, \textit{ssp534-over} and \textit{ssp585} held out additionally to \textit{ssp245}; evaluated on \textit{ssp585} and \textit{abrupt-4xCO2} (\textit{ssp534-over}: too few vetted runs) & high-end scenario extrapolation test \\
\bottomhline
\end{tabularx}

\vspace{3pt}
\footnotesize\parbox{\columnwidth}{\raggedright Variant suffixes: \mbox{-CE} = common-effect (joint cross-model forcing-response stage), \mbox{-PM} = perfect-model (per-CMIP6-model historical-only retraining for structural-uncertainty quantification), \mbox{-XS} = extrapolation-scenario (joint training with high-end and idealised scenarios held out). Predictor sets: the \emph{full} set adds the MAGICC-derived smoothed global indicators \path{tas_smoothed} and \path{rtmt_smoothed} on top of the model index, seasonality and forcing (ERF-component) predictors; the reduced set used by GCMagicc-PM and GCMagicc-XS omits these two MAGICC-derived indicators. Lag sets at the coarsest resolutions: $\mathcal{L}_1=\mathcal{L}_2=\mathcal{L}^{\star}$ for GCMagicc, GCMagicc-PM and GCMagicc-XS, and $\mathcal{L}_1=\mathcal{L}_2=\{1,\dots,47\}$ for GCMagicc-CE, with $\mathcal{L}^{\star}=\{1,2,3,4,5,6,\ 8,10,12,14,16,18,20,\ 24,27,30,33,36,\ 42,48,54,60,66,72\}$.}
\end{table}

\textbf{Stage 1: deterministic bias model (all versions).}
For $s>1$ the bias term $\textit{b}_s$ is a local neighbor upsampling rule with model- and month-dependent additive bias fields. Let $\textit{c}_t$ be the model index encoded in $\textit{x}^{\textit{model}}_t$ and let $\textit{m}_t\in\{1,\dots,12\}$ be the month encoded in $\textit{x}^{\textit{season}}_t$; for each high-resolution pixel $p\in\{1,\dots,P_s\}$ define $\mathcal{N}_s(p)\subseteq\{1,\dots,P_{s/2}\}$ as the set of $k=16$ nearest-neighbor pixels in the lower-resolution grid, where nearest neighbors are determined by Euclidean distance between unit vectors on $\mathbb{S}^2$. Then for each channel $f\in\{1,\dots,F\}$,
\begin{equation}
\widehat{\textit{b}}^{(s)}_{t,f,p}
=\sum_{j\in\mathcal{N}_s(p)} \alpha^{(s)}_{\textit{c}_t,f,p,j}
\Big(\textit{y}^{(s/2)}_{t,f,j}-\beta^{(s/2)}_{\textit{c}_t,\textit{m}_t,f,j}\Big)
+\beta^{(s)}_{\textit{c}_t,\textit{m}_t,f,p},
\label{eq:bias}
\end{equation}
where $\beta^{(s/2)}_{\textit{c},\textit{m},f,\cdot}$ and $\beta^{(s)}_{\textit{c},\textit{m},f,\cdot}$ are learnable bias fields on the low- and high-resolution grids, respectively, and $\alpha^{(s)}_{\textit{c},f,p,\cdot}$ are learned neighbor weights satisfying $\sum_{j\in\mathcal{N}_s(p)}\alpha^{(s)}_{\textit{c},f,p,j}=1$, implemented via a softmax over learned logits. For $s=1$ the bias is purely an additive model- and month-specific mean field over the 12 pixels, $\widehat{\textit{b}}^{(1)}_{t,f,p}=\beta^{(1)}_{\textit{c}_t,\textit{m}_t,f,p}$. For $s\ge 128$, the bias model drops dependence on the model index $\textit{c}_t$, the neighborhood size is reduced to $k=9$, and the generator uses $k=6$ neighbors.

\textbf{Stage 2: deterministic common-effect model (GCMagicc-CE only).}
The common-effect term $\textit{e}_s$, used only in GCMagicc-CE at $s\in\{1,2,4,8\}$, models a forcing-driven mean shift that is shared across all models by excluding $\textit{x}^{\textit{model}}_t$ from its inputs. It is implemented as a per-pixel residual multi-layer perceptron (MLP) shared across pixels, evaluated on the concatenation of $\textit{x}^{\textit{season}}_t$, $\textit{x}^{\textit{forcing}}_t$, pixel coordinates $(\theta_p,\phi_p)$, and the current per-pixel baseline state $\widehat{\textit{b}}^{(s)}_{t,\cdot,p}$. A concrete specification is
\begin{equation}
\widehat{\textit{e}}^{(s)}_{t,\cdot,p}
= \textit{h}_s\!\left(\textit{x}^{\textit{season}}_t,\textit{x}^{\textit{forcing}}_t,\theta_p,\phi_p,\widehat{\textit{b}}^{(s)}_{t,\cdot,p}\right),
\qquad
\widehat{\mu}^{(s)}_{t,\cdot,p}=\widehat{\textit{b}}^{(s)}_{t,\cdot,p}+\widehat{\textit{e}}^{(s)}_{t,\cdot,p},
\label{eq:effect}
\end{equation}
where $\textit{h}_s$ is shared across all pixels. As aforementioned, this stage is active only in GCMagicc-CE at $s\in\{1,2,4,8\}$; at $s\ge 16$ in GCMagicc-CE, and at every resolution in GCMagicc, GCMagicc-PM and GCMagicc-XS, it is omitted and the baseline mean reduces to $\widehat{\mu}^{(s)}_t=\widehat{\textit{b}}^{(s)}_t$.

\textbf{Stage 3: stochastic generator (all versions).}
The generator $\textit{g}_s$ adds stochastic residuals to the baseline mean $\widehat{\mu}^{(s)}_t$ and is used in all four versions. The multi-resolution stochastic generation across successive HEALPix levels follows the reverse Markov learning framework of \citet{shen2025reverseMarkov}. All $F$ channels, with one channel per geophysical variable (tas, pr, hurs, etc.), are modeled jointly. The overall form is
\begin{equation}
\widehat{\textit{y}}^{(s)}_t
= \widehat{\mu}^{(s)}_t + \textit{g}_s\!\left(\textit{y}^{(s/2)}_t,\textit{x}_t,\textit{z}_t\right),
\label{eq:gen}
\end{equation}
where for $s=1$ the argument $\textit{y}^{(s/2)}_t$ is omitted and $\textit{z}_t$ is specified by Eq.~(\ref{eq:outline_z_cases}). Internally, $\textit{g}_s$ collects features from the $k$ nearest lower-resolution neighbors of each pixel for $s>1$, applies two successive learned linear mixing maps per pixel, concatenates the result with a broadcast forcing embedding and noise fields, and feeds everything into a shared per-pixel MLP that predicts the additive residual.

\textbf{Multi-layer Perceptron (MLP) parameterization.}
All MLPs use ELU activations. Unless stated otherwise, each MLP has four linear layers (three hidden layers) and hidden width 32. The common-effect MLP $\textit{h}_s$ has the same depth but hidden width 64 at all resolutions where it is active. For $s\le 8$, the generator contains a forcing encoder MLP with four layers, hidden width 32, and five additional i.i.d.\ noise input dimensions, producing a forcing embedding of dimension 20 broadcast across pixels. At $s\ge 16$, the generator does not use the forcing predictors directly; the forcing information enters only through the bias and common-effect stages. At all resolutions, the generator includes an augmentation noise field of dimension 5 concatenated to the local neighborhood input, and a high-level latent noise field of dimension 5. For the dependent versions at $s\le 8$, the autoregressive replacement map $\textit{r}_s$ consumes auxiliary noise $\eta_t$ and lagged outputs over $\mathcal{L}_s$; for $s\ge 16$ temporal dependence is disabled and $\textit{z}_t=\varepsilon_t$. At $s\in\{1,2\}$, the generator also includes a per-model learned scaling of the first dimensions of the forcing embedding, allowing each model's response to forcings to differ in magnitude.

\subsection{Training}

\textbf{Optimization and losses.}
All models are trained with Adam using learning rate $10^{-4}$ \citep{kingma2015adam}. The bias and common-effect stages are trained with a mean-squared error loss. The stochastic generator is trained with the engression objective based on the energy score \citep{shenmeinshausen2025engression,gneiting_raftery_2007}. The common-effect models (where present) and the stochastic generators are trained for 500 epochs, and the bias models are trained for 10{,}000 epochs.

\textbf{Training data for GCMagicc and GCMagicc-CE.}
We train GCMagicc and GCMagicc-CE separately, but each one of them jointly on 32 CMIP6 models and the ERA5 reanalysis, using a broad cross-section of scenario experiments that varies by CMIP6 model (see Table~\ref{tab:runs} in the Supplement). The bias models $\textit{b}_s$ are trained on historical data only, and separately for each driving CMIP6 model. The common-effect models $\textit{e}_s$ (in GCMagicc-CE) and the stochastic generators $\textit{g}_s$ are trained across all available experiments listed above, with a subsampling scheme that retains at most five runs per model-scenario combination to balance the dataset. The scenario \textit{ssp245} is held out as an unseen-scenario test case in all training steps for GCMagicc and its variants. For the model UKESM1-0-LL and the MIROC family (MIROC-ES2H, MIROC-ES2L, MIROC6), all ScenarioMIP runs are held out and only their historical experiments are used for training, so that we can assess scenario-projection fidelity on these models when GCMagicc is run to emulate them, a setup analogous to the ERA5 case, for which, by definition, only historical training data exist (Sect.~\ref{sec:ERA5-splicing} in the Supplement).

\textbf{Training data for GCMagicc-PM.}
GCMagicc-PM uses the reduced predictor set described above. The baseline GCMagicc-PM emulator is trained on the ERA5 historical reanalysis. In addition, GCMagicc-PM is retrained independently on each CMIP6 model's historical run; these retrained instances are used in the perfect-model uncertainty experiment described in Sect.~\ref{sec:supp-perfect-model} in the Supplement and in the leave-one-family-out cross-validation (Sect.~\ref{sec:supp-emergent-constraint} in the Supplement).

\textbf{Training data for GCMagicc-XS.}
GCMagicc-XS uses the same reduced predictor set as GCMagicc-PM but is trained jointly on the same 32 CMIP6 models and ERA5 as GCMagicc and GCMagicc-CE, on the same multi-scenario data. Training uses \textit{1pctCO2}, \textit{abrupt-2xCO2}, \textit{hist-aer}, \textit{hist-GHG}, \textit{hist-nat}, \textit{historical}, \textit{ssp119}, \textit{ssp126}, \textit{ssp370}, \textit{ssp434} and \textit{ssp460}; the high-end and idealised experiments \textit{abrupt-4xCO2}, \textit{ssp534-over} and \textit{ssp585} are held out additionally to the standard \textit{ssp245} hold-out. At inference time, GCMagicc-XS is run forward on the held-out \textit{ssp585} experiment to check consistency under high-end forcing (Sect.~\ref{sec:supp-extrapolation-XS} in the Supplement).

\textbf{Computational efficiency and data volume.}
A single GCMagicc inference---one ensemble member covering 100 years at monthly resolution on a $1^{\circ}\times 1^{\circ}$ grid for all 10 variables---completes in 90--96\,s of wall-clock time on eight pinned CPU cores of an AMD EPYC 9555 processor, with no GPU (median 93\,s over two warm replicates; 90--182\,s across four replicates on a shared machine, the spread reflecting host contention rather than model variance). Roughly 40\,\% of that time is the HEALPix-to-latitude/longitude interpolation, so members retained on the native $\textit{nside}=64$ mesh cost about half as much, and per-member throughput saturates by about 16 cores, so ensembles are generated by running members concurrently rather than by adding cores to a single member. Peak resident memory is 140\,GB when the full 1200-month rollout at resolution nside64 is generated in one call, or 23\,GB when it is emitted in ten-year chunks, which is the configuration we recommend on memory-limited hardware. For the applications in this study we use 20--100 members per scenario; practical release size is governed primarily by storage and distribution.

\section{Evaluation of GCMagicc vs CMIP6}\label{val}

\subsection{Evaluation metrics}\label{sec:val-metrics}

We use energy distances to compare the full distributions of paired GCMagicc, CMIP6, and observational diagnostics rather than any single distributional moment. Complementary recent validation of a score-based climate emulator similarly evaluates probability distributions, cross-variable correlations, time of emergence, and tail behaviour \citep{bouabid2026score}; the diagnostic suites and normalization used there are not directly equivalent to those used here. To make results comparable across variables and diagnostic diagnostics, we define the energy-distance score $S=\mathrm{ED}(M,C)/\mathrm{ED}(C,Z)$, which normalizes the energy distance (ED) between GCMagicc (M) and its target CMIP6 model (C) by the energy distance between that target (C) and randomly selected other CMIP6 models (Z). Values of $S<1$ indicate that GCMagicc is closer to the target CMIP6 model than the CMIP6 models are to one another. For comparisons with observations, the unavailable observation--observation term is omitted, giving a modified distance with a non-zero perfect-match baseline; the full derivation and comparison-key definitions are provided in Sect.~\ref{sec:supp-energy-distances} in the Supplement.

\subsection{Evaluation findings}

CMIP6 ESM zonal means and global-mean time series (of non-\textit{tas} variables) are generally matched most closely by GCMagicc. The ten diagnostics are not equally independent of the training objective: the six that compare emulated fields directly evaluate a property closely related to the energy-score objective and are therefore consistency checks, while IndicatorFreq, ENSOTelecon, ENSOCharacteristics and HumidityCoupling carry the independent evidence (Sect.~\ref{sec:supp-validation-recipes} in the Supplement). Among the consistency checks, median energy-distance scores are 0.013 for the global-mean time series diagnostic and 0.11 for the zonal-mean diagnostic (Table~\ref{tab:validation-diagnostics-energy-distance} in the Supplement), that is, the GCMagicc--CMIP6 distance is a factor of roughly ten to eighty smaller than the distance between two random CMIP6 ESMs for the same scenario, depending on diagnostic and variable (Fig.~\ref{fig:validation-diagnostics}g,p). The zonal means, seasonal or annual bias maps of the 10th, 50th and 90th percentiles, and histograms of variables within IPCC AR6 regions are matched substantially better ($S\ll 1$) than CMIP6 inter-model differences (Fig.~\ref{fig:validation-diagnostics}d,j,m). For \textit{huss} HumidityCoupling the median score is 1.1, i.e. about the typical CMIP6 inter-model spread, and for \textit{pr} ENSO teleconnections it is 1.6, i.e. at the median GCMagicc is further from the target CMIP6 model than a randomly chosen other CMIP6 model is (Sect.~\ref{sec:supp-validation-recipes} and Table~\ref{tab:validation-diagnostics-energy-distance} in the Supplement); the inter-model spread is itself large for these diagnostics (Fig.~\ref{fig:validation-diagnostics}w,aa). Because GCMagicc is trained per CMIP6 model/ERA5 on historical and scenarios lower and higher than SSP2-4.5, comparison against the middle-of-the-road SSP2-4.5 hold-out scenario should be read as a fidelity check rather than as an extrapolation test. A qualitative consistency check under held-out high-end forcing is reported separately for GCMagicc-XS, for which \textit{ssp585} and \textit{abrupt-4xCO2} are held out from training (Sect.~\ref{sec:supp-extrapolation-XS} in the Supplement), and for the leave-one-family-out cross-validation (Sect.~\ref{sec:supp-emergent-constraint} in the Supplement).

\clearpage
\begin{figure*}[p]
  \centering
  \makebox[\textwidth][c]{\includegraphics[height=0.95\textheight,keepaspectratio]{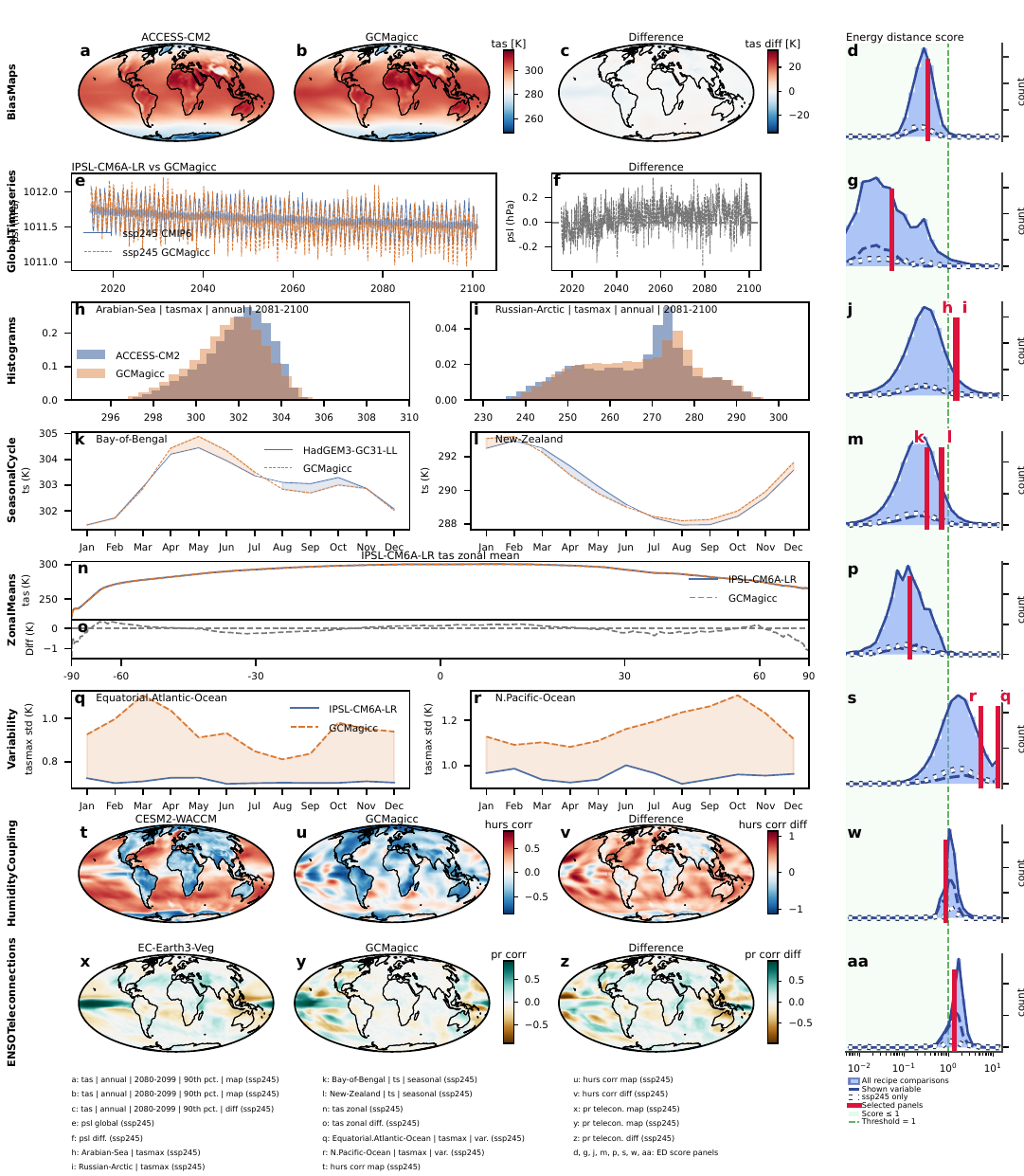}}
\end{figure*}
\clearpage
\begin{figure*}[p]
  \caption{\textbf{Evaluation diagnostics of GCMagicc against CMIP6.} \emph{Evaluation diagnostics for GCMagicc span a range of indicators, here shown in each row (see Sect.~\ref{sec:supp-validation-recipes} in the Supplement for detailed diagnostic definitions). The diagnostics shown in the rows are a random draw from the 1{,}265{,}222 records of the out-of-sample SSP2-4.5 hold-out comparison (Sect.~\ref{sec:supp-validation-recipes} in the Supplement). Relative energy distance scores are shown in the right column (see also Table~\ref{tab:validation-diagnostics-energy-distance} in the Supplement). The energy distance scores are shown for all undertaken fidelity checks (blue solid histogram), for this row's diagnostic (solid red vertical bar), the subgroup for the same variable (dashed histogram) and the subgroup for the hold-out scenario SSP2-4.5 (dotted line). The green dashed line marks an energy-distance score of one, where the GCMagicc--CMIP6 distance equals the distance between two random CMIP6 models; the pale green shading marks the region of scores below one, i.e.\ where GCMagicc matches the target model more closely than a random CMIP6 pair does. Ten additional figures with random draws of fidelity checks for the out-of-sample scenario SSP2-4.5 are provided in the Supplement (Figs.~\ref{fig:supp-validation-alt-001}--\ref{fig:supp-validation-alt-010}).}}
  \label{fig:validation-diagnostics}
\end{figure*}
\clearpage

For the out-of-sample scenario SSP2-4.5, we evaluate the diagnostic distributions that show how close GCMagicc is to the emulated CMIP6 model in comparison to how close two random CMIP6 models are (Figs.~\ref{fig:scoreedistc_ssp245_v101_part1of3}, \ref{fig:scoreedistc_ssp245_v101_part2of3} and \ref{fig:scoreedistc_ssp245_v101_part3of3} in the Supplement).
These methods are not comprehensive, but they intend to capture some of key elements of a future climate that will drive climate impacts, namely mean state, spatial patterns, regional specific seasonality, variability, distributional fidelity and tails, as well as climate mode frequency and persistency behaviour, dynamical fingerprints of teleconnections, and multivariate / compound consistency.

Despite the comprehensiveness of metrics, assessing GCM emulators is difficult because there is no single 'appropriate' set of metrics. Simple RMSE summaries and standard benchmarks such as ClimateBench can over-emphasise a particular characteristic that a given emulator is designed to match \citep{watsonparris_climatebench_2022}. Portrait diagrams based on relative RMSE provide a comparative view across models \citep{gleckler2008performance}, but a single aggregate can hide performance differences among fields. Proper distributional scores such as CRPS and its multivariate energy-score extension assess the forecast distribution rather than only a point summary \citep{hersbach2000crps,gneiting_raftery_2007}.

We therefore use energy-distances $E$, which are the multi-dimensional extension of CRPS, or energy-distance scores $S$ that compare emulator--ESM mismatch to the typical spread across ESMs (Sect.~\ref{sec:supp-energy-distances} in the Supplement). For the 10th, 50th and 90th percentile biases, GCMagicc is generally closer to a given CMIP6 model than a random other CMIP6 model across all variables. For ENSO teleconnections, GCMagicc is in most models no closer than another random ESM in RMSE match of the precipitation and temperature teleconnection patterns to the Ni\~no-3.4 index (Sect.~\ref{sec:supp-validation-recipes} in the Supplement). Global-mean time series are much more closely matched by GCMagicc than by a random other ESM, with the aforementioned caveat that the long-term \textit{tas} evolution is prescribed. Histograms in IPCC AR6 regions are generally well matched, while HumidityCoupling (covariance between near-surface temperature and humidity) is comparable to a random other ESM for \textit{hurs} and worse for \textit{huss}. Seasonal cycles and zonal means are closely matched. The intra-monthly variability is underestimated by GCMagicc; inter-monthly variability is reproduced because the seasonal cycle dominates that signal.

The specific examples of comparison diagnostics against one random draw of out-of-sample SSP2-4.5 comparisons between GCMagicc and CMIP6 ESMs (Figure \ref{fig:validation-diagnostics}) shows a close alignment for annual surface air temperatures 90 percentile of ACCESS-CM2 (Fig. \ref{fig:validation-diagnostics}a-c), with an energy distance slightly worse than the median compared to other bias comparisons (Fig. \ref{fig:validation-diagnostics}d), including for SSP2-4.5 and the \textit{tas} variable. The slightly decreasing trend of surface pressure \textit{psl} in IPSL-CM6A-LR is matched closely - being an example of global-mean variables of a particular ESM being mostly matched much more closely than a random other CMIP6 ESM would (Fig. \ref{fig:validation-diagnostics}e-g). The depicted histograms of aCCESS-CM2 monthly mean maximal temperature \textit{tasmax} distributions for the last twenty years of the 21st century in all grid cells of the Arabian Sea or the Russian Arctic (as in the remainder, those region names refer to the IPCC AR6 regions) are worse than the usual GCMagicc vs ESM fit (see red vertical bars towards higher values in \ref{fig:validation-diagnostics}j compared to the distributions) - which overall for histograms is mostly  better than the match between two random CMIP6 ESMs (Fig. \ref{fig:validation-diagnostics}h-j). The seasonal cycle of surface temperature \textit{ts} is one of the better-matching diagnostics, here shown for Bay of Bengal and New Zealand for the HadGEM3-GC31-LL model (Fig. \ref{fig:validation-diagnostics}k-m). The zonal mean is again very closely matched by GCMagicc for all variables and scenarios, here shown for the \textit{tas} zonal mean of the IPSL-CM6A-LR model (Fig. \ref{fig:validation-diagnostics}n-p). Even for the models for which we only use historical experiments to train, such as the UKESM1-0-LL model, the zonal means are better captured than for a random other CMIP6 ESM (Fig.~\ref{fig:supp-validation-alt-004} in the Supplement). The intra-monthly variability is one of the known limitations of GCMagicc, here shown for \textit{tasmax}, where GCMagicc shows higher variability for the Equatorial Atlantic Ocean and Northern Pacific compared to the IPSL-CM6A-LR example (Fig. \ref{fig:validation-diagnostics}q-s). This randomly picked example is opposite to many others, as usually GCMagicc shows a lower intra-monthly variability. The Humidity Coupling, here for the correlation between \textit{tas} and \textit{hurs} is in average only as well matched by GCMagicc as another random CMIP6 ESM would, so has in fact large divergences from the here picked example from CESM2-WACCM (Fig. \ref{fig:validation-diagnostics}t-w). Likewise, the \textit{pr} teleconnections with the El Niño 3.4 index are in this particular EC-Earth3-Veg vs GCMagicc ensemble of similar structure, but still show large differences in their correlation coefficients (Fig. \ref{fig:validation-diagnostics}x-z) - the shown example is representative of the average GCMagicc vs CMIP6 model match (Fig. \ref{fig:validation-diagnostics}aa).

In addition to these overall energy distance histograms per diagnostic, we also provide a detailed heatmap-like overview of the energy distances per variable and diagnostic. The energy distances are rank-ordered and the absolute energy-distance score $S=1$ sits at the 78.9th percentile of the GCMagicc vs CMIP6 ESM comparisons we show (Figs.~\ref{fig:scoreedistc_ssp245_v101_part1of3}, \ref{fig:scoreedistc_ssp245_v101_part2of3} and \ref{fig:scoreedistc_ssp245_v101_part3of3} in the Supplement). For all the diagnostics and variables with an energy distance score, the considered CMIP6 model is matched better by GCMagicc than by an average other CMIP6 model. In average, this also holds for the two ESM model families for which only the historical experiment was used for training: UKESM1-0-LL (58th percentile) and the MIROC family (45--46th percentile, both still below the 78.9th percentile that indicates $S=1$). The latter motivates the choice from here on of using GCMagicc in its observational setting, with the model-index predictor set to ERA5, for which by definition only historical training data exist.

\section{Applications}\label{sec:applications}

\subsection{Observationally and IPCC AR6 aligned projections}\label{sec:Obs-aligned}
ESM output faces two main challenges as a climate-impact driver: bias against observations and divergence from assessed future-trend ranges that take multiple lines of evidence into account. For example, for future surface air temperature 'tas' projection IPCC ranges were informed by paleoclimatic, observational, process understanding and ESM derived evidence (IPCC AR6 WG1 Chapter 4, \citealp{Lee2021IPCCAR6WG1Ch4}). ESM output is often not aligned with observational data, giving rise to bias-correction schemes such as ISIMIP3BASD \citep{lange_biascorrection}. Individual ESMs are also by construction often not aligned with the IPCC AR6 assessed warming ranges. ESMs mostly present initial condition runs within  ensembles of opportunity multi-model ranges (e.g. \citealp{tebaldi2007multimodel}); Individual ESMs sometimes do not even fall within the 5-95\% warming ranges assessed by IPCC. The median and upper range of the CMIP6 ensemble is up to a degree warmer than the IPCC AR6 assessed range for the higher scenarios, e.g.\ for SSP5-8.5 (`ssp585' in Fig.~\ref{fig:observational-alignment}b).

GCMagicc in its ERA5-spliced mode (Sect.~\ref{sec:ERA5-splicing} in the Supplement) addresses both issues by construction: predictor selection and the two-pass global-\textit{tas} correction align the global response with the MAGICC predictor distribution (which itself is calibrated to represent the IPCC AR6 findings), while the historical splice aligns the temperature predictor with ERA5. The overall finding is that GCMagicc - by construction - is able to produce warmings in line with the IPCC AR6 assessed ranges for the various SSP scenarios (compare the IPCC AR6 ranges and GCMagicc projections for the SSP scenarios in Fig.~\ref{fig:observational-alignment}b).

Subsampling to 20 members preserves the central and 5-95\% range of the AR6 predictor ensemble to within 0.22\,$^{\circ}$C across the tabulated scenarios (the largest difference between the 20-member AR6 and the 100-member 'AR7' columns is 0.39\,$^{\circ}$C, which combines subsampling with the AR6-to-CMIP7 ScenarioMIP change of predictor workflow; the workflow change alone accounts for up to 0.23\,$^{\circ}$C), the largest deviation between the 20-member and 100-member p5, median and p95 values in Table~\ref{tab:observational-alignment-panel-b-values} in the Supplement. The CMIP7 ScenarioMIP workflow adds a wetland methane feedback, which adds to warming \citep{sloughter_2025_methaneEmis}. The AR6 workflow is the default for comparability with assessed IPCC AR6 ranges; CMIP7 ScenarioMIP predictor distributions are shown for context. Note that this CMIP7 ScenarioMIP workflow is the interim distribution to produce GHG concentrations for the CMIP7 cycle, not the calibrated parameter distributions to match IPCC AR7 findings, which might or might not also include higher climate sensitivity and transient climate response distributions \citep[see e.g.,][]{Gyuleva2026TCR}.

The ERA5-only trained GCMagicc-PM projection could potentially provide a 'observational data' only climate projection that, if viable, would be an attractive avenue to pursue as many of the CMIP6 ESM biases towards observations could be circumvented. We tested this avenue and found that our ERA5-only trained GCMagicc-PM has a relatively narrow uncertainty when driven with just the radiative forcing predictors of the SSP scenarios (Fig.~\ref{fig:supp-emergent-lofo}d in the Supplement). That could be mistaken as an observational-data projection uncertainty that narrows future warming uncertainties considerably. However, when considering the perfect-model uncertainty, i.e. training GCMagicc-PM to a particular CMIP6 ESM only based on its historical data - like ERA5 - and then inferring the SSP scenarios from it, the mismatch in future projections is considerable and here illustratively combined with the GCMagicc-PM ERA5-aligned forecast (Fig.~\ref{fig:supp-emergent-lofo}b in the Supplement). That illustrative perfect model uncertainty range in projections then turns out to be comparable to the IPCC assessed range of future warming. Hence, we choose the alternative option of driving our ERA5-trained GCMagicc model with an expanded set of predictors, namely also global-mean \textit{tas} and global-mean \textit{rtnt} predictors from MAGICC - which are calibrated to represent the IPCC assessed ranges (Fig.~\ref{fig:observational-alignment}b).

\subsection{The regional example for IPCC AR6 alignment: Iran's surface air temperatures}\label{sec:Iran-regional-alignment}

We have shown that global-mean warming projections of GCMagicc are aligned with the IPCC AR6 assessed ranges. Here we show how this alignment affects regional projections. The IPCC AR6 did not produce 'assessed' projections at the regional level. Using Iran as an example, we first investigate 24 CMIP6 ESMs for the SSP2-4.5 scenario and consider their historical (1995-2014) and future (2081-2100) averages (Fig.~\ref{fig:observational-alignment}c). The median warming in Iran is 3.2\,$^{\circ}$C for CMIP6 ESMs and 3.1\,$^{\circ}$C for GCMagicc, with 5th-95th percentile ranges of 2.2-4.3\,$^{\circ}$C and 1.1-4.3\,$^{\circ}$C, respectively (Fig.~\ref{fig:observational-alignment}c1,c2). The distributions are not identical, but similar, indicating that GCMagicc captures the regional warming patterns of individual CMIP6 ESMs in Iran under the SSP2-4.5 scenario. When then running GCMagicc in the ERA5-aligned mode with probabilistic MAGICC predictors that represent the IPCC AR6 uncertainty ranges, the median warming in Iran is substantially reduced to only 2.3 (1.6-3.4)\,$^{\circ}$ in line with the global-mean IPCC assessed projection being lower than the median of CMIP6 ESMs (Fig.~\ref{fig:observational-alignment}c3 and b). This example illustrates how regional projections aligned with the global IPCC AR6 assessed ranges might be substantially different in mean warming (here almost one-third) compared to taking the projection medians and ranges from all available CMIP6 ESMs. 

\subsection{Comparing GCMagicc and CMIP6 ESMs with ERA5}\label{sec:ERA5-alignment}

We show energy distances (\ref{sec:val-metrics}) to represent the match of the emulator or ESM with the ERA5 reanalysis data for a range of diagnostics, here averaged per variable (see Figs.~\ref{fig:supp-edisto-observational}--\ref{fig:supp-edisto-observational-part3} in the Supplement for a disaggregated view of diagnostics and variables). The found average energy distances across all diagnostics, averaged by variable, are rank normalized from the 1st (blue) to the 100th (red) percentile across 649{,}058 observational energy distance comparisons (see Figure \ref{fig:observational-alignment}d and Figs.~\ref{fig:supp-edisto-observational}--\ref{fig:supp-edisto-observational-part3} in the Supplement). Under the pooled comparison-weighted ranking, which is the weighting displayed in panel d, the average match of an individual CMIP6 model varies between the 25th percentile (EC-Earth3-AerChem) and the 86th percentile (MIROC-ES2L, least well matching observations) of all CMIP6-ERA5 comparisons (Figure \ref{fig:observational-alignment}d). In comparison, the calibrated GCMagicc emulator achieves average percentile rankings of 17\% for the GCMagicc version and 8\% for the GCMagicc-CE version. Obviously, the emulator had the advantage of being directly calibrated to ERA5 data, while CMIP6 models only indirectly were aligned with observational evidence throughout their various models in the development phase of each ESM.

These rankings are computed over the eight diagnostic groups shown in Figs.~\ref{fig:supp-edisto-observational}--\ref{fig:supp-edisto-observational-part3} in the Supplement, which themselves display the variable-first weighting; the ENSOCharacteristics and IndicatorFreq groups, which cover the oscillation-index variables and frequency spectra identified as GCMagicc's main limitation (Section~\ref{limitations}), are not part of the ranking. These relative scores are dependent upon the relative weighting across the 649{,}058 energy distances. Under the variable-first ranking, which first averages for each variable and then averages across variables, the CMIP6 model range is 39\% to 63\%, with GCMagicc again being near the lower end at 41\% to 44\% for GCMagicc-CE and GCMagicc, respectively (the column percentages printed in Figs.~\ref{fig:supp-edisto-observational}--\ref{fig:supp-edisto-observational-part3} in the Supplement). Percentile rankings are conditional on the chosen diagnostics, and a diagnostic set that would weigh variability and oscillation frequency spectra more would see a deterioration of the GCMagicc-CE relative ranking percentile.

\clearpage
\begin{figure*}[p]
  \centering
  \makebox[\textwidth][c]{\includegraphics[height=0.95\textheight,keepaspectratio]{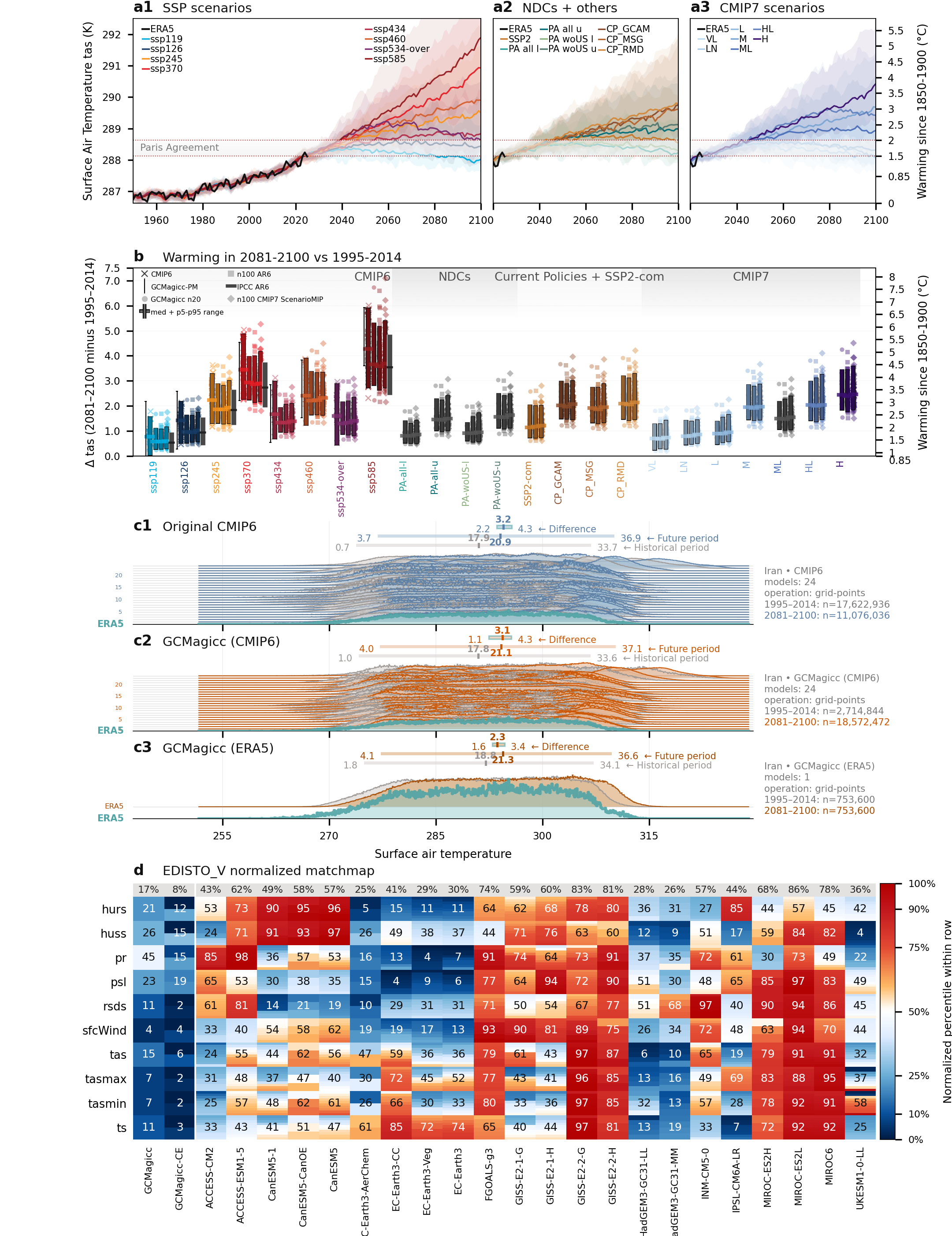}}
\end{figure*}
\clearpage
\begin{figure*}[p]
  \caption{\textbf{Alignment with observations and assessed warming}. \textbf{a1,} Global-mean temperature projections for the SSP scenarios; \textbf{a2,} current-policy and NDC-aligned scenarios; \textbf{a3,} CMIP7 scenarios. In panel b the left axis gives the 2081--2100 minus 1995--2014 change and the right axis the same values re-referenced to 1850--1900 by adding 0.85\,$^{\circ}$C, the IPCC AR6 central estimate of observed 1995--2014 warming \citep{IPCC_AR6_WG1_SPM_2021}; values quoted in the text and in Table~\ref{tab:observational-alignment-panel-b-values} in the Supplement are on the 1850--1900 reference. \textbf{b,} End-of-century (2081--2100) projections for the same scenarios. Where available, raw CMIP6 collections (crosses), IPCC AR6 assessed ranges (black bars), the observationally aligned GCMagicc ensemble (circles), and the AR6/CMIP7 ScenarioMIP MAGICC predictor distributions are shown. The perfect-model structural-uncertainty experiment is shown for the SSP scenarios (black think whiskers). \textbf{c,} Comparison of CMIP6 and GCMagicc calibrations over 1995--2014 and SSP2-4.5 in 2081--2100 for all grid cells in Iran and their respective distribution medians and widths, shown as histograms but also as 5-95th percentile ranges (horizontal bars) for area-weighted averages for the future and historical period and their the distribution of differences across grid-cells. \textbf{d,} Relative energy distance aggregated by variable: the row-wise normalised percentile of the observation-referenced energy distance for ten variables (rows) across GCMagicc, GCMagicc-CE and 22 CMIP6 models (columns); colour gives the within-row percentile from 0 (blue, closest to ERA5) to 100\% (red); the row of percentages above the matrix is the comparison-weighted column mean. Alignment with assessed warming is imposed by the MAGICC predictor distribution and the two-pass global-temperature correction; the remaining regional and multivariate diagnostics are evaluated independently.}
  \label{fig:observational-alignment}
\end{figure*}
\clearpage

\subsection{Scenario projections}\label{sec:variablematches}

We provide GCMagicc projections for all SSPs from the Coupled Model Intercomparison Project Phase-6 (CMIP6) cycle, the final CMIP7 pathways (which are called H (high)/ HL (high-low) / M (medium) / ML (medium-low)/ L (low)/ VL (very low)/ LN (low-negative)), current-policy scenarios from Phase~5.1 NGFS \citep{RichtersEtAl2026NGFSv51}, and NDC-aligned pathways reflecting the status of countries' NDC and LT-LEDS pledges (similar to \citep{meinshausen_realization_2022}) as of 30 June 2026 in low and high variants and with and without US participation in the Paris Agreement (Sect.~\ref{supp-NDCs-LTLEDs} in the Supplement). None of these scenario applications requires retraining of the released GCMagicc configuration (Fig.~\ref{fig:observational-alignment}a2,a3). The reason we choose this broad set of scenarios is that SSPs are currently the main anchor point for climate studies and the IPCC AR6 report, the CMIP7 pathways are likely going to be the new anchor point for the next IPCC AR7 report, and the NDC-aligned and current-policy pathways are the most relevant for near-term policy and climate-impact studies. The NDC-aligned and current-policy pathways are not part of the CMIP6 or CMIP7 scenario sets, so they are a test of GCMagicc's ability to emulate policy-relevant scenarios and were only recently created in the case of NDCs with a cutoff as of 30 June 2026.

As a first illustrative example, Fig.~\ref{fig:observational-alignment}a shows global-mean near-surface air temperature projections for the SSP scenarios, the NDC-aligned variants, and the current-policy pathways. The 2081--2100 warming range relative to 1850--1900, taken as the minimum of the 5th percentiles and the maximum of the 95th percentiles across the displayed scenarios, spans 1.2--6.1$^{\circ}$C for the SSP scenarios, 1.5--4.0$^{\circ}$C for the current policy and NDC-aligned scenarios, and 1.2--4.2$^{\circ}$C for the CMIP7 set of marker scenarios VL to H (see Table~\ref{tab:observational-alignment-panel-b-values} in the Supplement).

The new CMIP7 scenario generation reflects the new policy-realities, i.e. that on the lower end, the projected warming space is shifted slightly upwards, as the category of 'no and low overshooting of 1.5$^\circ$C' scenarios is not considered feasible anymore. The starker change is however at the top end of the scenarios. Given that the current policy scenarios render the previous high end scenarios (RCP 8.5 and SSP5-8.5 as well as SSP3-7.0) rather implausible - unless a major reversal of policies happens that act against the strong economic forces that are in favour of renewable energies - the former upper end is not considered plausible anymore. That can be taken as a success story of international climate action, while at the same time missing the opportunity to stay below 1.5$^\circ$C is a failure.

\begin{figure*}[p]
  \centering
  \includegraphics[width=\textwidth,height=0.79\textheight,keepaspectratio]{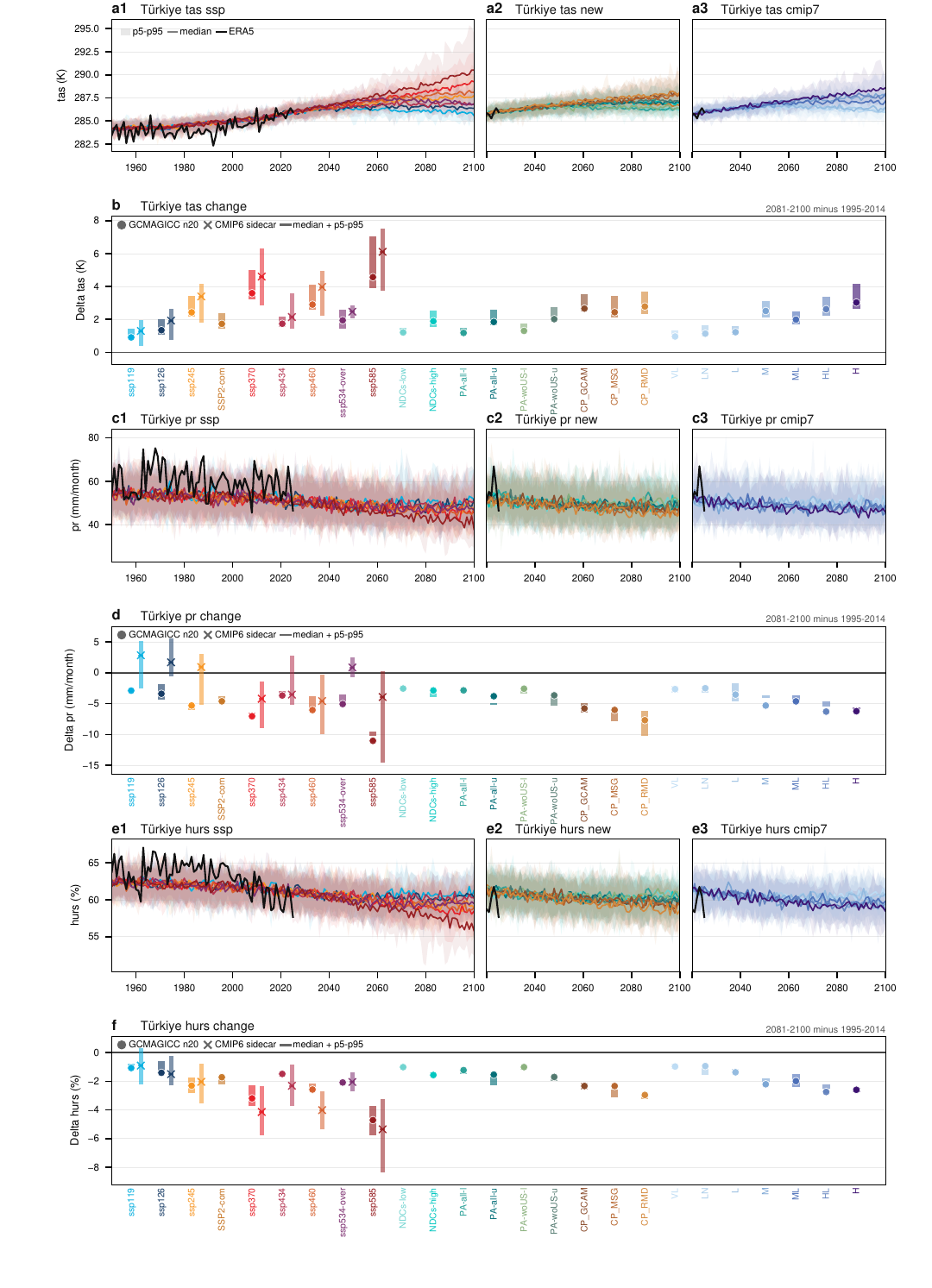}
  \caption{\textbf{Multi-variable regional scenario example for T\"urkiye.} Rows show annual near-surface air temperature (a,b), precipitation (c,d), and relative humidity (e,f). Columns show the eight CMIP6 SSP scenarios (left); SSP2-com, four NDC-aligned variants, and three current-policy pathways (middle); and all seven CMIP7 scenarios (right). Coloured lines are GCMagicc medians, light shading is the 5--95\% ensemble range, and the black line is ERA5 through 2025. The relative change between the 2081-2100 period and the 1995-2014 baseline is shown in panels b, d and f (dot medians and 5-95\% ranges) for GCMagicc, including the same diagnostic across CMIP6 models (cross and 5-95\% ranges)}
  \label{fig:turkiye-regional-scenarios}
\end{figure*}

We already considered surface air temperatures in Iran as a regional example. In addition, we here illustratively also consider surface air temperature, precipitation, and relative humidity over T\"urkiye, the host of COP31 (Fig.~\ref{fig:turkiye-regional-scenarios}). Across the 23 displayed pathways, the 2081-2100 median change relative to 1995-2014 spans 0.91--4.57$^{\circ}$C for temperature, $-2.50$ to $-11.01$~mm~month$^{-1}$ for precipitation, and $-0.94$ to $-4.72$ percentage points for relative humidity. For the warming, the SSP scenarios end of century values from GCMagicc are again lower than the raw CMIP6 ESM model range (Fig. \ref{fig:turkiye-regional-scenarios}b), in alignment with the expectation based on global-mean temperatures and the findings for Iran (cf Fig.~\ref{fig:observational-alignment}c). The 20-member ensemble averages are compared to the single member ERA5 reanalysis data and show a reasonable match, but seemingly worse performance for \textit{pr} and \textit{hurs} (Fig.~\ref{fig:turkiye-regional-scenarios}c1,e1). The precipitation projections for T\"urkiye in GCMagicc simulator a reduction throughout all SSP scenarios, while CMIP6 ESM scenarios show median projections for the end of the 21st century that imply a decrease for the higher scenarios SSP3-7.0, SSP4-3.4, SSP4-6.0, and most strongly for SSP5-8.5, while the median projection implies an increase in precipitation for the lower scenarios SSP1-1.9, SSP1-2.6, SSP2-4.5 and SSP5-3.4-over (Fig.~\ref{fig:turkiye-regional-scenarios}c2). GCMagicc projections show a more uniform decrease of precipitation across the board for the SSP and other scenarios and with a comparatively narrow uncertainty compared to CMIP6 ESMs. The ERA5 reanalysis data indicates a slight decrease over the historical period with large year to year variability. The ERA5 reanalysis for relative humidity \textit{hurs} shows a strong decrease from around year 2000, and GCMagicc and CMIP6 ESMs show a decrease until the end of the 21st century (Fig.~\ref{fig:turkiye-regional-scenarios}e1). GCMagicc decreases are slightly more moderate than the CMIP6 ESM median projected decrease, but more clearly within the range of CMIP6 ESMs than in the case of the \textit{pr} variable.

\subsection{Attribution of the late-2025 Iranian drought}\label{drought-attrib}
We follow the probabilistic event-attribution framework in which the likelihood of an observed event is compared with its likelihood in a counterfactual world without anthropogenic forcing, and the difference is expressed through occurrence probabilities and their ratios \citep{Stott2004Heatwave,Stott2016AttributionReview,NASEM2016Attribution}. The most directly comparable regional application is the multi-country drought attribution for Syria, Iraq and Iran of \citet{otto_2023}, which combines observations and model ensembles for this water-stressed region. Earlier work on the Fertile Crescent drought used a related framing \citep{Kelley_etal_2015}. Here the counterfactual is supplied by the natural-only GCMagicc ensemble driven by natural forcings alone, and the comparison ensembles are the historical+SSP2-4.5 and natural-only simulations by three CMIP6 ESMs that provided large ensembles.

\begin{figure*}[p]
  \centering
  \includegraphics[width=\textwidth,height=0.58\textheight,keepaspectratio]{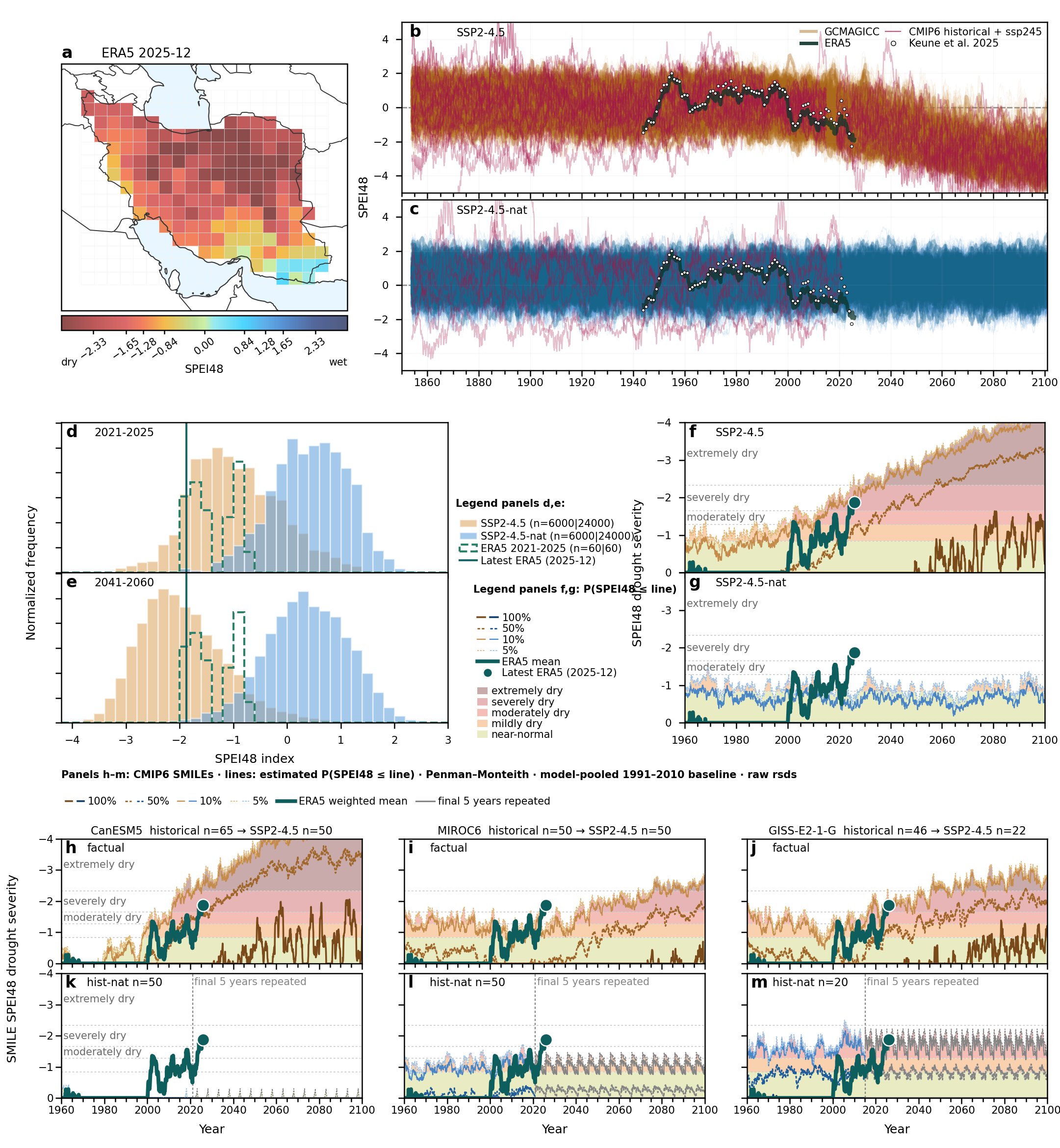}
  \caption{\textbf{Iranian Penman--Monteith SPEI-48 drought diagnostics and comparison with CMIP6 single-model initial-condition large ensembles.} \textbf{(a)} ERA5 SPEI-48 in December 2025. \textbf{(b,c)} Monthly cosine-latitude-weighted Iranian SPEI-48 for GCMagicc factual SSP2-4.5 and natural-only SSP2-4.5-nat ensembles, with ERA5 and CMIP6 historical/SSP2-4.5 context. Open circles in panel~(b) show the independently derived ERA5-Drought SPEI-48 product of \citet{keuneERA5DroughtGlobalDrought2025}, included over its observational period as a consistency check; it uses the same reanalysis but an independent implementation of the index, so agreement indicates that our SPEI-48 construction is not an artefact of our own processing chain. \textbf{(d,e)} Pooled monthly distributions for 2021--2025 and 2041--2060; the ERA5 distribution contains the 60 months of 2021--2025, and the solid green line marks the latest ERA5 threshold from December 2025. The GCMagicc predictors are ERA5-spliced only to the 2023 fade-out, so 24 of those 60 months lie outside the observationally anchored period. \textbf{(f,g)} Factual and natural-only monthly drought-severity trajectories with curves marking modeled SPEI-48 values at empirical below-threshold probabilities of 100\%, 50\%, 10\%, and 5\%. \textbf{(h--m)} The same probability curves for CanESM5, MIROC6, and GISS-E2-1-G historical-to-SSP2-4.5 and historical-natural trajectories. SMILE panels use model-pooled 1991--2010 calendar-month fits, raw surface downwelling shortwave radiation, and cosine-latitude weighting. Natural-only output ends in 2020 for CanESM5 and MIROC6 and in 2014 for GISS-E2-1-G; each member's final five complete years is repeated member-wise thereafter. These repeats are stationary extensions used both for display and for the post-endpoint diagnostics, and they supply no additional temporal degrees of freedom; titles report the live source inventory. Background bands indicate standard SPEI drought-severity classes.}
  \label{fig:iran-drought-attribution}
\end{figure*}

We use the late-2025 Iranian drought as a demanding multivariate application because SPEI-48 requires temporally stable precipitation and potential evapotranspiration drivers over several decades (Fig.~\ref{fig:iran-drought-attribution}). This is an illustrative capability evaluation, not a claim that one case study validates all possible impact applications.

We use FAO-56 Penman--Monteith potential evapotranspiration and cosine-latitude-weighted Iranian drought index SPEI-48 averaged over all Iranian 1x1-degree gridcells throughout. For each of the 60 ERA5 months in 2021--2025, the recent diagnostic estimates the probability of modeled SPEI-48 being at or below the observed threshold from the ensemble-member cross-section in the same year and calendar month. The future diagnostic compares each observed threshold with the corresponding calendar month's distribution pooled across members and 2041--2060. We report the median and 5--95\% range across the 60 month-matched probabilities, together with the probability obtained by pooling all modeled values in each window against the ERA5 derived SPEI48 value from December 2025 (Sect.~\ref{Drought-Methods} in the Supplement).

For GCMagicc in 2021--2025, the median month-matched probability of SPEI48 below ERA5-derived values is 29.00\% [13.00--57.05\%] in the full-forcing ensemble and 0.00\% [0.00--7.05\%] in the natural-only forcing ensemble. Against the December 2025 ERA5 threshold, pooling all 60 months gives probabilities of 11.25\% (675/6{,}000) in the full-forcing ensemble and 0.35\% (21/6{,}000) in the natural-only forcing ensemble. Under SSP2-4.5 in 2041--2060, the month-matched medians rise to 75.10\% [58.17--91.41\%] for the full-forcing ensemble and 1.40\% [0.25--6.39\%] for the natural-only forcing ensemble (Table~\ref{tab:drought-attribution-irn-v100-ensemble-comparison} in the Supplement). In our GCMagicc ensemble, we estimated that a similarly rare (10-15\% probability) drought state in just 25 years will not be categorised as "severely dry" but as "extremely dry" (Figure \ref{fig:iran-drought-attribution}f).

The results suggest that the drought state at the end of 2025 seems to have near-zero probability without human-induced climate change. Note that these are finite-ensemble empirical probabilities, though: a zero count is reported as 0\% and is not evidence that the event is physically impossible under natural-only conditions. Even under current human-induced climate change, it was a rare event with an estimated probability of only about 10-15\% in the 2021--2025 period. The results also suggest that the probability of such a drought state will increase substantially in the future. Under a scenario that resembles current-policy scenarios (SSP2-4.5), we estimate that in the period 2041-2060, such a drought state has a 3-in-4 probability, meaning that it counts as one of the 'wetter' states compared to the average. How societies can adapt to such drastic changes in water availability is unknown. Already in 2025, the issue of how to relocate millions of people, if the drought in Teheran were to continue, was being discussed. The active destruction of water infrastructure, especially water desalinisation plants \citep{shamim2026water}, and the potential risk for destruction in continuing of future conflict is a major risk for the region, and the results here suggest that such risks will be amplified by human-caused climate change.

We here further investigate the meteorological drivers of the Penman-Monteith drought state for the 21st century. We compare the individual drivers, like \textit{pr}, \textit{rsds}, \textit{sfcWind}, \textit{hurs}, \textit{psl}, across GCMagicc and GCMagicc-CE with the ones present in CMIP6 ESMs. In general, the CMIP6 ESM ensemble projects a wider range of outcomes than produced by either GCMagicc or GCMagicc-CE (Fig.~\ref{fig:iran-drought-attribution-sensitivity}a, b and c in the Supplement), but GCMagicc is well situated in the middle of the CMIP6 ESM ensemble range. Overall, it is apparent that evolution in surface air temperature related variables is the main driver of the drought intensification, as every other driver, in isolation produces either slightly positive (wetter) shifts for SPEI48 or near-zero ones (see grey-white sections of Fig.~\ref{fig:iran-drought-attribution-sensitivity} in the Supplement). Interestingly, the reduction in \textit{sfcWind} seems to persist and slightly increase as a driver for wetter conditions in a few models (GCMagicc-CE, CanESM5 and most of the CMIP6 models). If temperatures were held fixed (see 'clamp tas/tasmin/tasmax' in Fig.~\ref{fig:iran-drought-attribution-sensitivity} in the Supplement), GCMagicc and GCMagicc-CE project opposite signs of late 21st century drought evolution, approximately in line with the CMIP6 ESM spread.

\subsection{Comparison with conventional single-model initial-condition large ensembles (SMILEs)}
While most of the CMIP6 ESMs have only a few ensemble members (see the single member CMIP6 ESM examples in Figure \ref{fig:iran-drought-attribution}b and c), a few CMIP6 models provided large ensemble runs. We use those SMILEs to perform the same monthly Penman--Monteith diagnostic as with GCMagicc (Figure \ref{fig:iran-drought-attribution}h-m). We obtain recent empirical frequency probabilities of month-matched median probabilities under all-forcing/natural-only forcing runs of 44.00\%/0.00\% for CanESM5, 2.00\%/2.00\% for MIROC6, and 13.64\%/15.00\% for GISS-E2-1-G. The CanESM5 therefore shows qualitatively similar results to GCMagicc, whereas our analysis of MIROC6 and GISS-E2-1-G data indicates very low probabilities even under all-forcing conditions. The trend of future drought projections is increasing, yet, materially different across all three SMILES. The December 2025 drought state might be considered "extremely dry" under all SMILES if a similarly low probability event were picked in the future. For 2041--2060 the corresponding all-forcing/natural medians 2021-2025 drought months are 91.20\%/0.00\%, 24.55\%/2.40\%, and 49.55\%/16.00\% (Table~\ref{tab:drought-attribution-irn-v100-ensemble-comparison} in the Supplement), which means that under the CanESM5 model, the last 5 years of drought state would be considered exceptionally wet. The other two models still consider the recent drought state as a dry or average state. These latter two model ensembles of MIROC6 and GISS-E2-1-G seem to be implying a drought intensification over the 21st century, whereas the CanESM5 model seems to be more a middle-ground model across various CMIP6 ESMs (see first columns 'all evolving' in Fig.~\ref{fig:iran-drought-attribution-sensitivity}a, b and c in the Supplement).

\section{Limitations}\label{limitations}

GCMagicc's most consequential limitation for impact applications is its representation of monthly variability and of low-frequency climate oscillations: variability is over- or under-estimated depending on the variable, and the amplitude and frequency spectra of PDO, NAO and ENSO modes are not yet at parity with the range across CMIP6 ESMs (not shown), let alone with the observational record. SPEI-48 and other multi-year drought indices inherit this limitation, although the self-calibrating nature of the SPEI formulation, fitted against the baseline-period parameter distributions, limits the effect. Matching observational frequency-response spectra is generally a challenge for climate emulators and to our knowledge is not yet achieved by other comparable approaches \citep{tebaldi_emulators_2025}; obtaining ENSO teleconnection patterns at the level of the typical CMIP6 inter-model spread (Fig.~\ref{fig:validation-diagnostics}aa) is therefore a step rather than a solution.

The remaining limitations fall into two categories. First, the training data are imperfect: ESM-specific internally active radiative forcings are not all known; the number of CMIP6 ESMs spanning the full idealised, single-forcing and ScenarioMIP experiment set is small; regridding loses fine-grid detail and interpolates coarse models; and ensemble-member counts are uneven. Second, the output space is constrained to ten monthly variables without vertical, interior-ocean, or soil-state output and without convection-resolving processes. Drought results additionally depend on the short 1991--2010 fitting baseline, generalised logistic (GLO) tail estimation, cosine-latitude regional aggregation, Penman--Monteith assumptions, forcing construction, model dependence, and reanalysis uncertainty. SPEI-48 autocorrelation does not invalidate a probability estimated across independent ensemble members at a fixed year and month, but it prevents interpreting the 60 monthly diagnostics or the nominal member-times-month pool as independent observations. 

\conclusions\label{conc}
We have introduced \emph{GCMagicc}, a hybrid machine-learning emulator that couples the simple climate model MAGICC to a multi-variable, spatially explicit generator trained on CMIP6 and ERA5, with global-mean temperature aligned to observations and, by predictor construction, to the IPCC AR6 assessed warming distribution, and regional and multivariate fidelity evaluated rather than imposed. GCMagicc produces $1^{\circ}\times 1^{\circ}$ monthly ensembles whose multivariate relationships remain coherent across variables that jointly shape impacts. Rather than replacing Earth system models, GCMagicc complements them, preserving the role of ESMs for process discovery and feedback analysis while providing a computationally efficient pathway to translate the evolving scenario landscape into impact-relevant regional hazards at ensemble scale \citep{tebaldi_emulators_2025,kendon_mlemulators_2025,watsonparris_climatebench_2022}.

The combined capability matters. The Iranian drought example shows increasing future risk in GCMagicc and all three SMILEs, even though the time-evolving probabilities of these drought states are model-dependent. Producing the comparison requires GCMagicc's multivariate output, decomposed forcings, observation alignment, and large stochastic ensembles. As scenario ranges for CMIP7 narrow at both the high and the low ends compared to the SSP scenarios and ESM ensembles for new policy-relevant scenarios remain limited, machine-learning emulators that can scan the assessed uncertainty distribution across multiple variables and across forcing combinations within or near the envelope they were trained on become increasingly important for regional climate-risk assessment \citep{ipcc_ar6_wgi_ts,ngfs_scenarios_v5}.

\clearpage
\setcounter{section}{0}
\setcounter{subsection}{0}
\setcounter{figure}{0}
\setcounter{table}{0}
\setcounter{equation}{0}
\renewcommand{\thesection}{S\arabic{section}}
\renewcommand{\thesubsection}{\thesection.\arabic{subsection}}
\renewcommand{\thesubsubsection}{\thesubsection.\arabic{subsubsection}}
\renewcommand{\thefigure}{S\arabic{figure}}
\renewcommand{\thetable}{S\arabic{table}}
\renewcommand{\theequation}{S\arabic{equation}}
\section*{Supplementary Information}
\addcontentsline{toc}{section}{Supplementary Information}

\section{CMIP6 training dataset}\label{CMIP6-training-dataset}

The model-scenario coverage and ensemble-member counts used for training are listed in Table~\ref{tab:runs}.

\begin{table*}[p]
\centering
\begingroup
\renewcommand{\arraystretch}{0.62}
\caption{Number of ensemble runs available in the CMIP6 archive for each model-scenario combination. A dash (--) indicates no runs are available for that combination. Entries in parentheses are available but held out from training, namely all ScenarioMIP runs of UKESM1-0-LL and of the MIROC family, for which only the historical, single-forcing and idealised runs enter the training set. The table also does not list \textit{ssp245}, which is held out from training for every model as the unseen-scenario test case.}
\label{tab:runs}
\scriptsize
\setlength{\tabcolsep}{2pt}
\resizebox{0.92\textwidth}{!}{%
\begin{tabular}{l|ccc|ccc|c|ccccccc}
\tophline
& \rotatebox{70}{\textit{1pctCO2}}
& \rotatebox{70}{\textit{abrupt-2x}}
& \rotatebox{70}{\textit{abrupt-4x}}
& \rotatebox{70}{\textit{hist-GHG}}
& \rotatebox{70}{\textit{hist-aer}}
& \rotatebox{70}{\textit{hist-nat}}
& \rotatebox{70}{\textit{historical}}
& \rotatebox{70}{\textit{ssp119}}
& \rotatebox{70}{\textit{ssp126}}
& \rotatebox{70}{\textit{ssp370}}
& \rotatebox{70}{\textit{ssp434}}
& \rotatebox{70}{\textit{ssp460}}
& \rotatebox{70}{\textit{ssp534-over}}
& \rotatebox{70}{\textit{ssp585}} \\
\middlehline
ACCESS-CM2        & 1 &-- & 1 & 3 & 3 & 3 &10 &-- & 9 & 5 &-- &-- & 1 & 9 \\
ACCESS-ESM1-5     & 1 &-- & 2 & 3 & 3 & 3 &37 &-- &30 &35 &-- &-- & 1 &30 \\
CanESM5           & 6 & 1 & 2 &50 &30 &50 &50 &50 &50 &50 & 5 & 5 & 5 &50 \\
CanESM5-1         & 2 &-- & 2 &-- &-- &-- &50 & 1 & 1 & 1 &-- & 1 &-- &16 \\
CanESM5-CanOE     & 1 &-- &-- &-- &-- &-- & 3 &-- & 3 & 3 &-- &-- &-- & 3 \\
CESM2             &-- &-- &-- & 3 & 2 & 3 & 1 &-- & 5 & 5 &-- &-- &-- & 5 \\
CESM2-WACCM       &-- &-- &-- &-- &-- &-- & 1 &-- & 1 &-- &-- &-- & 1 & 4 \\
EC-Earth3         &-- &-- & 2 &-- &-- &-- &17 & 2 & 9 & 8 &-- &-- &-- & 7 \\
EC-Earth3-AerChem & 1 &-- & 1 &-- &-- &-- & 2 &-- &-- & 2 &-- &-- &-- &-- \\
EC-Earth3-CC      & 1 &-- & 1 &-- &-- &-- & 9 &-- &-- &-- &-- &-- &-- & 1 \\
EC-Earth3-Veg     & 1 &-- & 1 &-- &-- &-- & 9 & 2 & 6 & 5 &-- &-- &-- & 7 \\
EC-Earth3-Veg-LR  & 1 &-- &-- &-- &-- &-- & 1 & 3 & 3 & 3 &-- &-- &-- & 3 \\
FGOALS-g3         & 2 &-- & 1 &-- &-- &-- & 6 & 1 & 3 & 5 & 1 & 1 & 1 & 4 \\
GFDL-CM4          & 1 &-- & 1 &-- &-- & 2 & 1 &-- &-- &-- &-- &-- &-- & 1 \\
GFDL-ESM4         & 1 &-- & 1 &-- &-- &-- & 1 & 1 & 1 & 1 &-- &-- &-- & 1 \\
GISS-E2-1-G       & 5 & 4 & 5 &10 &10 &20 &47 & 6 &14 &27 & 7 & 8 &11 &15 \\
GISS-E2-1-G-CC    &-- &-- &-- &-- &-- &-- & 1 &-- & 1 &-- &-- &-- &-- &-- \\
GISS-E2-1-H       & 2 & 3 & 3 &-- &-- &-- &25 & 2 &10 & 6 & 2 & 2 & 6 &10 \\
GISS-E2-2-G       & 1 & 2 & 2 &-- &-- &-- &11 &-- &-- & 4 &-- &-- & 1 & 5 \\
GISS-E2-2-H       & 1 & 1 & 1 &-- &-- &-- & 3 &-- &-- &-- &-- &-- &-- &-- \\
HadGEM3-GC31-LL   & 4 & 1 & 1 & 5 & 5 &10 & 5 &-- & 1 &-- &-- &-- &-- & 4 \\
HadGEM3-GC31-MM   & 1 &-- & 1 &-- &-- &-- & 4 &-- & 1 &-- &-- &-- &-- & 4 \\
INM-CM4-8         & 1 &-- & 1 &-- &-- &-- & 1 &-- & 1 & 1 &-- &-- &-- & 1 \\
INM-CM5-0         & 1 &-- & 1 &-- &-- &-- &10 &-- & 1 & 5 &-- &-- &-- & 1 \\
IPSL-CM6A-LR      & 1 & 1 &11 & 6 & 6 & 5 &31 & 6 & 5 &11 & 2 & 7 &-- & 6 \\
IPSL-CM6A-LR-INCA &-- &-- & 1 &-- &-- &-- & 1 &-- &-- &-- &-- &-- &-- &-- \\
IPSL-CM6A-MR1     & 1 &-- & 1 &-- &-- &-- & 1 &-- &-- &-- &-- &-- &-- &-- \\
MIROC-ES2H        &-- &-- & 3 &-- &-- &-- & 3 &-- &-- &-- &-- &-- &-- &-- \\
MIROC-ES2L        & 1 &-- & 1 &-- &-- &-- &31 &(10) & (7) &(10) &-- &-- &-- & (9) \\
MIROC6            & 1 & 1 & 1 &10 &10 &50 &50 &(50) &(50) &(50) & (1) & (1) & (1) &(50) \\
SAM0-UNICON       & 1 &-- & 1 &-- &-- &-- & 1 &-- &-- &-- &-- &-- &-- &-- \\
UKESM1-0-LL       & 4 &-- & 1 &-- &-- &-- &19 & (5) &(16) &(16) & (5) &-- & (5) & (5) \\
\middlehline
ERA5              &\multicolumn{14}{c}{1 run (historical reanalysis)} \\
\bottomhline
\end{tabular}}
\endgroup
\end{table*}

\clearpage
\begingroup
\footnotesize
\renewcommand{\arraystretch}{0.82}
\setlength{\tabcolsep}{4pt}
\begin{longtable}{lll}
\caption{CMIP6 data-citation references for the model output used in this study, at the
model/MIP granularity defined by the CMIP6 Data Citation Guidelines. Each DOI resolves to the
WDCC/DKRZ citation landing page for the corresponding
\texttt{CMIP6.\textit{activity\_id}.\textit{institution\_id}.\textit{source\_id}} collection.
Data creators are given in the reference list of this Supplement as the first creator et al.; the complete creator lists are on the DOI landing pages.}
\label{tab:cmip6-dois} \\
\tophline
Source ID & Activity & DOI \\
\middlehline
\endfirsthead
\tophline
Source ID & Activity & DOI \\
\middlehline
\endhead
ACCESS-CM2 & CMIP & \href{https://doi.org/10.22033/ESGF/CMIP6.2281}{10.22033/ESGF/CMIP6.2281} \\
ACCESS-CM2 & DAMIP & \href{https://doi.org/10.22033/ESGF/CMIP6.14361}{10.22033/ESGF/CMIP6.14361} \\
ACCESS-CM2 & ScenarioMIP & \href{https://doi.org/10.22033/ESGF/CMIP6.2285}{10.22033/ESGF/CMIP6.2285} \\
ACCESS-ESM1-5 & CMIP & \href{https://doi.org/10.22033/ESGF/CMIP6.2288}{10.22033/ESGF/CMIP6.2288} \\
ACCESS-ESM1-5 & DAMIP & \href{https://doi.org/10.22033/ESGF/CMIP6.14362}{10.22033/ESGF/CMIP6.14362} \\
ACCESS-ESM1-5 & ScenarioMIP & \href{https://doi.org/10.22033/ESGF/CMIP6.2291}{10.22033/ESGF/CMIP6.2291} \\
CanESM5 & CFMIP & \href{https://doi.org/10.22033/ESGF/CMIP6.1302}{10.22033/ESGF/CMIP6.1302} \\
CanESM5 & CMIP & \href{https://doi.org/10.22033/ESGF/CMIP6.1303}{10.22033/ESGF/CMIP6.1303} \\
CanESM5 & DAMIP & \href{https://doi.org/10.22033/ESGF/CMIP6.1305}{10.22033/ESGF/CMIP6.1305} \\
CanESM5 & ScenarioMIP & \href{https://doi.org/10.22033/ESGF/CMIP6.1317}{10.22033/ESGF/CMIP6.1317} \\
CanESM5-1 & CMIP & \href{https://doi.org/10.22033/ESGF/CMIP6.17184}{10.22033/ESGF/CMIP6.17184} \\
CanESM5-1 & ScenarioMIP & \href{https://doi.org/10.22033/ESGF/CMIP6.17199}{10.22033/ESGF/CMIP6.17199} \\
CanESM5-CanOE & CMIP & \href{https://doi.org/10.22033/ESGF/CMIP6.10205}{10.22033/ESGF/CMIP6.10205} \\
CanESM5-CanOE & ScenarioMIP & \href{https://doi.org/10.22033/ESGF/CMIP6.10207}{10.22033/ESGF/CMIP6.10207} \\
CESM2 & CMIP & \href{https://doi.org/10.22033/ESGF/CMIP6.2185}{10.22033/ESGF/CMIP6.2185} \\
CESM2 & DAMIP & \href{https://doi.org/10.22033/ESGF/CMIP6.2187}{10.22033/ESGF/CMIP6.2187} \\
CESM2 & ScenarioMIP & \href{https://doi.org/10.22033/ESGF/CMIP6.2201}{10.22033/ESGF/CMIP6.2201} \\
CESM2-WACCM & CMIP & \href{https://doi.org/10.22033/ESGF/CMIP6.10024}{10.22033/ESGF/CMIP6.10024} \\
CESM2-WACCM & ScenarioMIP & \href{https://doi.org/10.22033/ESGF/CMIP6.10026}{10.22033/ESGF/CMIP6.10026} \\
EC-Earth3 & CMIP & \href{https://doi.org/10.22033/ESGF/CMIP6.181}{10.22033/ESGF/CMIP6.181} \\
EC-Earth3 & ScenarioMIP & \href{https://doi.org/10.22033/ESGF/CMIP6.251}{10.22033/ESGF/CMIP6.251} \\
EC-Earth3-AerChem & CMIP & \href{https://doi.org/10.22033/ESGF/CMIP6.639}{10.22033/ESGF/CMIP6.639} \\
EC-Earth3-AerChem & ScenarioMIP & \href{https://doi.org/10.22033/ESGF/CMIP6.724}{10.22033/ESGF/CMIP6.724} \\
EC-Earth3-CC & CMIP & \href{https://doi.org/10.22033/ESGF/CMIP6.640}{10.22033/ESGF/CMIP6.640} \\
EC-Earth3-CC & ScenarioMIP & \href{https://doi.org/10.22033/ESGF/CMIP6.15327}{10.22033/ESGF/CMIP6.15327} \\
EC-Earth3-Veg & CMIP & \href{https://doi.org/10.22033/ESGF/CMIP6.642}{10.22033/ESGF/CMIP6.642} \\
EC-Earth3-Veg & ScenarioMIP & \href{https://doi.org/10.22033/ESGF/CMIP6.727}{10.22033/ESGF/CMIP6.727} \\
EC-Earth3-Veg-LR & CMIP & \href{https://doi.org/10.22033/ESGF/CMIP6.643}{10.22033/ESGF/CMIP6.643} \\
EC-Earth3-Veg-LR & ScenarioMIP & \href{https://doi.org/10.22033/ESGF/CMIP6.728}{10.22033/ESGF/CMIP6.728} \\
FGOALS-g3 & CMIP & \href{https://doi.org/10.22033/ESGF/CMIP6.1783}{10.22033/ESGF/CMIP6.1783} \\
FGOALS-g3 & ScenarioMIP & \href{https://doi.org/10.22033/ESGF/CMIP6.2056}{10.22033/ESGF/CMIP6.2056} \\
GFDL-CM4 & CMIP & \href{https://doi.org/10.22033/ESGF/CMIP6.1402}{10.22033/ESGF/CMIP6.1402} \\
GFDL-CM4 & DAMIP & \href{https://doi.org/10.22033/ESGF/CMIP6.11383}{10.22033/ESGF/CMIP6.11383} \\
GFDL-CM4 & ScenarioMIP & \href{https://doi.org/10.22033/ESGF/CMIP6.9242}{10.22033/ESGF/CMIP6.9242} \\
GFDL-ESM4 & CMIP & \href{https://doi.org/10.22033/ESGF/CMIP6.1407}{10.22033/ESGF/CMIP6.1407} \\
GFDL-ESM4 & ScenarioMIP & \href{https://doi.org/10.22033/ESGF/CMIP6.1414}{10.22033/ESGF/CMIP6.1414} \\
GISS-E2-1-G & CFMIP & \href{https://doi.org/10.22033/ESGF/CMIP6.2061}{10.22033/ESGF/CMIP6.2061} \\
GISS-E2-1-G & CMIP & \href{https://doi.org/10.22033/ESGF/CMIP6.1400}{10.22033/ESGF/CMIP6.1400} \\
GISS-E2-1-G & DAMIP & \href{https://doi.org/10.22033/ESGF/CMIP6.2062}{10.22033/ESGF/CMIP6.2062} \\
GISS-E2-1-G & ScenarioMIP & \href{https://doi.org/10.22033/ESGF/CMIP6.2074}{10.22033/ESGF/CMIP6.2074} \\
GISS-E2-1-G-CC & CMIP & \href{https://doi.org/10.22033/ESGF/CMIP6.11657}{10.22033/ESGF/CMIP6.11657} \\
GISS-E2-1-G-CC & ScenarioMIP & \href{https://doi.org/10.22033/ESGF/CMIP6.17581}{10.22033/ESGF/CMIP6.17581} \\
GISS-E2-1-H & CFMIP & \href{https://doi.org/10.22033/ESGF/CMIP6.13941}{10.22033/ESGF/CMIP6.13941} \\
GISS-E2-1-H & CMIP & \href{https://doi.org/10.22033/ESGF/CMIP6.1421}{10.22033/ESGF/CMIP6.1421} \\
GISS-E2-1-H & ScenarioMIP & \href{https://doi.org/10.22033/ESGF/CMIP6.2080}{10.22033/ESGF/CMIP6.2080} \\
GISS-E2-2-G & CFMIP & \href{https://doi.org/10.22033/ESGF/CMIP6.11659}{10.22033/ESGF/CMIP6.11659} \\
GISS-E2-2-G & CMIP & \href{https://doi.org/10.22033/ESGF/CMIP6.2081}{10.22033/ESGF/CMIP6.2081} \\
GISS-E2-2-G & ScenarioMIP & \href{https://doi.org/10.22033/ESGF/CMIP6.11671}{10.22033/ESGF/CMIP6.11671} \\
GISS-E2-2-H & CFMIP & \href{https://doi.org/10.22033/ESGF/CMIP6.17582}{10.22033/ESGF/CMIP6.17582} \\
GISS-E2-2-H & CMIP & \href{https://doi.org/10.22033/ESGF/CMIP6.15861}{10.22033/ESGF/CMIP6.15861} \\
HadGEM3-GC31-LL & CFMIP & \href{https://doi.org/10.22033/ESGF/CMIP6.435}{10.22033/ESGF/CMIP6.435} \\
HadGEM3-GC31-LL & CMIP & \href{https://doi.org/10.22033/ESGF/CMIP6.419}{10.22033/ESGF/CMIP6.419} \\
HadGEM3-GC31-LL & DAMIP & \href{https://doi.org/10.22033/ESGF/CMIP6.471}{10.22033/ESGF/CMIP6.471} \\
HadGEM3-GC31-LL & ScenarioMIP & \href{https://doi.org/10.22033/ESGF/CMIP6.10845}{10.22033/ESGF/CMIP6.10845} \\
HadGEM3-GC31-MM & CMIP & \href{https://doi.org/10.22033/ESGF/CMIP6.420}{10.22033/ESGF/CMIP6.420} \\
HadGEM3-GC31-MM & ScenarioMIP & \href{https://doi.org/10.22033/ESGF/CMIP6.10846}{10.22033/ESGF/CMIP6.10846} \\
INM-CM4-8 & CMIP & \href{https://doi.org/10.22033/ESGF/CMIP6.1422}{10.22033/ESGF/CMIP6.1422} \\
INM-CM4-8 & ScenarioMIP & \href{https://doi.org/10.22033/ESGF/CMIP6.12321}{10.22033/ESGF/CMIP6.12321} \\
INM-CM5-0 & CMIP & \href{https://doi.org/10.22033/ESGF/CMIP6.1423}{10.22033/ESGF/CMIP6.1423} \\
INM-CM5-0 & ScenarioMIP & \href{https://doi.org/10.22033/ESGF/CMIP6.12322}{10.22033/ESGF/CMIP6.12322} \\
IPSL-CM6A-LR & CFMIP & \href{https://doi.org/10.22033/ESGF/CMIP6.1522}{10.22033/ESGF/CMIP6.1522} \\
IPSL-CM6A-LR & CMIP & \href{https://doi.org/10.22033/ESGF/CMIP6.1534}{10.22033/ESGF/CMIP6.1534} \\
IPSL-CM6A-LR & DAMIP & \href{https://doi.org/10.22033/ESGF/CMIP6.13801}{10.22033/ESGF/CMIP6.13801} \\
IPSL-CM6A-LR & ScenarioMIP & \href{https://doi.org/10.22033/ESGF/CMIP6.1532}{10.22033/ESGF/CMIP6.1532} \\
IPSL-CM6A-LR-INCA & CMIP & \href{https://doi.org/10.22033/ESGF/CMIP6.13582}{10.22033/ESGF/CMIP6.13582} \\
IPSL-CM6A-MR1 & CMIP & \href{https://doi.org/10.22033/ESGF/CMIP6.15922}{10.22033/ESGF/CMIP6.15922} \\
MIROC-ES2H & CMIP & \href{https://doi.org/10.22033/ESGF/CMIP6.901}{10.22033/ESGF/CMIP6.901} \\
MIROC-ES2L & CMIP & \href{https://doi.org/10.22033/ESGF/CMIP6.902}{10.22033/ESGF/CMIP6.902} \\
MIROC-ES2L & ScenarioMIP & \href{https://doi.org/10.22033/ESGF/CMIP6.936}{10.22033/ESGF/CMIP6.936} \\
MIROC6 & CFMIP & \href{https://doi.org/10.22033/ESGF/CMIP6.885}{10.22033/ESGF/CMIP6.885} \\
MIROC6 & CMIP & \href{https://doi.org/10.22033/ESGF/CMIP6.881}{10.22033/ESGF/CMIP6.881} \\
MIROC6 & DAMIP & \href{https://doi.org/10.22033/ESGF/CMIP6.894}{10.22033/ESGF/CMIP6.894} \\
MIROC6 & ScenarioMIP & \href{https://doi.org/10.22033/ESGF/CMIP6.898}{10.22033/ESGF/CMIP6.898} \\
SAM0-UNICON & CMIP & \href{https://doi.org/10.22033/ESGF/CMIP6.1489}{10.22033/ESGF/CMIP6.1489} \\
UKESM1-0-LL & CMIP & \href{https://doi.org/10.22033/ESGF/CMIP6.1569}{10.22033/ESGF/CMIP6.1569} \\
UKESM1-0-LL & ScenarioMIP & \href{https://doi.org/10.22033/ESGF/CMIP6.1567}{10.22033/ESGF/CMIP6.1567} \\
\bottomhline
\end{longtable}
\endgroup

\nocite{cmip6_access_cm2_cmip,cmip6_access_cm2_damip,cmip6_access_cm2_scenariomip,cmip6_access_esm1_5_cmip,cmip6_access_esm1_5_damip,cmip6_access_esm1_5_scenariomip,cmip6_canesm5_cfmip,cmip6_canesm5_cmip,cmip6_canesm5_damip,cmip6_canesm5_scenariomip,cmip6_canesm5_1_cmip,cmip6_canesm5_1_scenariomip,cmip6_canesm5_canoe_cmip,cmip6_canesm5_canoe_scenariomip,cmip6_cesm2_cmip,cmip6_cesm2_damip,cmip6_cesm2_scenariomip,cmip6_cesm2_waccm_cmip,cmip6_cesm2_waccm_scenariomip,cmip6_ec_earth3_cmip,cmip6_ec_earth3_scenariomip,cmip6_ec_earth3_aerchem_cmip,cmip6_ec_earth3_aerchem_scenariomip,cmip6_ec_earth3_cc_cmip,cmip6_ec_earth3_cc_scenariomip,cmip6_ec_earth3_veg_cmip,cmip6_ec_earth3_veg_scenariomip,cmip6_ec_earth3_veg_lr_cmip,cmip6_ec_earth3_veg_lr_scenariomip,cmip6_fgoals_g3_cmip,cmip6_fgoals_g3_scenariomip,cmip6_gfdl_cm4_cmip,cmip6_gfdl_cm4_damip,cmip6_gfdl_cm4_scenariomip,cmip6_gfdl_esm4_cmip,cmip6_gfdl_esm4_scenariomip,cmip6_giss_e2_1_g_cfmip,cmip6_giss_e2_1_g_cmip,cmip6_giss_e2_1_g_damip,cmip6_giss_e2_1_g_scenariomip,cmip6_giss_e2_1_g_cc_cmip,cmip6_giss_e2_1_g_cc_scenariomip,cmip6_giss_e2_1_h_cfmip,cmip6_giss_e2_1_h_cmip,cmip6_giss_e2_1_h_scenariomip,cmip6_giss_e2_2_g_cfmip,cmip6_giss_e2_2_g_cmip,cmip6_giss_e2_2_g_scenariomip,cmip6_giss_e2_2_h_cfmip,cmip6_giss_e2_2_h_cmip,cmip6_hadgem3_gc31_ll_cfmip,cmip6_hadgem3_gc31_ll_cmip,cmip6_hadgem3_gc31_ll_damip,cmip6_hadgem3_gc31_ll_scenariomip,cmip6_hadgem3_gc31_mm_cmip,cmip6_hadgem3_gc31_mm_scenariomip,cmip6_inm_cm4_8_cmip,cmip6_inm_cm4_8_scenariomip,cmip6_inm_cm5_0_cmip,cmip6_inm_cm5_0_scenariomip,cmip6_ipsl_cm6a_lr_cfmip,cmip6_ipsl_cm6a_lr_cmip,cmip6_ipsl_cm6a_lr_damip,cmip6_ipsl_cm6a_lr_scenariomip,cmip6_ipsl_cm6a_lr_inca_cmip,cmip6_ipsl_cm6a_mr1_cmip,cmip6_miroc_es2h_cmip,cmip6_miroc_es2l_cmip,cmip6_miroc_es2l_scenariomip,cmip6_miroc6_cfmip,cmip6_miroc6_cmip,cmip6_miroc6_damip,cmip6_miroc6_scenariomip,cmip6_sam0_unicon_cmip,cmip6_ukesm1_0_ll_cmip,cmip6_ukesm1_0_ll_scenariomip}

\begin{table}[ht]
\centering
\begingroup
\renewcommand{\arraystretch}{1}
\small
\caption{Affine normalization coefficients per target variable.}
\label{tab:supp-affine}
\begin{tabular}{llrr}
\tophline
Short name & Long name & $a_v$ & $b_v$ \\
\middlehline
\textit{psl}     & Sea Level Pressure                     & $10^{-3}$          & $-100$ \\
\textit{tas}     & Near-Surface Air Temperature           & $10^{-1}$          & $-30$  \\
\textit{pr}      & Precipitation                          & $2\!\cdot\!10^{4}$ & $-5$   \\
\textit{sfcWind} & Near-Surface Wind Speed                & $1/3$              & $-2$   \\
\textit{ts}      & Surface Temperature                    & $10^{-1}$          & $-30$  \\
\textit{tasmin}  & Daily Min.\ Near-Surface Air Temp.     & $10^{-1}$          & $-30$  \\
\textit{tasmax}  & Daily Max.\ Near-Surface Air Temp.     & $10^{-1}$          & $-30$  \\
\textit{rsds}    & Surface Downwelling Shortwave Radiation & $10^{-1}$          & $-15$  \\
\textit{hurs}    & Near-Surface Relative Humidity         & $10^{-1}$          & $-5$   \\
\textit{huss}    & Near-Surface Specific Humidity         & $10^{2}$           & $-1$   \\
\bottomhline
\end{tabular}
\endgroup
\end{table}

\section{Predictors: MAGICC setup and AR6/CMIP7 ScenarioMIP workflows}\label{sec:supp-magicc-setup}\label{sec:supp-gcmagicc-predictors}

We use the simple climate model \textsc{MAGICC} to generate probabilistic, scenario-specific global predictors (temperature, energy imbalance/heat uptake, and forcing components) that condition the GCMagicc regional emulator, enabling IPCC-consistent uncertainty propagation at low computational cost \citep{meinshausen_magicc_2011,magicc_website,ipcc_ar6_wgi_ts}.

\subsection{MAGICC version, provenance, and calibration}
We generate global predictors with \textsc{MAGICC} v7.5.3 driven by the AR6 probabilistic parameter set --- the AR6 drawn set \texttt{magicc-ar6-0fd0f62-f023edb} --- as configured by the \texttt{AR6} workflow default \citep{meinshausen_magicc_2011}. \textsc{MAGICC} is an emulator of coupled climate--carbon-cycle behaviour, with a physically based energy-balance climate module calibrated against more complex models and observations \citep{meinshausen_magicc_2011}. We use the AR6-consistent probabilistic parameter ensemble distributed with \textsc{MAGICC} ($M=600$ members) to represent uncertainty in global transient response and related parameters, consistent with the assessed warming constraints in IPCC AR6 Working Group I \citep{ipcc_ar6_wgi_ts}.

The complete training and inference sequence is shown in Fig.~\ref{fig:training-inference-workflow} in the main manuscript; the following subsections document each predictor-construction and inference stage.

\subsection{Energy-balance core and predictor time series}
\textsc{MAGICC}'s climate module can be summarised as a low-order energy-balance model. MAGICC has a 50-layer upwelling-diffusion-entrainment ocean parameterisation \citep{meinshausen_magicc_2011}. In our implementation, \textsc{MAGICC} produces per-scenario, per-ensemble-member global-mean surface air temperature anomaly $T_i(t)$ and a global imbalance or heat-uptake predictor $N_i(t)$, together with ERF components, which are then used as conditioning variables for GCMagicc. Previous studies sometimes also have used emulator chains coupling global simple climate models to spatial emulators, mostly however focused on global-mean temperature only as the coupling variable \citep{beusch_magicc_mesmer_2022}, although an ocean heat uptake variant is explored in their Supplementary.

In GCMagicc and GCMagicc-CE we train the model using smoothed global surface-air-temperature and energy-balance predictors alongside the ERF components. The surface-air-temperature predictor is aligned over the historical period (1940--2024) with ERA5. The 600 MAGICC predictor members are subsampled to either $n=20$ or $n=100$ to preserve the central and 5--95\% range. A deterministic two-pass correction, described below, aligns the smoothed global GCMagicc \textit{tas} output with its predictor; it is not an iterative convergence loop.

\subsection{Predictor variable set exported from MAGICC}
For each scenario $s$ and probabilistic member $i \in \{1,\dots,M\}$, we construct a monthly predictor vector $\mathbf{x}_{s,i}(t,m)$ comprising:

\subsection{(i) Smoothed global predictors.}
We export a smoothed global temperature predictor $T^{(W)}_{s,i}(t)$ and a smoothed imbalance or heat-uptake predictor $N^{(W)}_{s,i}(t)$, denoted in code as \texttt{tas\_smoothed} and \texttt{rtmt\_smoothed}. For training, the naturally variable CMIP6 monthly global-mean series are filtered with a 21-year-span LOWESS smoother. For inference, the already smooth MAGICC series are filtered with a centred 20-year (240-month) running mean after linear extension by 20 years at each endpoint, followed by trimming to the requested interval. Both operations remove interannual variability so that internal variability is supplied by the trained stochastic component rather than by predictor noise. Conditioning on both $T^{(W)}$ and $N^{(W)}$ provides a compact representation of disequilibrium states, where two members may share the same $T$ but differ in $N$, which is useful when mapping scenarios with different forcing trajectories, such as overshoot pathways into regional patterns \citep{beusch_magicc_mesmer_2022,tebaldi_emulators_2025}.

\subsection{(ii) Forcing components.}
We export ERF components in W m$^{-2}$ sufficient to distinguish forcing composition and disequilibrium pathways beyond global-mean temperature alone. The exported set comprises \texttt{CO2\_ERF}, \texttt{GHG\_ERF}, \texttt{aer\_ERF}, \texttt{totalO3\_ERF}, \texttt{stratO3\_ERF}, \texttt{sol\_ERF}, \texttt{volc\_ERF}, \texttt{nat\_ERF}, and \texttt{other\_ERF}, where $F_{\mathrm{nat}}(t) = F_{\mathrm{sol}}(t) + F_{\mathrm{volc}}(t)$.
Here \texttt{other\_ERF} denotes the residual between total ERF and the explicitly tracked components.

\begin{figure}[htbp]
\centering
\includegraphics[width=\textwidth]{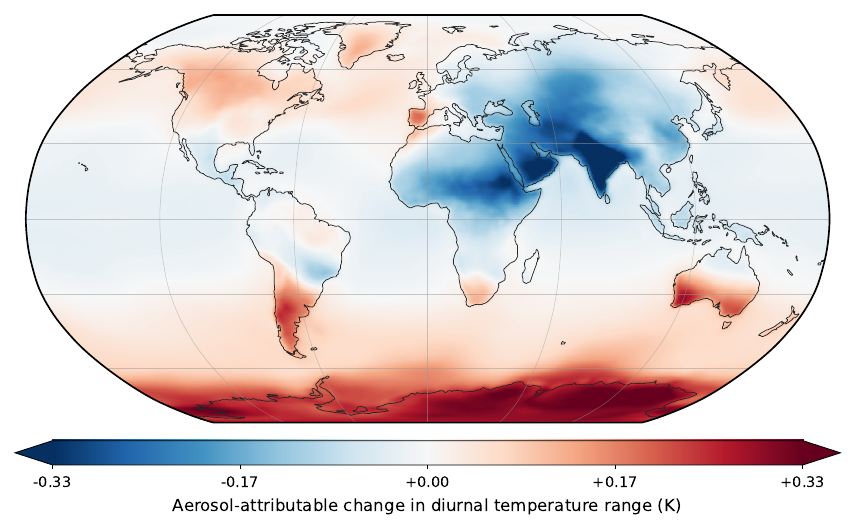}
\caption{\textbf{Aerosol-covariate pattern in the diurnal temperature range diagnosed by GCMagicc.}
Ensemble mean over the 20 Monte Carlo members ($\textit{mc}=0$--$19$, $\textit{nsim}=100$ each) of
$\Delta_{\mathrm{mean}} = \overline{y}_{\mathrm{factual}} - \overline{y}_{\mathrm{counterfactual}}$ for
\textit{tasmax}~$-$~\textit{tasmin} at $\textit{nside}=64$, where the counterfactual sets the aerosol
effective-radiative-forcing predictor \texttt{aer\_ERF} to zero and leaves all other predictors at their
factual values. Blue indicates a reduced diurnal temperature range under historical aerosol forcing.
The strongest suppression falls over the Indo-Gangetic plain (approximately $-0.44$\,K at
$78^{\circ}$E, $25^{\circ}$N), eastern China ($-0.18$\,K) and the Saharan and Arabian dust belt
($-0.16$\,K), consistent with a reduction in surface shortwave radiation damping daytime maxima more
than nighttime minima. Area-weighted regional means are $-0.110 \pm 0.011$\,K over South and East Asia
and $-0.129 \pm 0.024$\,K over the Sahara ($\pm$ one standard error across members); the
absolute-magnitude response is 3.9 times larger over land than over ocean between $60^{\circ}$S and
$70^{\circ}$N. Because \texttt{aer\_ERF} is a single global forcing time series rather than a spatial
field, the map is the emulator's learned spatial response to a change in global aerosol forcing and need
not coincide with the emission distribution. The global mean of the field is $1\times10^{-5}$\,K by
construction: the diurnal temperature range is a difference of two temperature variables, so a spatially
uniform cooling cancels and only the differential daytime--nighttime response remains. The positive
localized positive values over the Southern Ocean, parts of the Sahara and Antarctica (including values of $+0.27 \pm 0.03$\,K) illustrate the limitations of the approach, because GCMagicc can learn spurious covariate signals associated with the aerosol time series. The figure shows an average across per-member
$\Delta_{\mathrm{mean}}$ maps}
\label{fig:supp-aerosol-pattern}
\end{figure}

For the natural-only SSP2-4.5 experiment used in the drought application, anthropogenic ERF components are zeroed while the archived MAGICC natural-forcing output supplies \texttt{sol\_ERF} and \texttt{volc\_ERF}. Neither ERA5 splicing nor the two-pass global-temperature bias correction is applied to the natural-only predictors: MAGICC's natural-only global \texttt{tas} and \texttt{rtmt} anomalies are by construction close to zero.
For the aerosol-perturbation counterfactual, the aerosol ERF predictor is set to zero while all other forcings are held at their final historical values. As illustration, we provide an aerosol-pattern difference map $\Delta_{\mathrm{mean}}$ (by default \textit{tasmax} $-$ \textit{tasmin}). For each of 20 categorical model-index settings (\(c=0,\ldots,19\)), the emulator is sampled 100 times under both full-forcing and aerosol-off forcing. The difference between the two 100-realization means is calculated for each setting, and Fig.~\ref{fig:supp-aerosol-pattern} shows the equally weighted mean of the resulting 20 model-index-specific maps.

The intervention is a \emph{fixed-forcing epoch} contrast rather than a transient run. The predictor
matrix is taken from the normalised ERA5 historical cube (not an SSP scenario); all forcing predictors
are frozen at their values in the final time step of that record
(\texttt{x[:,\,4:]\,=\,x[-1,\,4:]}) and the analysed window is the last 120 months, so the seasonal cycle
is averaged out while the forcing state is held fixed. The factual and counterfactual runs differ in the
single \texttt{aer\_ERF} column and in nothing else.

Two properties of this design bound how the resulting map should be read. First, \texttt{aer\_ERF} is a
single global forcing time series, not a spatial field, so the intervention perturbs one global number
and the spatial structure of the response is entirely the emulator's learned covariance between global
aerosol forcing and regional diurnal temperature range over the training period \citep{Stjern2020}. Aerosol
effective radiative forcing is defined over the anthropogenic perturbation relative to pre-industrial
conditions, so natural mineral dust and sea salt remain in the background state of both the factual and
the counterfactual run and are not removed by zeroing the channel. Any apparent signal over a
dust-dominated region is consequently not a dust response but a region whose diurnal temperature range
co-varies with the global anthropogenic aerosol trajectory in the training data; regions with a large
climatological diurnal range are the most exposed to this confounding, and the arid subtropical values
in Fig.~\ref{fig:supp-aerosol-pattern} should be read in that light rather than as an attributable
local aerosol effect.

The signals over the Indo-Gangetic plain and eastern China are, by contrast, consistent with the
observational literature, in which persistent aerosol loading over northern India has been associated
with declining surface solar radiation and a falling diurnal temperature range
\citep{Mall2021DTRIndia}, and anthropogenic aerosol forcing has been identified as a contributor to the
observed diurnal-range decline over eastern China \citep{Liu2016DTRAsia}. That the Indian signal here
exceeds the Chinese one is expected for a recent forcing epoch: aerosol optical depth over eastern China
has declined markedly since about 2010 under air-quality policy while South Asian loading has continued
to rise \citep{Ratnam2021AerosolDipole}.

\subsection{(iii) Monthly seasonality features.}
To preserve the seasonality of the regional emulator while using smoothed global predictors, we include month-of-year features: an integer month index $m \in \{1,\dots,12\}$ and its circular embedding
\begin{equation}
\sin_m = \sin\left(2\pi m/12\right), \qquad \cos_m = \cos\left(2\pi m/12\right).
\end{equation}
The full conditioning vector combines long-term forced evolution ($T^{(W)}$, $N^{(W)}$, and ERFs) with a deterministic seasonal phase.

\subsection{(iv) Ensemble sampling and subsampling (600 $\to$ 20/100)}
For each scenario, \textsc{MAGICC} yields $M=600$ probabilistic predictor realisations $\{\mathbf{x}_{s,i}\}_{i=1}^{M}$. For the manuscript's primary ensembles we subsample to $n \in \{20,100\}$ members for computational tractability while preserving the median end-of-century warming. The selection is median-anchored and stratified. Writing $\Delta T_i = \overline{T_i}(2081\text{--}2100) - \overline{T_i}(1995\text{--}2014)$ for the end-of-century warming of member $i$, the procedure is:
\begin{enumerate}
\item order the $M$ members by $\Delta T_i$;
\item retain the two central members, whose mean is the full-ensemble median for even $M$;
\item draw $(n-2)/2$ further members from below and $(n-2)/2$ from above that central pair, at evenly spaced ranks within each half.
\end{enumerate}
Because $M$ and $n$ are both even, the subsample median equals the full-ensemble median exactly rather than approximately. Even spacing in rank preserves the shape of the distribution as well as its centre. Preserving distributional characteristics rather than only the mean matters because the assessed ranges are themselves distributional statements \citep{ipcc_ar6_wgi_ts}, and because distributional fidelity is the relevant criterion in emulator evaluation for risk applications \citep{watsonparris_climatebench_2022}. Note though that this procedure only considers distributions in the marginal distribution of surface air temperature and not in the joint distribution of all predictors, so the subsample is not guaranteed to preserve the full multivariate distribution.

\subsection{(v) ERA5 alignment and replication modes} \label{sec:ERA5-splicing}
To ensure observational anchoring of historical predictors, we provide an ERA5-aligned mode in which the historical portion of the global predictors is nudged or spliced to match ERA5-derived global-mean temperature anomalies \citep{era5_dataset}. The splice is a two-sided crossfade: ERA5 is faded in over the five years from 1945 and faded out over the single year ending in 2023, with ERA5 used directly in between and the MAGICC series shifted to the ERA5 mean over the adjoining window outside it. Shift alignment is used for \textit{tas\_smoothed}, \textit{rtmt\_smoothed}. In the description below, let $t_0$ denote the last month of the fade-out window, i.e.\ the last month in which reanalysis information enters the predictors. For each probabilistic member $i$, we define an anomaly-splicing operator
\begin{equation}
T_i^{\mathrm{ERA5}}(t)=
\begin{cases}
T_{\mathrm{ERA5}}(t), & t \le t_0, \\
T_{\mathrm{ERA5}}(t_0) + \left[T_i(t) - T_i(t_0)\right], & t > t_0,
\end{cases}
\end{equation}
The same shift splice is applied to \textit{rtmt\_smoothed}. Its historical anchor is the global ERA5 net top-of-atmosphere radiation series after the preprocessing stored as \textit{rtmt\_smoothed}; the vetted source spans January 1940--December 2025. In the replication setting, the categorical \textit{model\_index} is set to the ERA5 category (index 0), allowing an ERA5-flavoured historical climate while retaining the multi-model training distribution for future responses \citep{era5_dataset,gcmagicc_software}.

\subsection{(vi) Two-pass global-temperature correction}
Because the regional emulator is trained on gridded fields, its implied global-mean temperature may differ slightly from the conditioning predictor. The released workflow therefore performs exactly two emulator passes. In pass 1, monthly gridded \textit{tas} is reduced to annual cosine-latitude-weighted global means and smoothed with a centred 21-year window. For each year $y$,
\begin{equation}
\Delta_y = \overline{T}^{(1)}_{\mathrm{GCMagicc},21}(y)-T_{\mathrm{predictor}}(y).
\end{equation}
In pass 2, the normalised temperature-predictor column is replaced by
\begin{equation}
x'_{\texttt{tas\_smoothed}}(t)=x_{\texttt{tas\_smoothed}}(t)-\Delta_{y(t)}/10,
\end{equation}
and the emulator is run once more with the same ensemble seed. The divisor 10 reverses the predictor normalisation. No other predictor or output field is directly modified.

\subsection{Limitations and assumptions}
First, the ERF decomposition is only as reliable as the MAGICC configuration and supplied forcing series; scenario-level forcing is a proxy for broad forcing evolution and does not reconstruct each ESM's diagnosed effective forcing. Second, low-pass filtering deliberately removes predictor-scale variability, leaving internal variability to the learned stochastic component and potentially underrepresenting very-low-frequency modes \citep{tebaldi_emulators_2025,watsonparris_climatebench_2022}. Third, ERA5 alignment depends on reanalysis uncertainty and the chosen splice boundary. Finally, the two-pass correction enforces global-mean temperature consistency only; precipitation--energy consistency and other global constraints remain emergent properties. See also the discussion of limitations in the main manuscript.

\section{Perfect-model GCMagicc-PM and extrapolation GCMagicc-XS experiments}

\subsection{Perfect-model uncertainty experiment (GCMagicc-PM)}\label{sec:supp-perfect-model}
To quantify the structural uncertainty of GCMagicc-PM (the variant without the MAGICC-derived smoothed global \textit{tas} and \textit{rtmt} indicators), we repeat the GCMagicc-PM training in a perfect-model setup. For each CMIP6 model in turn, a fresh GCMagicc-PM instance is trained on that model's historical run alone, treating that CMIP6 model as ground truth. Each retrained instance is then run forward under future forcing and its end-of-century warming is compared to the corresponding CMIP6 model's own end-of-century projection. The spread of these deviations across CMIP6 models defines the perfect-model uncertainty spread. Combining this spread with the ERA5-informed central estimate from GCMagicc-PM yields ranges that are consistent with, and no narrower than, the IPCC AR6 assessed range for the middle and upper end of the SSP scenarios (see the small black vertical bars in panel b of Fig.~\ref{fig:observational-alignment}).

\subsection{Held-out high-end scenario extrapolation (GCMagicc-XS)}\label{sec:supp-extrapolation-XS}
The GCMagicc-XS variant (Methods, Table~\ref{tab:gcmagicc-variants}) uses the same reduced predictor set as GCMagicc-PM (no MAGICC-derived smoothed global indicators) and is trained jointly across the CMIP6 models and ERA5. In the configuration reported here, GCMagicc-XS is trained on \textit{historical}, \textit{hist-GHG}, \textit{hist-aer}, \textit{hist-nat}, \textit{ssp119}, \textit{ssp126}, \textit{ssp370}, \textit{ssp434}, \textit{ssp460} and the idealised \textit{1pctCO2} and \textit{abrupt-2xCO2}; the high-end scenarios \textit{ssp534-over} and \textit{ssp585} and the idealised \textit{abrupt-4xCO2} are held out from training. Of these, \textit{ssp585} and \textit{abrupt-4xCO2} could be evaluated; \textit{ssp534-over} is provided by too few CMIP6 models to appear in the normalised evaluation set and is therefore part of the held-out design.

\begin{figure}[H]
  \centering
  \includegraphics[width=\linewidth,height=0.70\textheight,keepaspectratio]{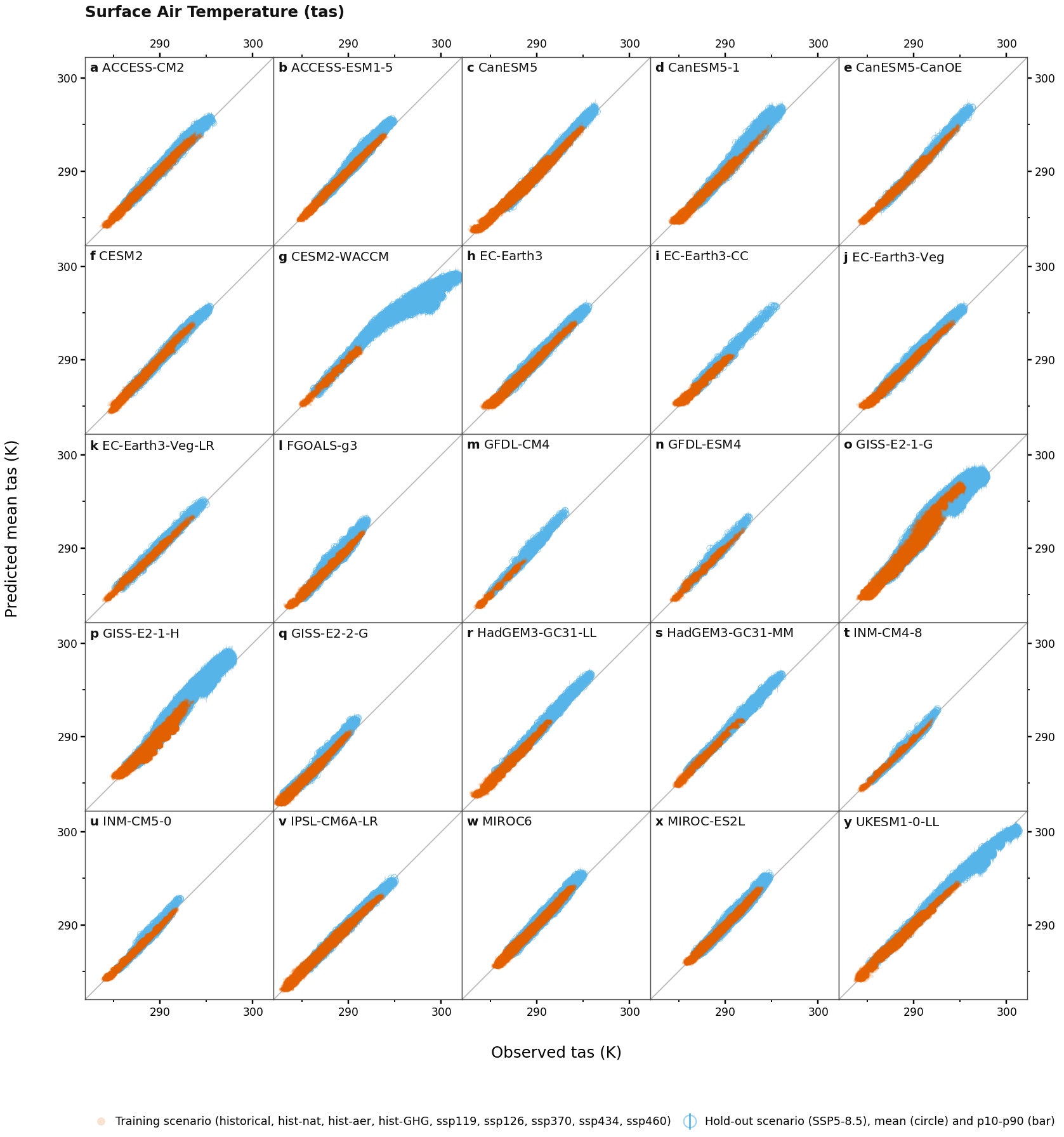}
  \caption{\textbf{GCMagicc-XS per-model near-surface air temperature (\textit{tas}): observed versus predicted.} Twenty-five per-model panels, lettered a--y, one for each CMIP6 driving model for which \textit{ssp585} runs are available (Table~\ref{tab:runs}); the remaining seven models of the training set have no \textit{ssp585} runs and therefore no hold-out case to display. Within each panel, the orange markers show the pooled in-distribution training scenarios (\textit{historical}, \textit{hist-nat}, \textit{hist-aer}, \textit{hist-GHG}, \textit{ssp119}, \textit{ssp126}, \textit{ssp370}, \textit{ssp434}, \textit{ssp460}) and the light blue markers the held-out \textit{SSP5-8.5}, given as the mean (circle) with a p10--p90 bar. The grey diagonal is the 1:1 line, and both axes are monthly global-mean \textit{tas} in K. The panels demonstrate that GCMagicc-XS remains in the correct physical range and reproduces the seasonal cycle under a held-out high-end forcing. Each point pairs a single CMIP6 member with an ensemble-mean prediction, and the spread along the 1:1 line is due to both the seasonal cycle and high-low scenario variations.}
  \label{fig:supp-extrapolation-XS-tas}
\end{figure}

\begin{figure}[H]
  \centering
  \includegraphics[width=\linewidth,height=0.85\textheight,keepaspectratio]{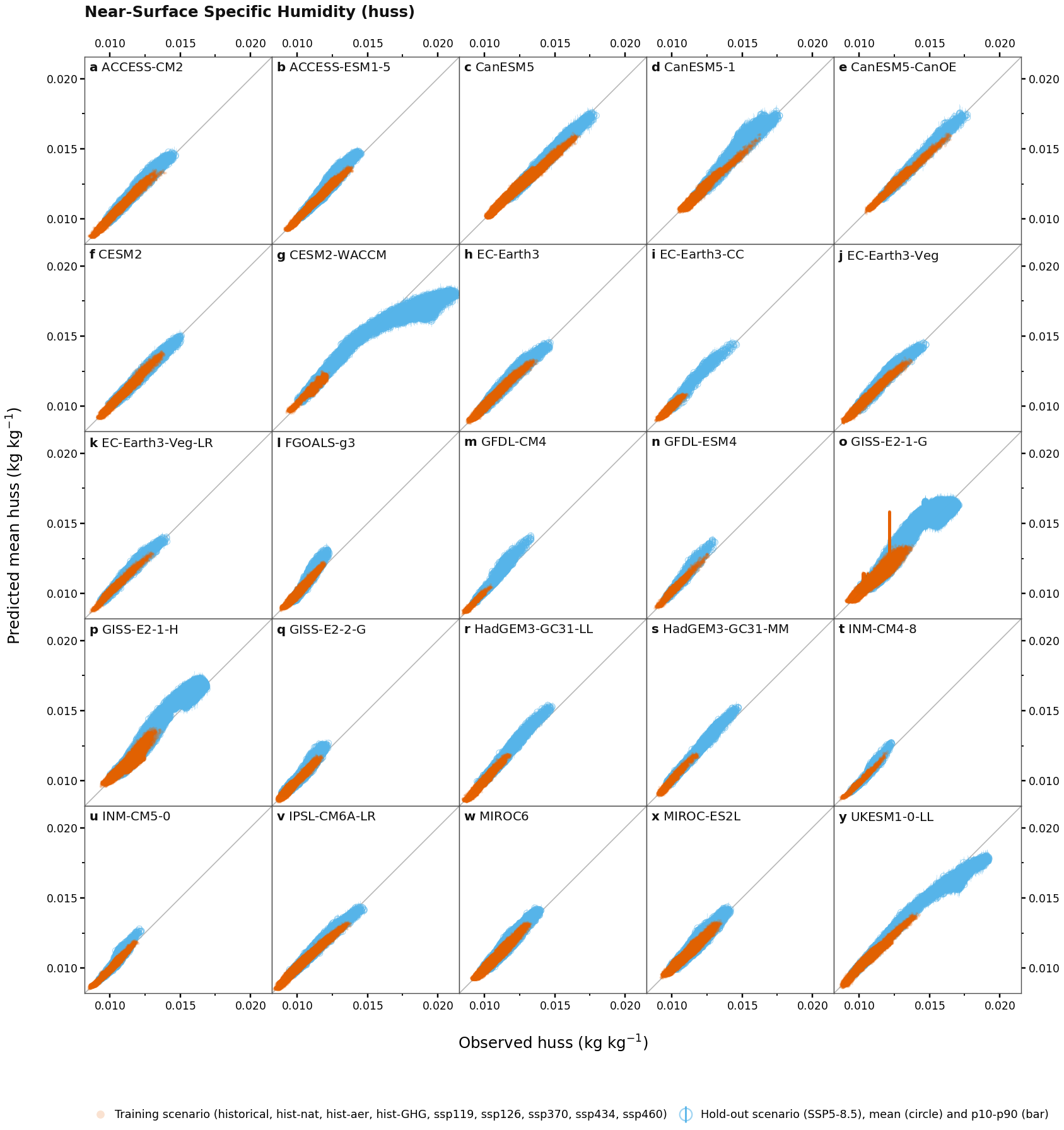}
  \caption{\textbf{GCMagicc-XS per-model near-surface specific humidity (\textit{huss}): observed versus predicted.} As Fig.~\ref{fig:supp-extrapolation-XS-tas} but for near-surface specific humidity.}
  \label{fig:supp-extrapolation-XS-huss}
\end{figure}

\begin{figure}[H]
  \centering
  \includegraphics[width=\linewidth,height=0.85\textheight,keepaspectratio]{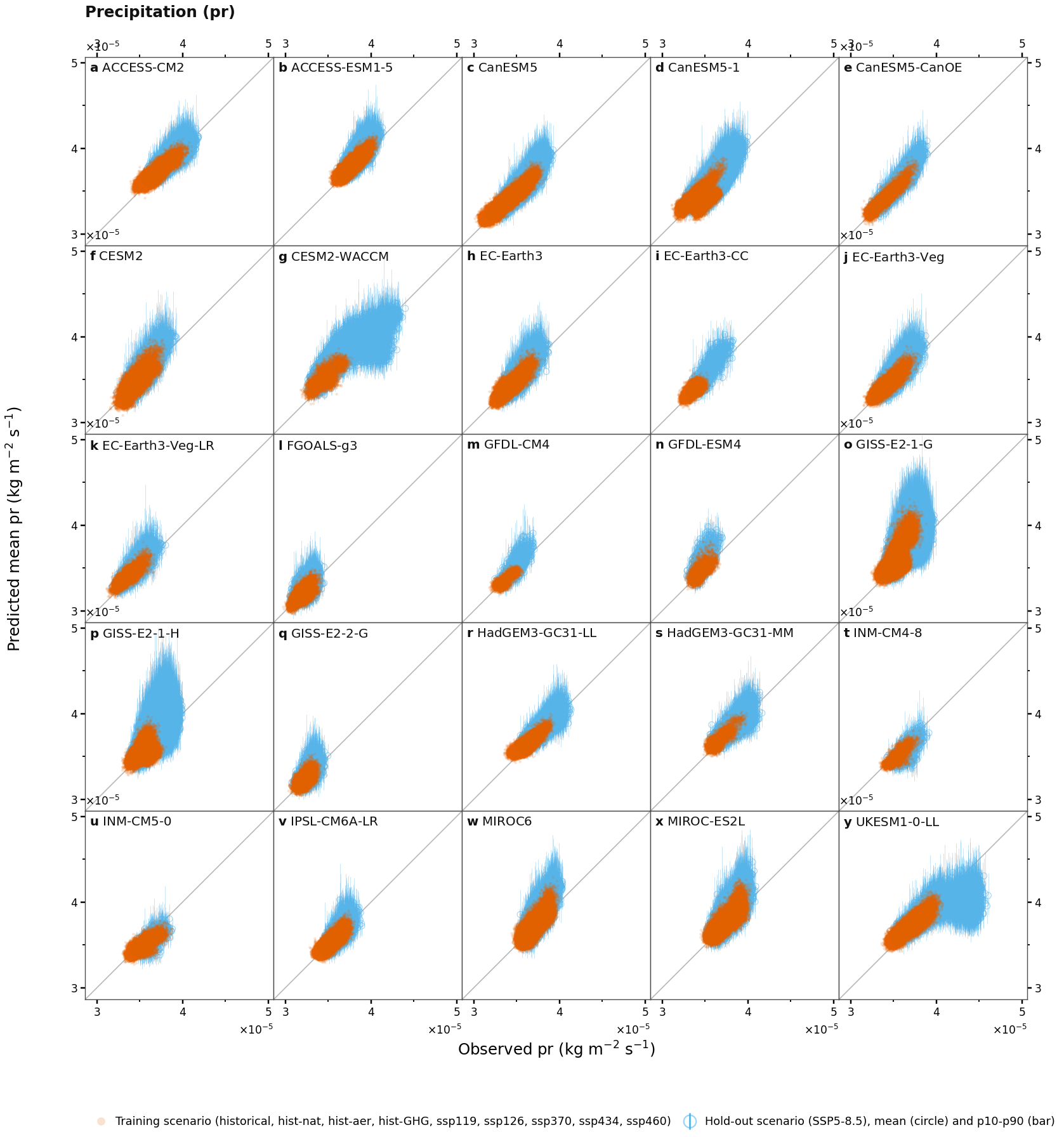}
  \caption{\textbf{GCMagicc-XS per-model precipitation (\textit{pr}): observed versus predicted.} As Fig.~\ref{fig:supp-extrapolation-XS-tas} but for precipitation.}
  \label{fig:supp-extrapolation-XS-pr}
\end{figure}

\begin{figure}[H]
  \centering
  \includegraphics[width=\linewidth,height=0.85\textheight,keepaspectratio]{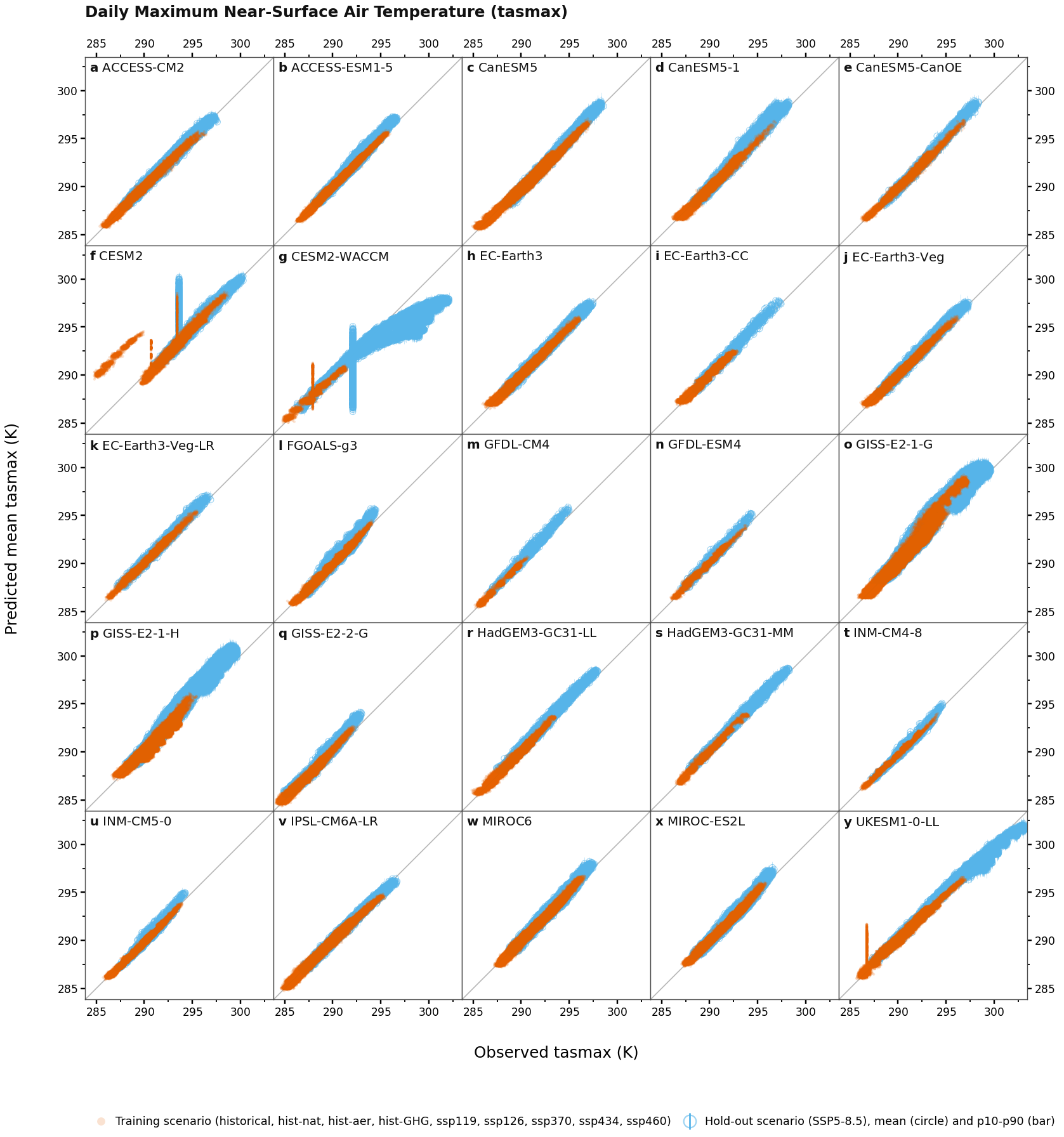}
  \caption{\textbf{GCMagicc-XS per-model daily maximum near-surface air temperature (\textit{tasmax}): observed versus predicted.} As Fig.~\ref{fig:supp-extrapolation-XS-tas} but for daily maximum near-surface air temperature. Note that some data issues related to the CESM2, CESM2-WACCM and UKESM1-0-LL data and calibrations are visible in the \textit{tasmax} and \textit{tasmin} panels.}
  \label{fig:supp-extrapolation-XS-tasmax}
\end{figure}

\begin{figure}[H]
  \centering
  \includegraphics[width=\linewidth,height=0.85\textheight,keepaspectratio]{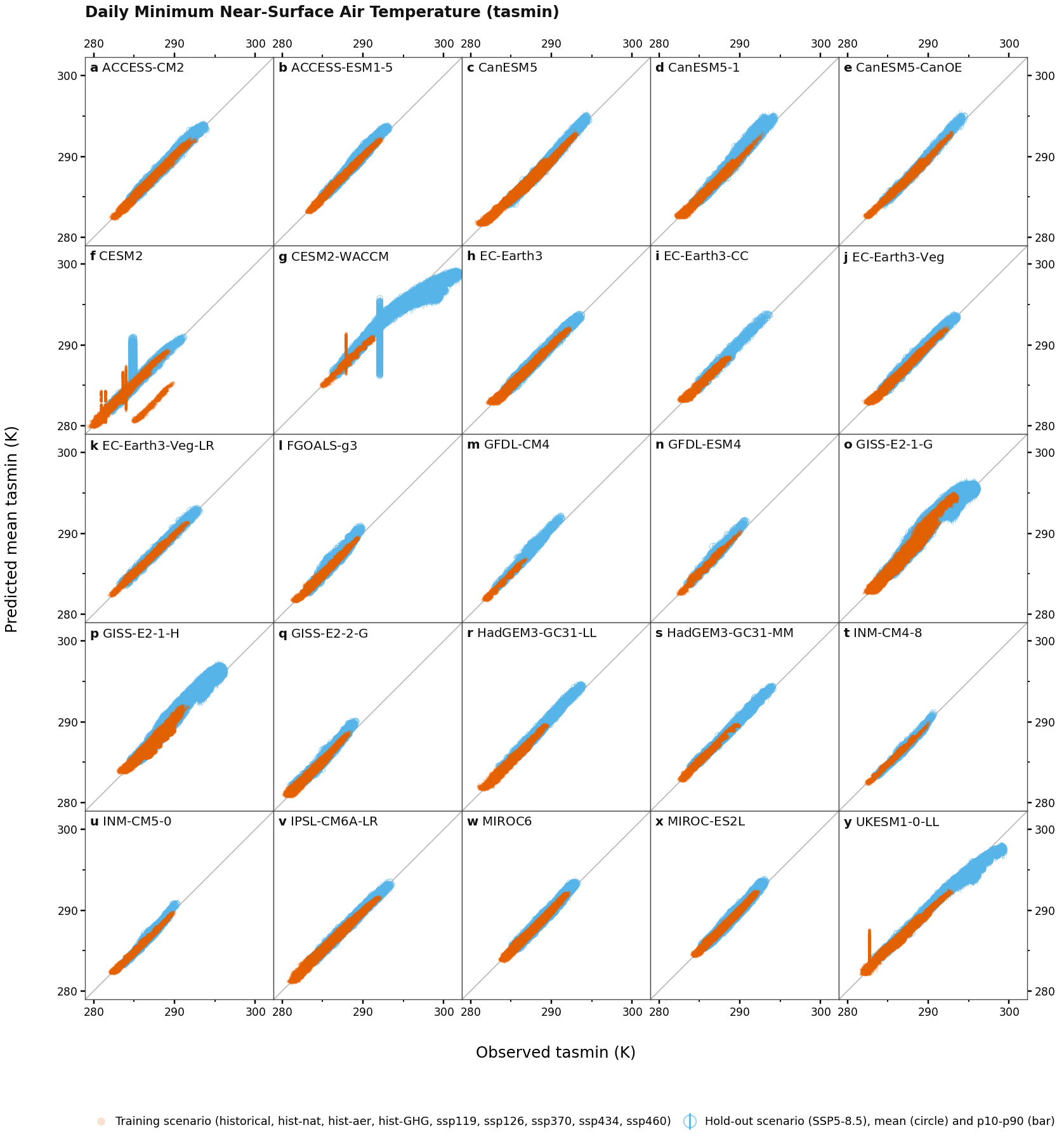}
  \caption{\textbf{GCMagicc-XS per-model daily minimum near-surface air temperature (\textit{tasmin}): observed versus predicted.} As Fig.~\ref{fig:supp-extrapolation-XS-tas} but for daily minimum near-surface air temperature. Note that some data issues related to the CESM2, CESM2-WACCM and UKESM1-0-LL data and calibrations are visible in the \textit{tasmax} and \textit{tasmin} panels.}
  \label{fig:supp-extrapolation-XS-tasmin}
\end{figure}

\begin{figure}[H]
  \centering
  \includegraphics[width=\linewidth,height=0.85\textheight,keepaspectratio]{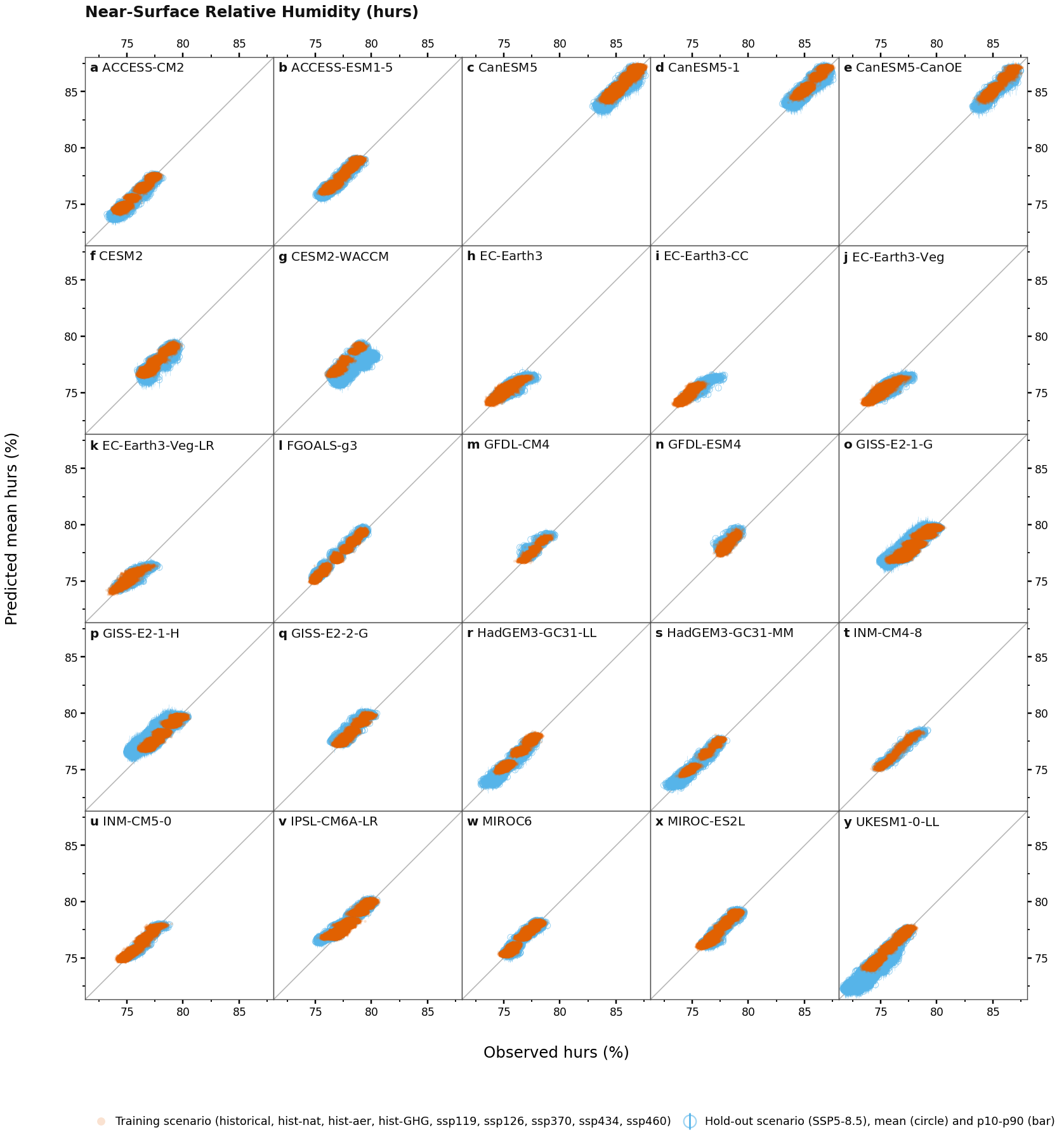}
  \caption{\textbf{GCMagicc-XS per-model near-surface relative humidity (\textit{hurs}): observed versus predicted.} As Fig.~\ref{fig:supp-extrapolation-XS-tas} but for near-surface relative humidity.}
  \label{fig:supp-extrapolation-XS-hurs}
\end{figure}

\begin{figure}[H]
  \centering
  \includegraphics[width=\linewidth,height=0.85\textheight,keepaspectratio]{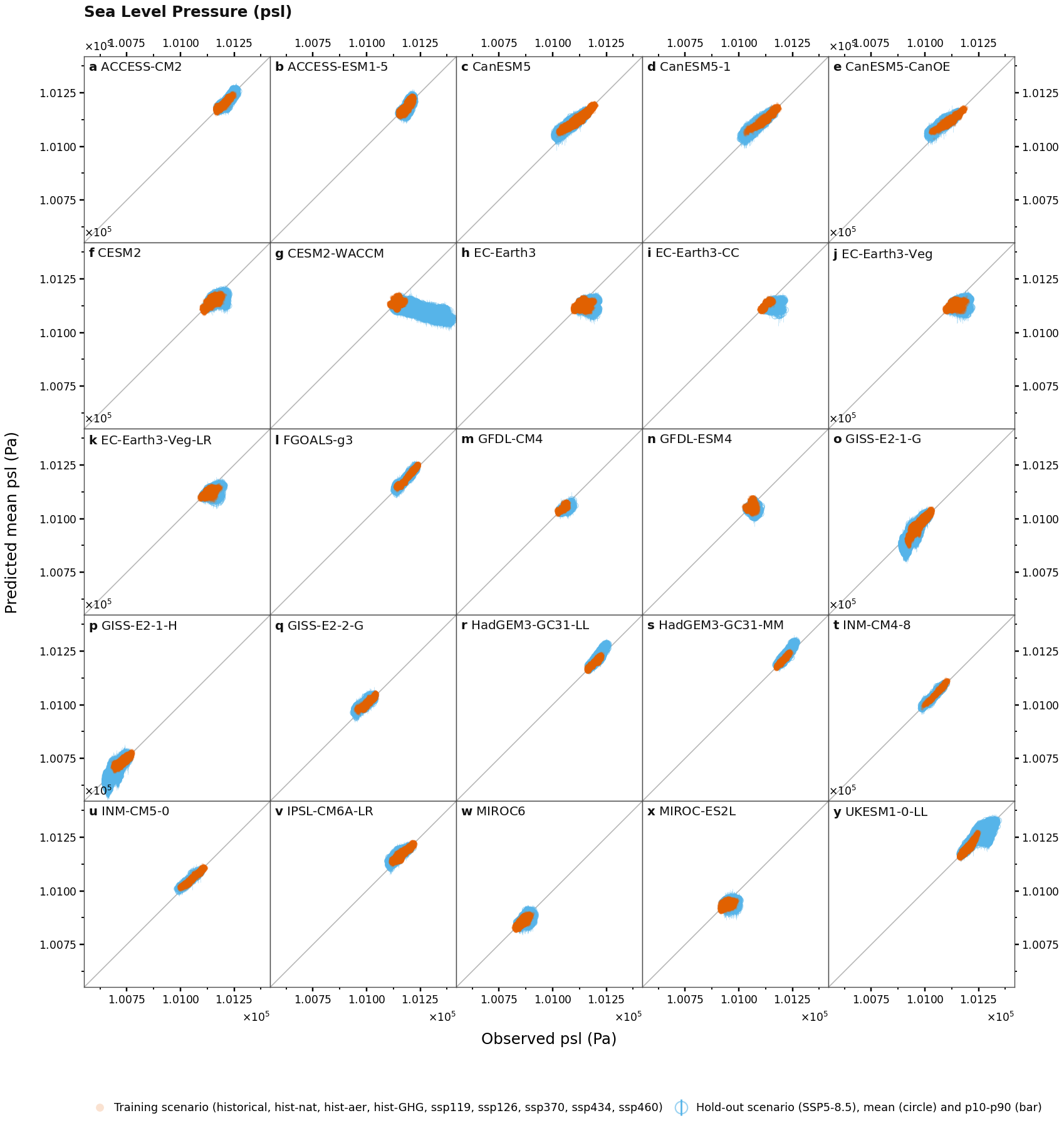}
  \caption{\textbf{GCMagicc-XS per-model sea-level pressure (\textit{psl}): observed versus predicted.} As Fig.~\ref{fig:supp-extrapolation-XS-tas} but for sea-level pressure.}
  \label{fig:supp-extrapolation-XS-psl}
\end{figure}

\begin{figure}[H]
  \centering
  \includegraphics[width=\linewidth,height=0.85\textheight,keepaspectratio]{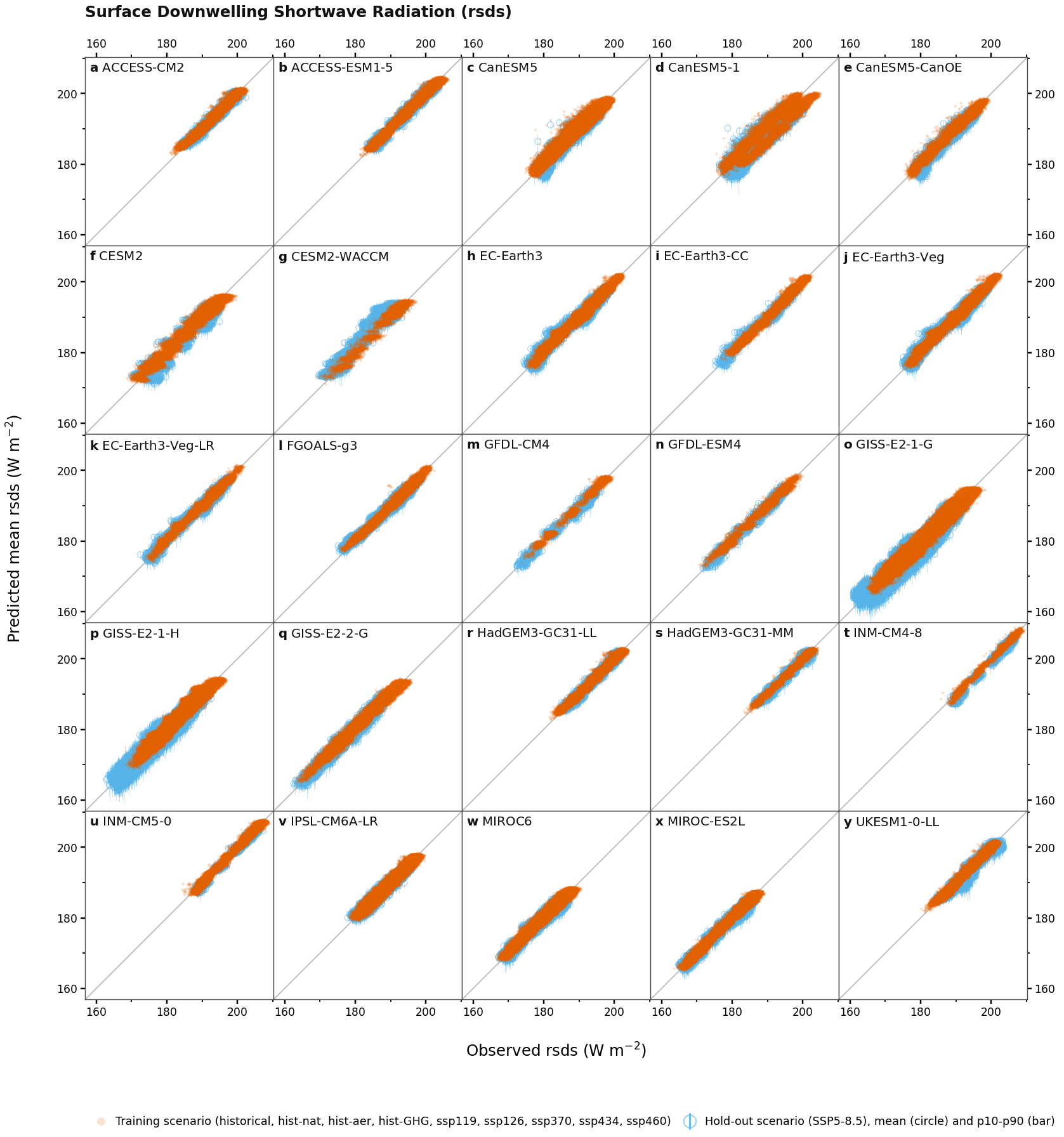}
  \caption{\textbf{GCMagicc-XS per-model surface downwelling shortwave radiation (\textit{rsds}): observed versus predicted.} As Fig.~\ref{fig:supp-extrapolation-XS-tas} but for surface downwelling shortwave radiation.}
  \label{fig:supp-extrapolation-XS-rsds}
\end{figure}

\begin{figure}[H]
  \centering
  \includegraphics[width=\linewidth,height=0.85\textheight,keepaspectratio]{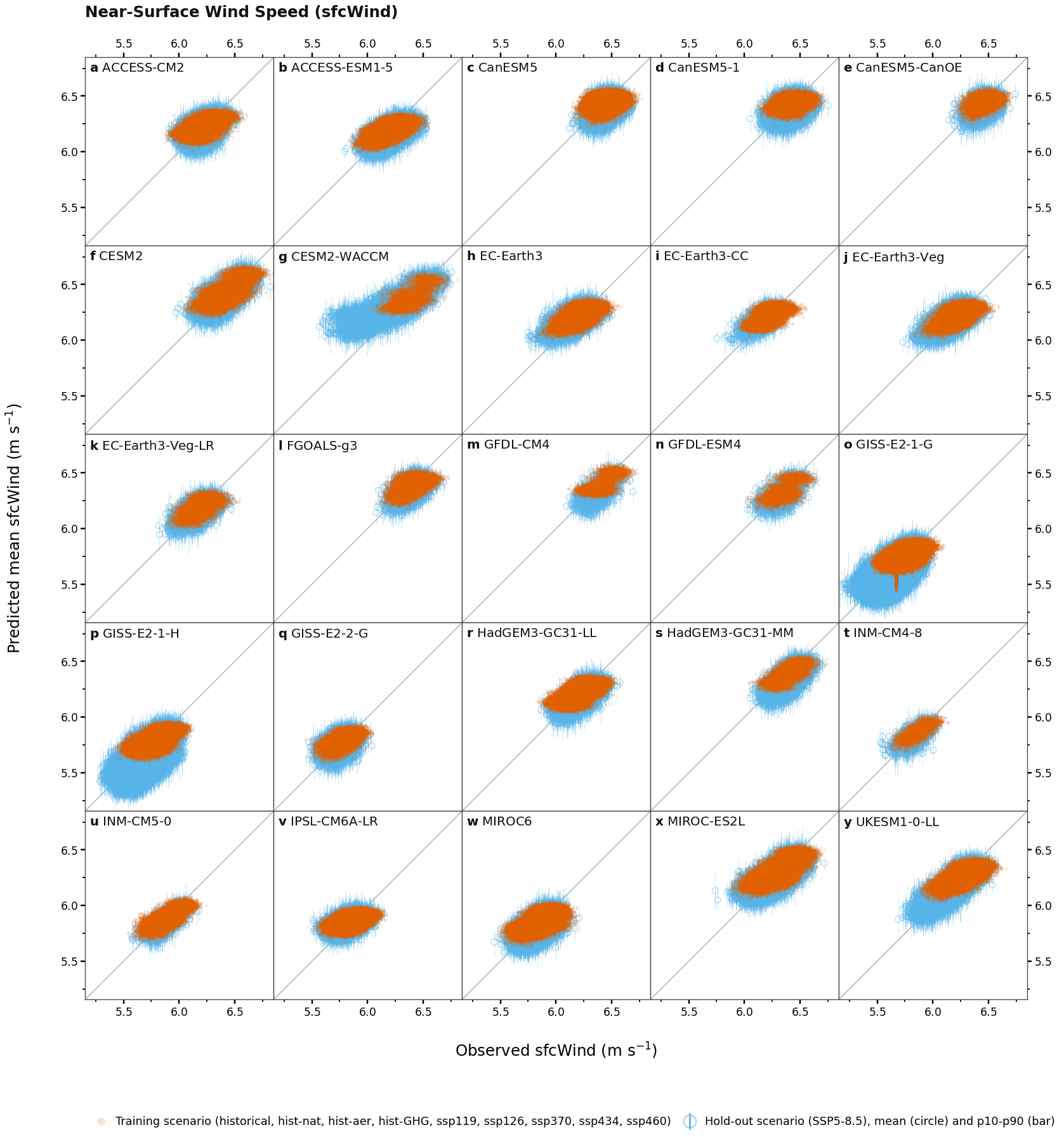}
  \caption{\textbf{GCMagicc-XS per-model near-surface wind speed (\textit{sfcWind}): observed versus predicted.} As Fig.~\ref{fig:supp-extrapolation-XS-tas} but for near-surface wind speed.}
  \label{fig:supp-extrapolation-XS-sfcWind}
\end{figure}

\begin{figure}[H]
  \centering
  \includegraphics[width=\linewidth,height=0.85\textheight,keepaspectratio]{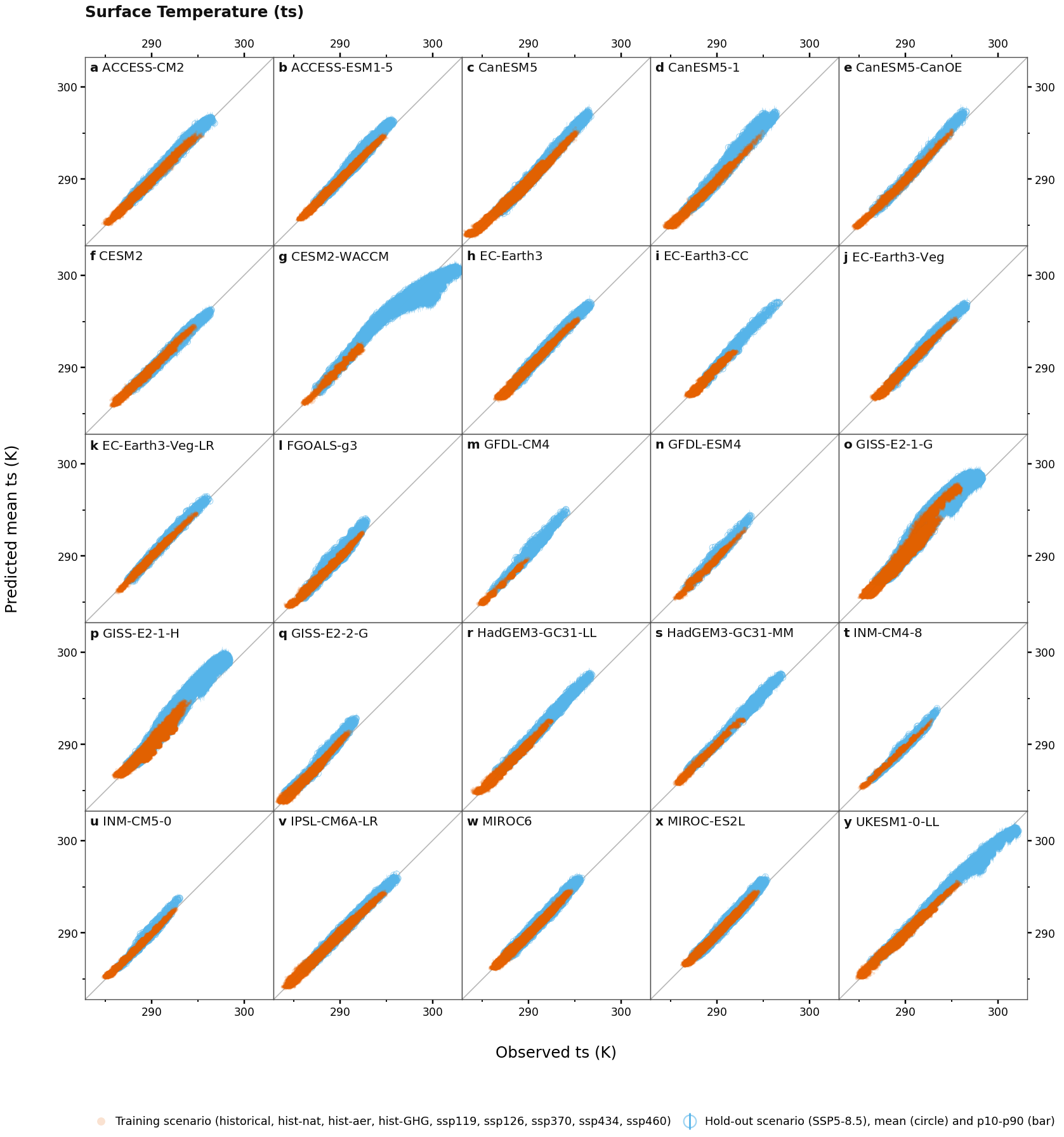}
  \caption{\textbf{GCMagicc-XS per-model surface temperature (\textit{ts}): observed versus predicted.} As Fig.~\ref{fig:supp-extrapolation-XS-tas} but for surface temperature.}
  \label{fig:supp-extrapolation-XS-ts}
\end{figure}

We focus on surface air temperature (Fig.~\ref{fig:supp-extrapolation-XS-tas}) and near-surface specific humidity (Fig.~\ref{fig:supp-extrapolation-XS-huss}) because these are the variables that show the largest relative change under \textit{ssp585} relative to the training-scenario range and therefore provide the most informative qualitative consistency check under high-end forcing; analogous per-model observed-versus-predicted panels are shown for precipitation (Fig.~\ref{fig:supp-extrapolation-XS-pr}), daily maximum near-surface air temperature (Fig.~\ref{fig:supp-extrapolation-XS-tasmax}), daily minimum near-surface air temperature (Fig.~\ref{fig:supp-extrapolation-XS-tasmin}), near-surface relative humidity (Fig.~\ref{fig:supp-extrapolation-XS-hurs}), sea-level pressure (Fig.~\ref{fig:supp-extrapolation-XS-psl}), surface downwelling shortwave radiation (Fig.~\ref{fig:supp-extrapolation-XS-rsds}), near-surface wind speed (Fig.~\ref{fig:supp-extrapolation-XS-sfcWind}) and surface temperature (Fig.~\ref{fig:supp-extrapolation-XS-ts}). Each per-model panel shows the pooled in-distribution training scenarios (historical and SSP families, in orange) and the held-out \textit{ssp585} (in light blue) on the same pair of axes. Each plotted point is one monthly global-mean value for one CMIP6 ensemble member (up to 10 per model--scenario) paired with its GCMagicc-XS ensemble-mean prediction; the visible cloud width along the 1:1 line therefore reflects both the seasonal cycle and the CMIP6 inter-member spread within that driving model. Skill is near uniform across driving models with one exception: CESM2-WACCM is flagged as an RMSE outlier on seven of ten variables under held-out \textit{ssp585}, the only model so flagged more than three times.

\clearpage

\clearpage

\clearpage

\clearpage

\clearpage

\clearpage

\clearpage

\clearpage

\clearpage

\section{Comparison of GCMagicc versus CMIP6 ESMs}
\label{sec:supp-validation-recipes}

Systematic evaluation frameworks for climate emulators are emerging. Here we use ten benchmark diagnostics. These are not all independent of the machine-learning training objective: the stochastic generator is trained with an engression objective based on the energy score, and the diagnostics that compare the emulated fields themselves (GlobalTimeseries, BiasMaps, SeasonalCycle, ZonalMeans, Variability and Histograms) evaluate a closely related functional, so a score below one on those diagnostics is a consistency check rather than independent validation. The diagnostics built on derived diagnostics (IndicatorFreq, ENSOTelecon, ENSOCharacteristics and HumidityCoupling) are not targeted by the training objective and carry the independent evidence. The frozen database contains 4,539,079 \textit{gofnc} records across GCMagicc and GCMagicc-CE, of which 3,974,527 belong to the ten diagnostic domains documented here. The database includes 1,265,222 records for the SSP2-4.5 hold-out comparison.

\subsection{Fidelity metric diagnostics to compare GCMagicc vs CMIP6}

The validation goodness-of-fit statistics we employ are $L_2$ distances between a specific CMIP6 run (identified by \textit{source\_id} (model), \textit{experiment\_id} (scenario), and \textit{member\_id} (ensemble member, capped at 4 per scenario)) and a comparator. We use ten diagnostic families across complementary aspects of the climate state. Per-diagnostic record counts in the consolidated database are reported in Table~\ref{tab:supp-recipe-counts} and descriptions are provided below.

All diagnostics first align File A and File B on their common coordinates before computing statistics. Let $\Omega$ denote that common support and let $w_i$ denote cosine-latitude area weights where spatial weighting is used, with $w_i=1$ for unweighted statistics. We write
\begin{equation}
\mathrm{Dev}(A,B) = \frac{\sum_{i \in \Omega} w_i (A_i - B_i)}{\sum_{i \in \Omega} w_i},
\qquad
\mathrm{RMSE}(A,B) = \left( \frac{\sum_{i \in \Omega} w_i (A_i - B_i)^2}{\sum_{i \in \Omega} w_i} \right)^{1/2},
\label{eq:supp-dev-rmse}
\end{equation}
and, for histogram-based diagnostics,
\begin{equation}
\|h_A - h_B\|_{L^2} = \left( \sum_b \left[h_A(z_b) - h_B(z_b)\right]^2 \Delta z_b \right)^{1/2}.
\label{eq:supp-l2}
\end{equation}
Temporal means and quantiles are always taken over the diagnostic-specific window encoded in the relevant record. Pearson correlations use the standard covariance normalisation on the aligned monthly series.

\subsubsection{GlobalTimeseries}
For each variable the diagnostic forms the global cosine-latitude-weighted mean time series, aligns the two series in time, and computes the root-mean-square difference. Global-annual-mean \textit{tas} is a predictor in the default GCMagicc setup and is therefore not informative for this metric (other than as a consistency check), but the other variables (\textit{hurs}, \textit{huss}, \textit{pr}, \textit{psl}, \textit{rsds}, \textit{sfcWind}) show informative global monthly-resolved evolution over the 21st century.

\subsubsection{BiasMaps}
BiasMaps investigates regional biases of the gridded $1^{\circ}\times 1^{\circ}$ output, for both median patterns and tails (10th and 90th percentiles), with RMSE computed per month or season, variable, 21-year window, scenario and percentile. The diagnostic compares seasonal or annual spatial fields after selecting the diagnostic window and then aggregating over time with the configured mean, median, or percentile. If $F^{(s,q)}_{A}(x)$ and $F^{(s,q)}_{B}(x)$ denote the season-$s$, aggregation-$q$ fields at grid cell $x$, the stored difference field is $D(x)=F^{(s,q)}_{A}(x)-F^{(s,q)}_{B}(x)$. The $L^2$ norm is the RMSE obtained by applying Eq.~\ref{eq:supp-dev-rmse} to $D(x)$ with cosine-latitude weights. Default windows depend on scenario class (historical, SSP, overlap, or trailing 20-year windows).

\subsubsection{SeasonalCycle}
SeasonalCycle computes monthly climatologies for each IPCC AR6 region and compares the resulting 12-month curves, since phase and amplitude of the seasonal cycle matter for agriculture, hydrology and energy demand. The principal statistic is \textit{Seasonal\_RMSE}, the RMSE across the 12 monthly climatological values.

\subsubsection{ZonalMeans}
ZonalMeans forms monthly climatologies, averages them zonally over longitude, and then averages over months to obtain annual zonal profiles for File A and File B. The frozen diagnostic stores \textit{ZonalClim\_RMSE} as the unweighted RMSE across the aligned latitude coordinates of those two annual profiles, alongside the difference in cosine-latitude-weighted high-minus-low-latitude gradients.

\subsubsection{Variability}
Variability is analogous to SeasonalCycle but is applied to monthly standard deviation climatologies rather than monthly means. Before computing standard deviations, each grid point is detrended separately for each calendar month using a 21-year running mean with 10-year trend-preserving edge treatment.

\subsubsection{Histograms}
Histograms pool all valid values from the chosen time window within each IPCC AR6 region, using the common spatial grid after alignment. The diagnostic first determines a common set of histogram edges from the pooled minima and maxima across both files, then computes regional density histograms $h_A$ and $h_B$ on those common bins. It stores the Kolmogorov-Smirnov distance $\sup_z |F_A(z)-F_B(z)|$ and the histogram $L^2$ distance given by Eq.~\ref{eq:supp-l2}, the latter being used for the energy distances.

\subsubsection{HumidityCoupling}\label{recipe:humidity-coupling}
Humidity coupling evaluates the co-variability between global-mean temperature and either relative humidity (\textit{hurs}) or specific humidity (\textit{huss}). The diagnostics computes grid-cell correlations between monthly \textit{tas} anomalies and humidity anomalies, forms the correlation-difference map, and provides the weighted mean and weighted RMSE of that map, the latter again being used for the energy distances.

\subsubsection{IndicatorFreq}
IndicatorFreq analyses six monthly climate indices: Ni\~no-3.4, Ni\~no-3, Ni\~no-4, SOI, PDO, and NAO. Ni\~no indices are built from area-weighted tropical Pacific SST or near-surface temperature anomalies; SOI is the Darwin-minus-Tahiti sea-level-pressure anomaly difference; PDO is the first principal component from an SVD of North Pacific SST anomalies weighted by $\sqrt{\cos \phi}$; and NAO is the Azores-minus-Iceland sea-level-pressure anomaly difference. All six indices use the same post-processing: a centred 3-month running mean followed by removal of low-frequency variability with a fourth-order Butterworth low-pass filter with a 30-year cutoff, subtracted from the smoothed series. The diagnostic estimates power spectral density with Welch's method, a Hann window, and $n_{\mathrm{perseg}}=\min(256,\lfloor N/2 \rfloor)$ for series length $N \ge 32$ (otherwise the full series length), interpolates $\log_{10}$ PSD to 150 common log-spaced frequencies, and we use the resulting RMSE between CMIP6 and GCMagicc for the energy distances.

\subsubsection{ENSOTelecon}
ENSO teleconnections are diagnosed by regressing DJF anomalies of each target variable (tas, pr) on the 1-year lagged OND Ni\~no-3.4 index, producing a slope map for File A and for File B. The anomaly preprocessing uses smoothing and detrending prior to the seasonal regression. The RMSE is computed from the map difference after restricting latitude to $[-60^\circ,60^\circ]$ and applying cosine-latitude weighting.

\subsubsection{ENSOCharacteristics}
ENSO characteristics are derived from a Ni\~no-3.4 index built from monthly anomalies over $5^\circ$S--$5^\circ$N and $190^\circ$E--$240^\circ$E, followed by linear detrending and triangular 5-month smoothing. The lifecycle diagnostic regresses each lagged monthly anomaly on December Ni\~no-3.4 anomalies over lags $-36,\ldots,+36$ months; \textit{ENSO\_lifecycle\_RMSE} is the RMSE between the two resulting slope vectors over common lags. Event duration is the number of contiguous lags around zero for which the absolute lifecycle slope exceeds $0.25$, and the diagnostic stores both the signed duration difference and its absolute value as \textit{ENSO\_duration\_RMSE}. ENSO amplitude is the sample standard deviation of DJF means of the smoothed Ni\~no-3.4 series, again stored both as a signed difference and as \textit{ENSO\_amplitude\_RMSE}.


\begin{table}[t]
\centering
\begingroup
\renewcommand{\arraystretch}{1}
\footnotesize
\caption{Number of CMIP6 ESM and GCMagicc comparisons we performed. The database holds 4,539,079 \textit{gofnc} records in total, of which 3,974,527 fall in these ten diagnostic domains, 1,265,222 belong to the SSP2-4.5 hold-out comparison, and 649,058 enter the observation-referenced displays of Figs.~\ref{fig:supp-edisto-observational}--\ref{fig:supp-edisto-observational-part3}. Note that the below counts comprise RMSE, but also amplitude differences, phase-shifts, and absolute differences.}
\label{tab:supp-recipe-counts}
\begin{tabular}{lrr}
\tophline
Diagnostic & GCMagicc & GCMagicc-CE \\
\middlehline
BiasMaps & 156,270 & 91,770 \\
ENSOCharacteristics & 2,950 & 2,045 \\
ENSOTelecon & 2,092 & 1,316 \\
GlobalTimeseries & 10,298 & 6,618 \\
Histograms & 597,284 & 343,244 \\
HumidityCoupling & 2,368 & 1,636 \\
IndicatorFreq & 10,638 & 5,004 \\
SeasonalCycle & 892,446 & 507,906 \\
Variability & 890,706 & 431,520 \\
ZonalMeans & 10,238 & 8,178 \\
\middlehline
\textbf{Total} & 2,575,290 & 1,399,237 \\
\bottomhline
\end{tabular}
\endgroup
\end{table}

\subsection{Validation: Energy distances and energy distance scores}\label{sec:supp-energy-distances}

The database comparison keys used here are 'goodness-of-fit' GOF statistics with key letters, as explained in the following table.

\begin{table}[htbp]
    \centering
    \caption{Database comparison keys and corresponding data-source comparisons.
    Here, \texttt{N} denotes a GCMagicc member or configuration in the legacy
    database code, \texttt{C} the paired CMIP6 model/member, \texttt{Z} a
    different randomly selected CMIP6 model, and \texttt{O} an observational
    or reanalysis product.}
    \label{tab:database-comparison-keys}
    \begin{tabular}{@{}ll@{}}
        \toprule
        \textbf{Key} & \textbf{Comparison} \\
        \midrule
        \texttt{GOFNC} & Paired GCMagicc and CMIP6 outputs \\
        \texttt{GOFNN} & GCMagicc members or configurations \\
        \texttt{GOFCC} & Initial-condition members of the same CMIP6 model \\
        \texttt{GOFCZ} & Target CMIP6 model and another CMIP6 model \\
        \texttt{GOFZZ} & Members within the other CMIP6 models \\
        \texttt{GOFOC} & Observations/reanalysis and CMIP6 \\
        \texttt{GOFON} & Observations/reanalysis and GCMagicc \\
        \bottomrule
    \end{tabular}
\end{table}

For random vectors $X$ and $Y$ with independent copies $X'$ and $Y'$, the energy distance between the distribution $P_X$ and $P_Y$ of $X$ and $Y$ respectively is
\[
\mathrm{ED}(P_X,P_Y)=2\,\mathbb{E}\|X-Y\|_2-\mathbb{E}\|X-X'\|_2-\mathbb{E}\|Y-Y'\|_2.
\]
By construction, $\mathrm{ED}$ reflects distributional agreement rather than matching a single moment.

While these energy distances all approach zero when the emulator matches the original ESM distribution, their magnitudes are not directly comparable across diagnostics or variables. We therefore define the energy-distance score as
\[
S(M,C;Z)=\frac{\mathrm{ED}(M,C)}{\mathrm{ED}(C,Z)},
\]
where $\mathrm{ED}(M,C)$ is the energy distance between the calibrated GCMagicc $M$ and a CMIP6 model $C$, and $\mathrm{ED}(C,Z)$ is the energy distance between that CMIP6 model $C$ and a random selection of other CMIP6 models $Z$ for the same experiment. $S<1$ indicates that the emulator is closer to a given ESM than the ESMs are to each other for that metric and experiment (green area in the last column in Figure~\ref{fig:validation-diagnostics}).

For observational data $O_i,i=1,\ldots,n$ with empirical distribution $P_{n,O}$ and model data $M_i,i=1,\ldots,n$ with empirical distribution $P_{n,M}$ and an independent set of draws $M'_i, i=1,\ldots,n$ generated from the same model, we get with the definition above
\[
\mathrm{ED}(P_{n,O},P_{n,M})= n^{-1}\sum_{i=1}^n \,\big( 2 \| O_i - M_i\|_2 - \|M_i - M'_i\|_2 -  \|O_i - O'_i\|_2\big),
\]
Since the real world provides only a single realisation but the  term $\|O_i - O'_i\|_2$ is independent of the model $M$, it can safely be dropped to get
\[
\mathrm{ED}_{\mathrm{obs}}(P_{n,O},P_{n,M})= n^{-1}\sum_{i=1}^n \,\big( 2 \| O_i - M_i\|_2 - \|M_i - M'_i\|_2\big),
\]
 Because there is no second observational ensemble member, this quantity does not approach zero under a perfect distributional match; instead it asymptotically approaches $\mathbb{E} \| O - O'\|_2$, where $O$ and $O'$ denote two hypothetical ensemble members of the real world (Figure~\ref{fig:observational-alignment}).

\subsection{Extended data tables and alternate draws of Figure~\ref{fig:validation-diagnostics}}

The following data tables provide the row-specific selections, energy-distance-score summaries, and 10 alternate random draws of Figure~\ref{fig:validation-diagnostics} in the main manuscript. Each draw is an independent random selection of metric diagnostics from the same pool, hence directly comparable to the main text Figure~\ref{fig:validation-diagnostics}. The alternate draws are numbered 001--010.

\begin{table}[!ht]
\centering
\caption{Row-specific selections and validation metrics corresponding to the retained rows in Figure~\ref{fig:validation-diagnostics}. Energy-distance scores computed from the smallest samples ($n=1$ to $n=4$) can take negative values under the estimator used here; these are retained rather than truncated. Given the close agreement between ensemble members at these sample sizes this does not materially affect the relative comparisons reported.}
\label{tab:validation-diagnostics-row-selections}
\fontsize{6}{6.2}\selectfont
\setlength{\tabcolsep}{2pt}
\renewcommand{\arraystretch}{0.72}
\begin{tabularx}{\textwidth}{@{}l l >{\raggedright\arraybackslash}X >{\raggedright\arraybackslash}X@{}}
\tophline
Panel & diagnostic & Selected case & Metrics summary \\
\middlehline
a-d & BiasMaps & source\_id: ACCESS-CM2 \newline member\_id: r1i1p1f1 \newline experiment: ssp245 \newline variable: tas \newline metrictype: Map\_RMSE\_2080to2099\_annual\_perc90 & SCOREEDISTC: 0.339; GOFNC: 0.973254 (n=1); GOFNN: 0.578135 (n=2); GOFCC: 0.309362 (n=3); GOFCZ: 1.88792 (Z n=4); GOFZZ: 0.340901 (Z n=4); Z coverage: 4/5 (partial available case); EDISTNC: 1.059; EDISTCZ: 3.126 (n=4) \\
e-g & GlobalTimeseries & source\_id: IPSL-CM6A-LR \newline member\_id: r11i1p1f1 \newline variable: psl \newline scenario: ssp245 \newline metrictype: RMSE\_global & SCOREEDISTC: mean=0.053 (n=1); GOFNC: 11.8997 (n=1); GOFNN: 12.8161 (n=1); GOFCC: 4.99083 (n=3); GOFCZ: 62.3034 (Z n=5); GOFZZ: 5.86809 (Z n=5); Z coverage: 5/5 (complete); EDISTNC: 5.993; EDISTCZ: 113.748 (n=5) \\
h-j & Histograms & source\_id: ACCESS-CM2 \newline member\_id: r1i1p1f1 \newline variable: tasmax \newline scenario: ssp245 \newline region 1: ARABIAN-SEA; metrictype: Histo\_L2\_ARABIAN-SEA\_2081to2100 \newline region 2: RUSSIAN-ARCTIC; metrictype: Histo\_L2\_RUSSIAN-ARCTIC\_2081to2100 \newline shown panels: h = Arabian-Sea; i = Russian-Arctic & ARABIAN-SEA: SCOREEDISTC: 1.434; GOFNC: 0.106665 (n=1); GOFNN: 0.0641687 (n=2); GOFCC: 0.0146257 (n=3); GOFCZ: 0.0597889 (Z n=3); ARABIAN-SEA: GOFZZ: 0.0111461 (Z n=3); Z coverage: 3/4 (partial available case); EDISTNC: 0.135; EDISTCZ: 0.094 (n=3); RUSSIAN-ARCTIC: SCOREEDISTC: 1.563; GOFNC: 0.0518836 (n=1); GOFNN: 0.0154951 (n=2); GOFCC: 0.00777645 (n=3); GOFCZ: 0.034306 (Z n=3); RUSSIAN-ARCTIC: GOFZZ: 0.00932748 (Z n=3); Z coverage: 3/4 (partial available case); EDISTNC: 0.080; EDISTCZ: 0.052 (n=3) \\
k-m & SeasonalCycle & source\_id: HadGEM3-GC31-LL \newline member\_id: r1i1p1f3 \newline variable: ts \newline scenario: ssp245 \newline region 1: BAY-OF-BENGAL; metrictype: Seasonal\_RMSE\_BAY-OF-BENGAL\_2081to2100 \newline region 2: NEW-ZEALAND; metrictype: Seasonal\_RMSE\_NEW-ZEALAND\_2081to2100 \newline shown panels: k = Bay-of-Bengal; l = New-Zealand & BAY-OF-BENGAL: SCOREEDISTC: 0.329; GOFNC: 0.318403 (n=1); GOFNN: 0.229299 (n=1); GOFCC: 0.0937754 (n=3); GOFCZ: 0.556747 (Z n=3); BAY-OF-BENGAL: GOFZZ: 0.0648674 (Z n=3); Z coverage: 3/4 (partial available case); EDISTNC: 0.314; EDISTCZ: 0.955 (n=3); NEW-ZEALAND: SCOREEDISTC: 0.702; GOFNC: 0.488596 (n=1); GOFNN: 0.280022 (n=1); GOFCC: 0.130764 (n=3); GOFCZ: 0.521203 (Z n=3); NEW-ZEALAND: GOFZZ: 0.104618 (Z n=3); Z coverage: 3/4 (partial available case); EDISTNC: 0.566; EDISTCZ: 0.807 (n=3) \\
n-p & ZonalMeans & source\_id: IPSL-CM6A-LR \newline member\_id: r10i1p1f1 \newline variable: tas \newline scenario: ssp245 \newline metrictype: ZonalClim\_RMSE\_tas\_full & SCOREEDISTC: 0.138; GOFNC: 0.374453 (n=1); GOFNN: 0.054541 (n=1); GOFCC: 0.150019 (n=3); GOFCZ: 2.31859 (Z n=5); GOFZZ: 0.54091 (Z n=5); Z coverage: 5/5 (complete); EDISTNC: 0.544; EDISTCZ: 3.946 (n=5) \\
q-s & Variability & source\_id: IPSL-CM6A-LR \newline member\_id: r11i1p1f1 \newline variable: tasmax \newline scenario: ssp245 \newline region 1: EQUATORIAL-ATLANTIC-OCEAN; metrictype: Variability\_RMSE\_EQUATORIAL-ATLANTIC-OCEAN\_full \newline region 2: N-PACIFIC-OCEAN; metrictype: Variability\_RMSE\_N-PACIFIC-OCEAN\_full \newline shown panels: q = Equatorial.Atlantic-Ocean; r = N.Pacific-Ocean & EQUATORIAL-ATLANTIC-OCEAN: SCOREEDISTC: 12.874; GOFNC: 0.418542 (n=1); GOFNN: 0.0530612 (n=1); GOFCC: 0.026107 (n=1); GOFCZ: 0.0632225 (Z n=5); EQUATORIAL-ATLANTIC-OCEAN: GOFZZ: 0.0414671 (Z n=5); Z coverage: 5/5 (complete); EDISTNC: 0.758; EDISTCZ: 0.059 (n=5); N-PACIFIC-OCEAN: SCOREEDISTC: 5.301; GOFNC: 0.281592 (n=1); GOFNN: 0.0227166 (n=1); GOFCC: 0.0196799 (n=1); GOFCZ: 0.0750647 (Z n=5); N-PACIFIC-OCEAN: GOFZZ: 0.0322043 (Z n=5); Z coverage: 5/5 (complete); EDISTNC: 0.521; EDISTCZ: 0.098 (n=5) \\
t-w & HumidityCoupling & source\_id: CESM2-WACCM \newline member\_id: r2i1p1f1 \newline experiment: ssp245 \newline variable: hurs \newline metrictype: TempHum\_RMSE\_Diff\_hurs\_full & SCOREEDISTC: 0.888; GOFNC: 0.372297 (n=1); GOFNN: 0.297419 (n=1); GOFCC: 0.04765 (n=3); GOFCZ: 0.28067 (Z n=5); GOFZZ: 0.0637613 (Z n=5); Z coverage: 5/5 (complete); EDISTNC: 0.400; EDISTCZ: 0.450 (n=5) \\
x-aa & ENSOTeleconnections & source\_id: EC-Earth3-Veg \newline member\_id: r12i1p1f1 \newline experiment: ssp245 \newline variable: pr \newline metrictype: Map\_RMSE\_2015to2100\_annual & SCOREEDISTC: 1.357; GOFNC: 0.356968 (n=1); GOFNN: 0.311549 (n=1); GOFCC: 0.139724 (n=3); GOFCZ: 0.235148 (Z n=4); GOFZZ: 0.13702 (Z n=4); Z coverage: 4/5 (partial available case); EDISTNC: 0.263; EDISTCZ: 0.194 (n=4) \\
\bottomhline
\end{tabularx}
\end{table}

\begin{table*}[t]
\centering
\caption{Energy-distance score summaries corresponding to the histogram panels in Figure~\ref{fig:validation-diagnostics}. Entries give the median (10th--90th percentile range) and sample size. Three diagnostics have median scores above one over all records: Variability (1.6), HumidityCoupling (1.1) and ENSOTelecon (1.6). For these diagnostics GCMagicc is, at the median, further from the target CMIP6 model than a randomly chosen other CMIP6 model is, so the statement that GCMagicc matches every considered CMIP6 model better than an average other CMIP6 model holds for the mean percentile rank across the displayed diagnostic set and not for the individual diagnostics.}
\label{tab:validation-diagnostics-energy-distance}
\scriptsize
\setlength{\tabcolsep}{2pt}
\begin{tabularx}{\textwidth}{@{}l l c *{3}{>{\centering\arraybackslash}X} *{3}{r} >{\raggedright\arraybackslash}X@{}}
\tophline
Panel & diagnostic & Variable subset & \multicolumn{3}{c}{Energy-distance score} & \multicolumn{3}{c}{$n$} & Selected example \\
& & & all & var-subset & ssp245 & all & var-subset & ssp245 & \\
\middlehline
d & BiasMaps & tas & 0.27 (0.13--0.53) & 0.26 (0.13--0.45) & 0.22 (0.12--0.44) & 33750 & 3375 & 4050 & 0.34 \\
g & GlobalTimeseries & psl & 0.013 (-0.078--0.21) & 0.022 (0.0078--0.068) & 0.0047 (-0.12--0.15) & 2480 & 248 & 270 & 0.053 \\
j & Histograms & tasmax & 0.26 (0.025--0.97) & 0.24 (-0.013--0.85) & 0.21 (-0.0079--0.82) & 124120 & 12412 & 15080 & h: 1.4; i: 1.6 \\
m & SeasonalCycle & ts & 0.18 (0.018--0.66) & 0.18 (0.011--0.62) & 0.14 (-0.0067--0.51) & 107880 & 10788 & 15080 & k: 0.33; l: 0.70 \\
p & ZonalMeans & tas & 0.11 (0.021--0.37) & 0.11 (-0.0063--0.31) & 0.084 (0.0076--0.28) & 2440 & 244 & 270 & 0.14 \\
s & Variability & tasmax & 1.6 (0.37--6.1) & 1.8 (0.25--6.9) & 1.4 (0.13--5) & 89262 & 8932 & 15602 & q: 13; r: 5.3 \\
w & HumidityCoupling & hurs & 1.1 (0.73--1.6) & 1.1 (0.69--1.6) & 1 (0.63--1.6) & 496 & 248 & 54 & 0.89 \\
aa & ENSOTelecon & pr & 1.6 (0.85--2.3) & 1.4 (0.78--2.1) & 1.2 (0.34--2.7) & 424 & 212 & 52 & 1.4 \\
\bottomhline
\end{tabularx}
\end{table*}

\clearpage
\begin{figure}[H]
  \centering
  \includegraphics[width=\linewidth,height=0.85\textheight,keepaspectratio]{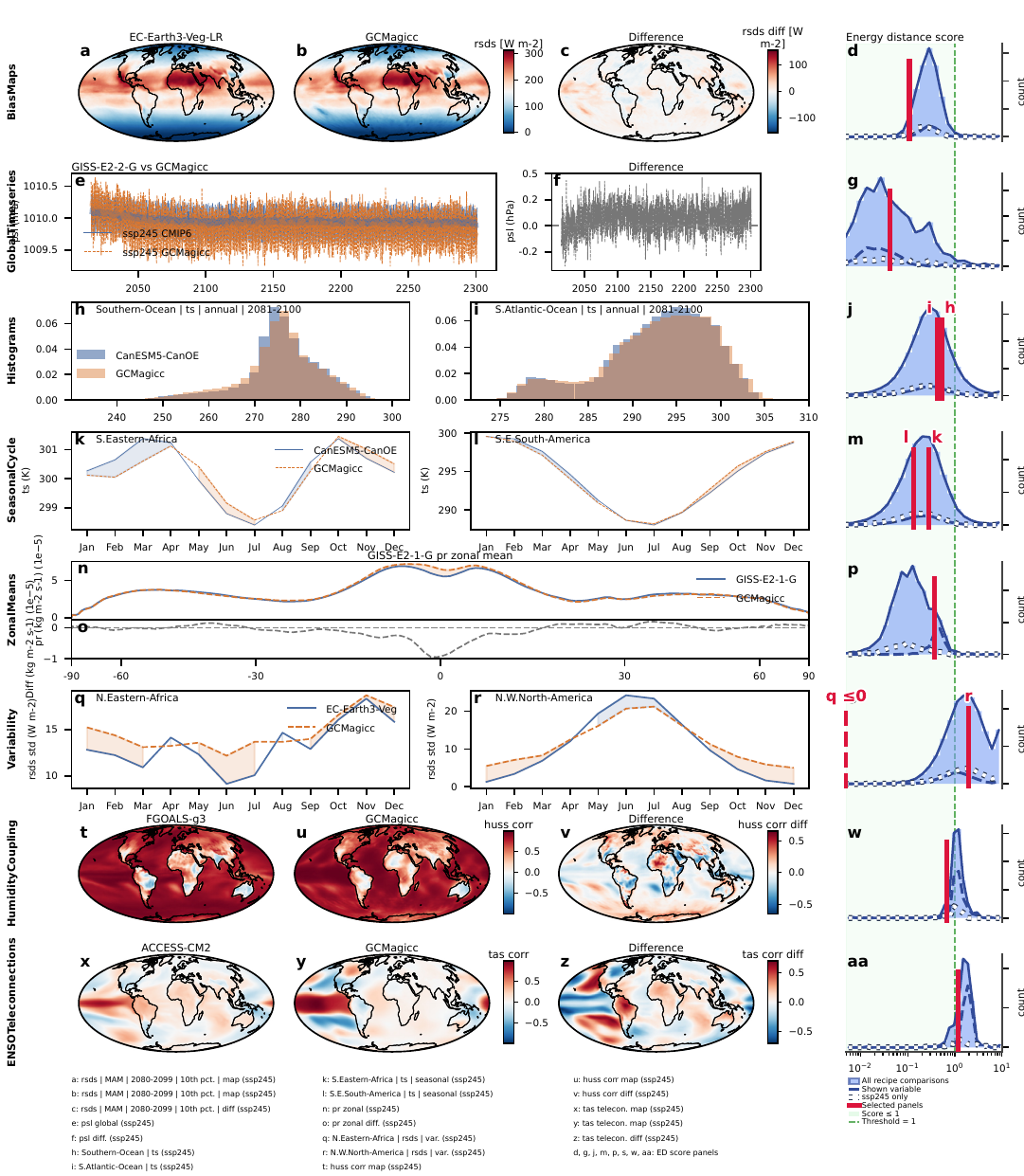}
  \caption{\textbf{Fidelity evaluation of GCMagicc vs CMIP6.} \emph{As Figure~\ref{fig:validation-diagnostics} in the manuscript, but random alternate draws of metric diagnostics}}
  \label{fig:supp-validation-alt-001}
\end{figure}

\clearpage
\begin{figure}[H]
  \centering
  \includegraphics[width=\linewidth,height=0.85\textheight,keepaspectratio]{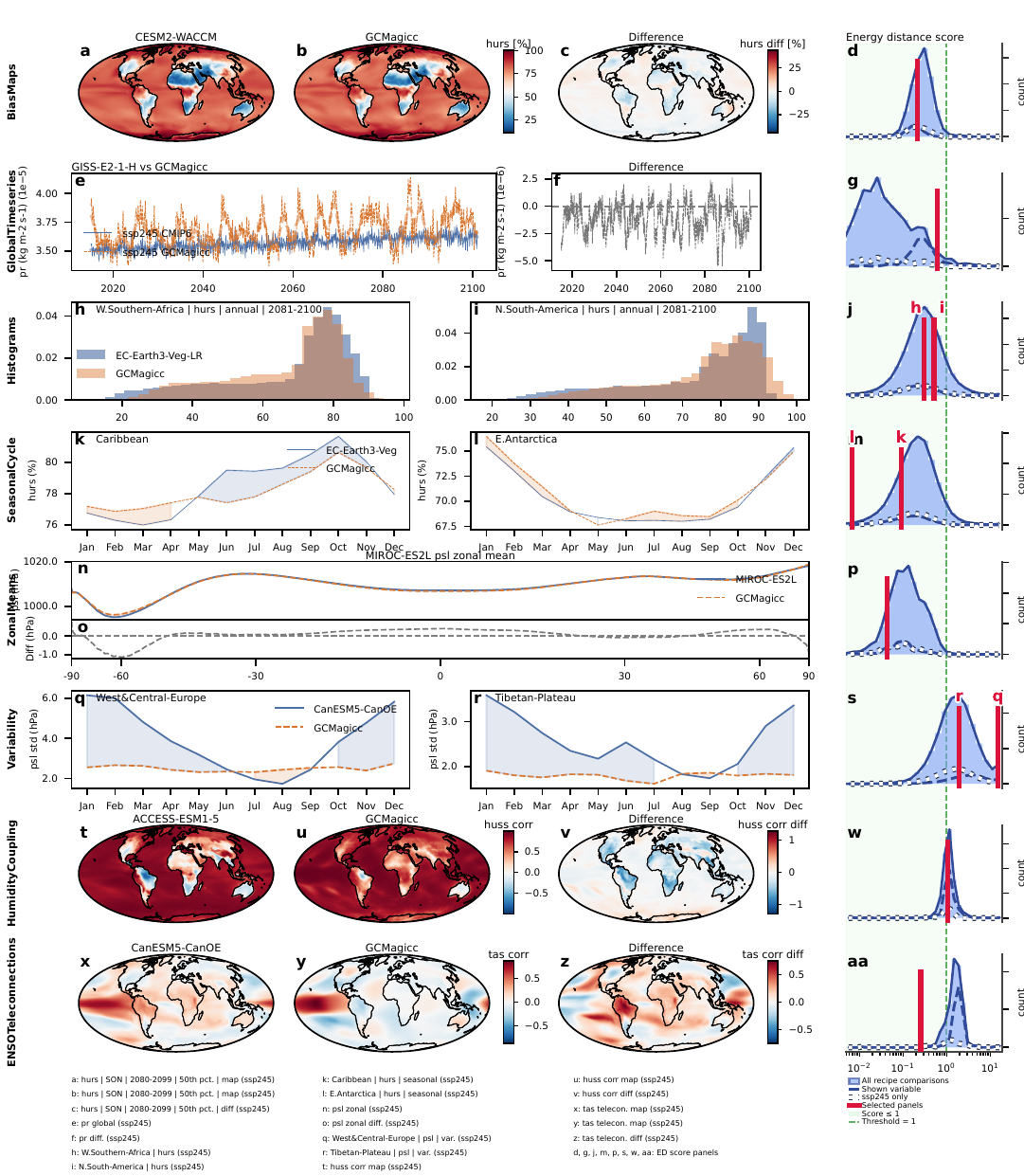}
  \caption{\textbf{Fidelity evaluation of GCMagicc vs CMIP6.} \emph{As Figure~\ref{fig:validation-diagnostics} in the manuscript, but random alternate draws of metric diagnostics}}
  \label{fig:supp-validation-alt-002}
\end{figure}

\clearpage
\begin{figure}[H]
  \centering
  \includegraphics[width=\linewidth,height=0.85\textheight,keepaspectratio]{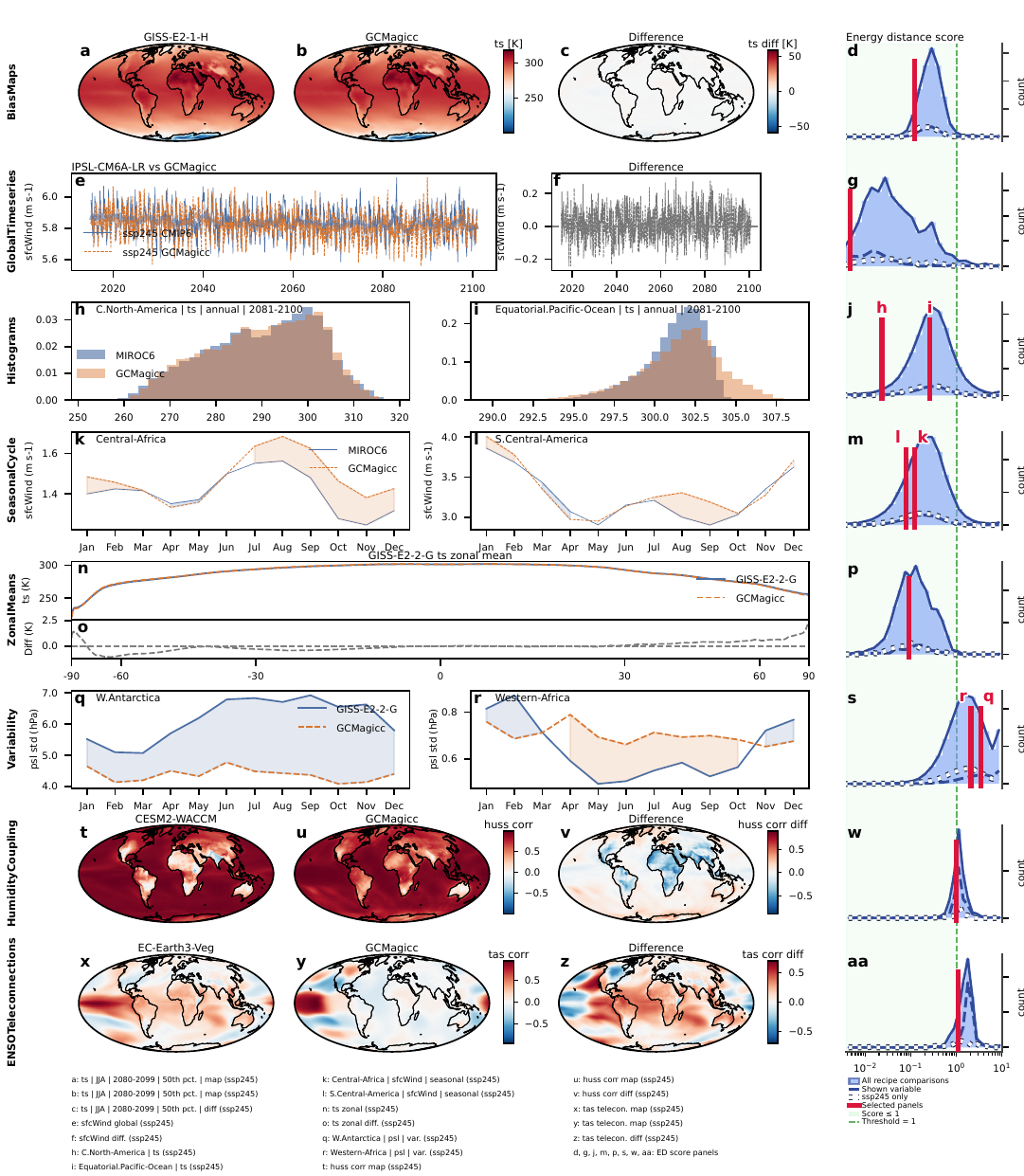}
  \caption{\textbf{Fidelity evaluation of GCMagicc vs CMIP6.} \emph{As Figure~\ref{fig:validation-diagnostics} in the manuscript, but random alternate draws of metric diagnostics}}
  \label{fig:supp-validation-alt-003}
\end{figure}

\clearpage
\begin{figure}[H]
  \centering
  \includegraphics[width=\linewidth,height=0.85\textheight,keepaspectratio]{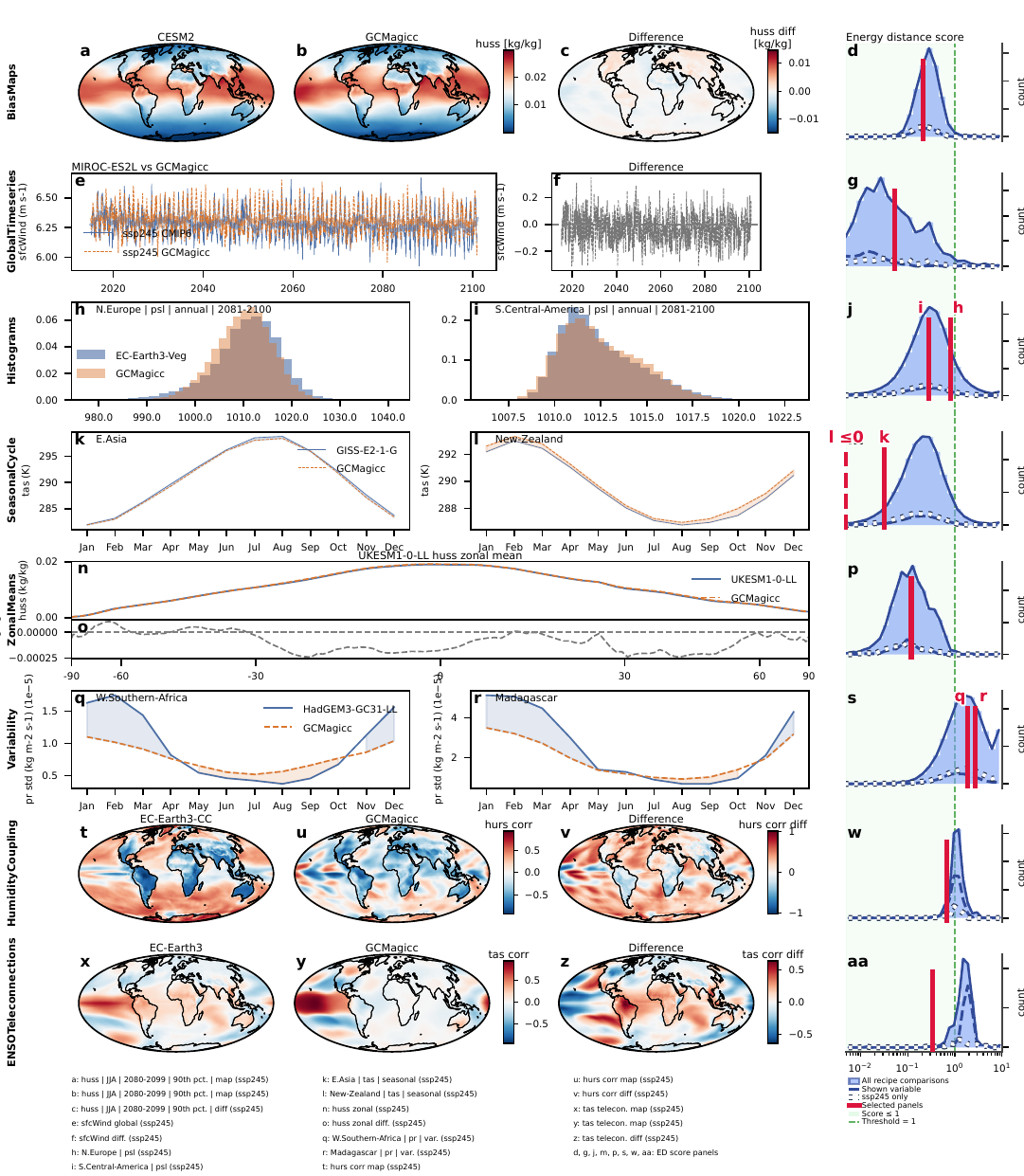}
  \caption{\textbf{Fidelity evaluation of GCMagicc vs CMIP6.} \emph{As Figure~\ref{fig:validation-diagnostics} in the manuscript, but random alternate draws of metric diagnostics}}
  \label{fig:supp-validation-alt-004}
\end{figure}

\clearpage
\begin{figure}[H]
  \centering
  \includegraphics[width=\linewidth,height=0.85\textheight,keepaspectratio]{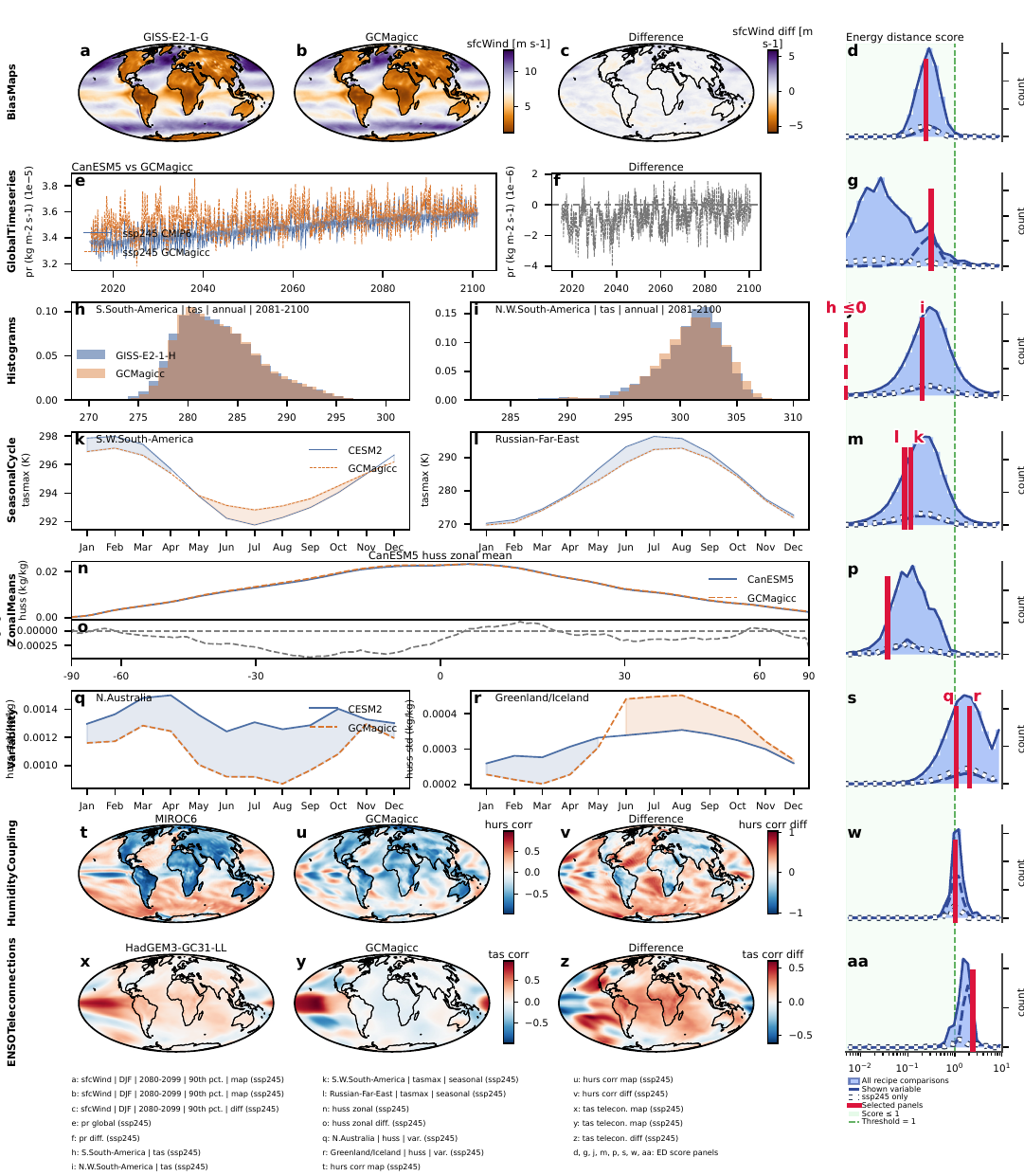}
  \caption{\textbf{Fidelity evaluation of GCMagicc vs CMIP6.} \emph{As Figure~\ref{fig:validation-diagnostics} in the manuscript, but random alternate draws of metric diagnostics}}
  \label{fig:supp-validation-alt-005}
\end{figure}

\clearpage
\begin{figure}[H]
  \centering
  \includegraphics[width=\linewidth,height=0.85\textheight,keepaspectratio]{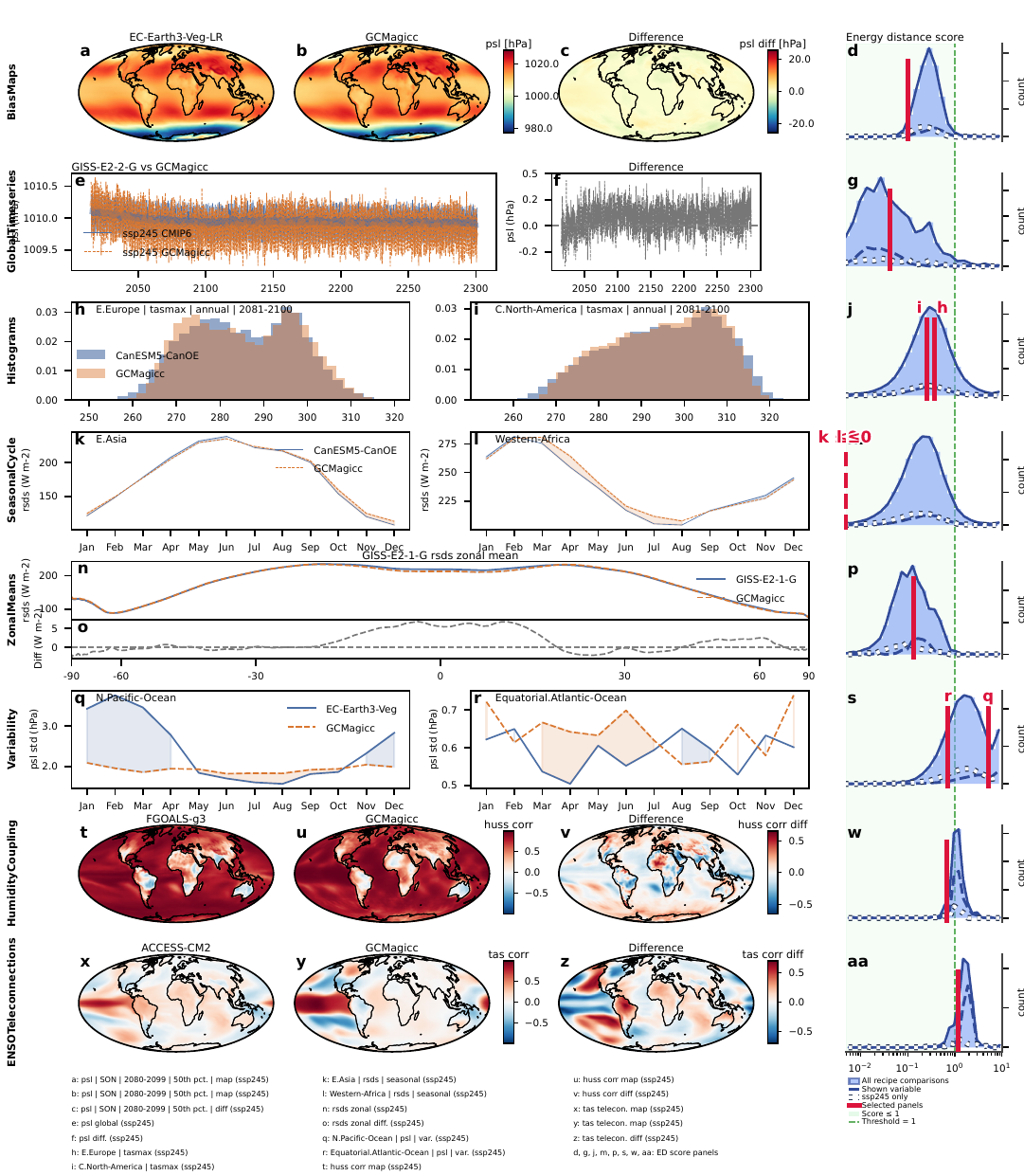}
  \caption{\textbf{Fidelity evaluation of GCMagicc vs CMIP6.} \emph{As Figure~\ref{fig:validation-diagnostics} in the manuscript, but random alternate draws of metric diagnostics}}
  \label{fig:supp-validation-alt-006}
\end{figure}

\clearpage
\begin{figure}[H]
  \centering
  \includegraphics[width=\linewidth,height=0.85\textheight,keepaspectratio]{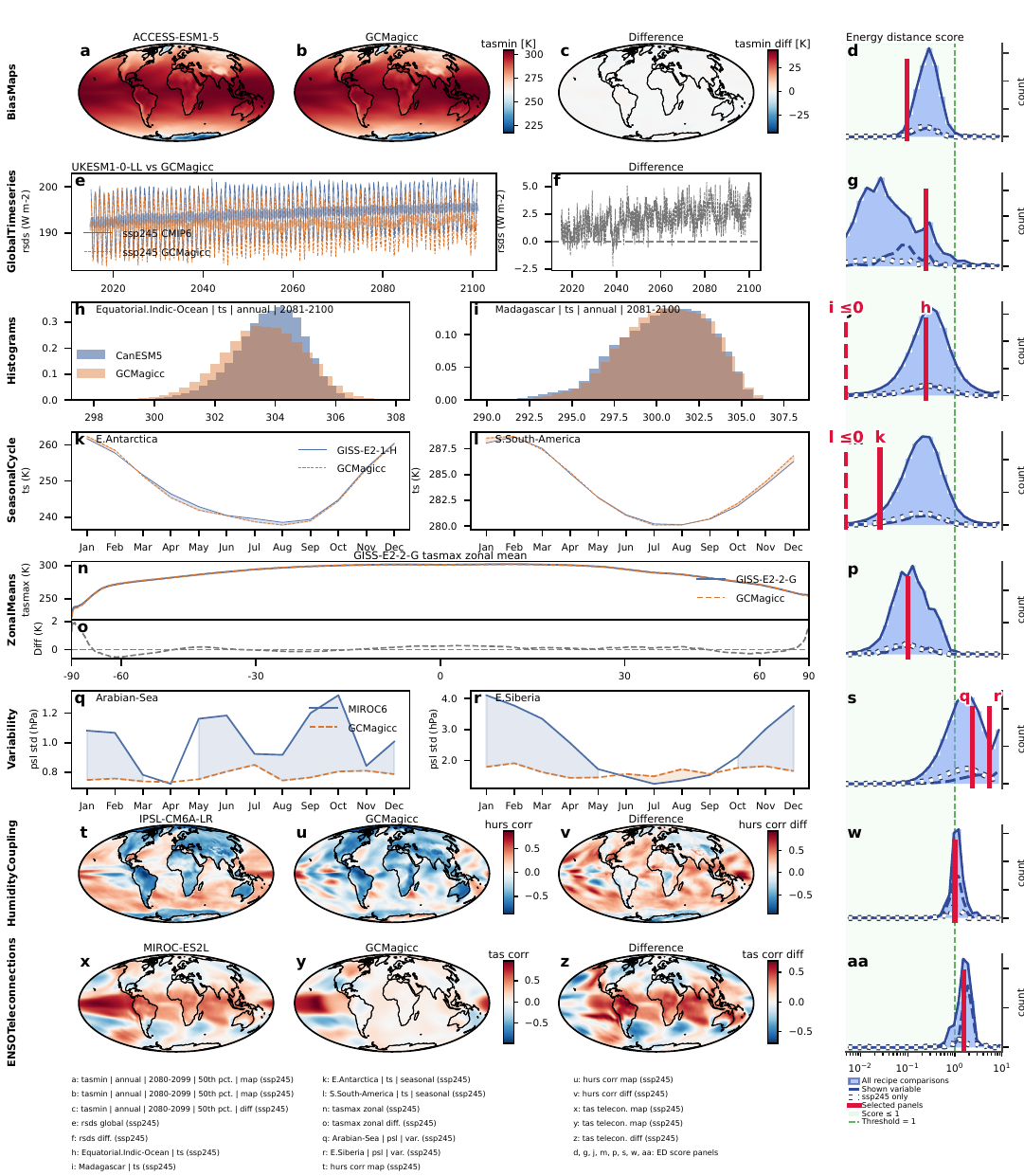}
  \caption{\textbf{Fidelity evaluation of GCMagicc vs CMIP6.} \emph{As Figure~\ref{fig:validation-diagnostics} in the manuscript, but a random alternate draw of metric diagnostics.}}
  \label{fig:supp-validation-alt-007}
\end{figure}

\clearpage
\begin{figure}[H]
  \centering
  \includegraphics[width=\linewidth,height=0.85\textheight,keepaspectratio]{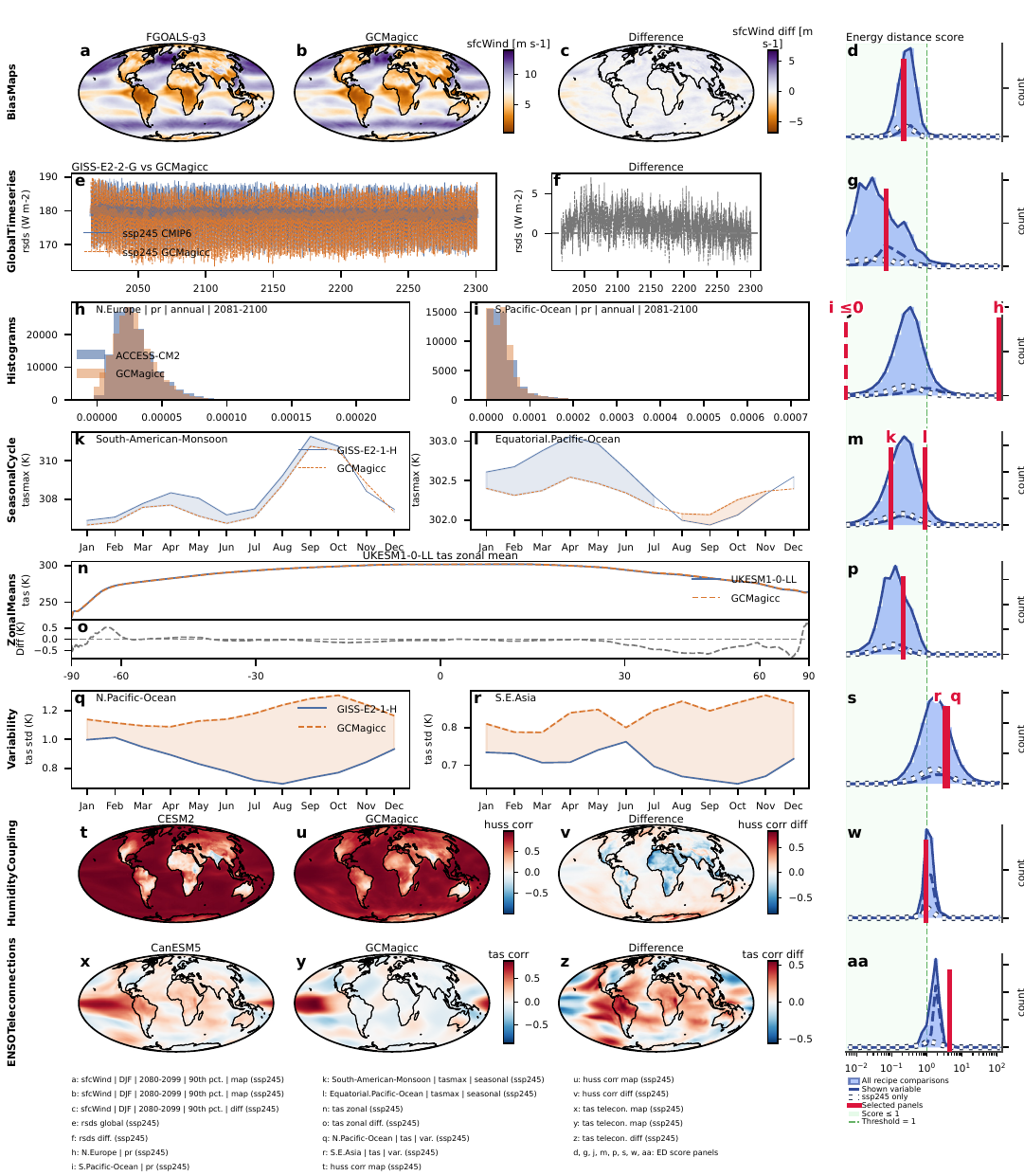}
  \caption{\textbf{Fidelity evaluation of GCMagicc vs CMIP6.} \emph{As Figure~\ref{fig:validation-diagnostics} in the manuscript, but random alternate draws of metric diagnostics}}
  \label{fig:supp-validation-alt-008}
\end{figure}

\clearpage
\begin{figure}[H]
  \centering
  \includegraphics[width=\linewidth,height=0.85\textheight,keepaspectratio]{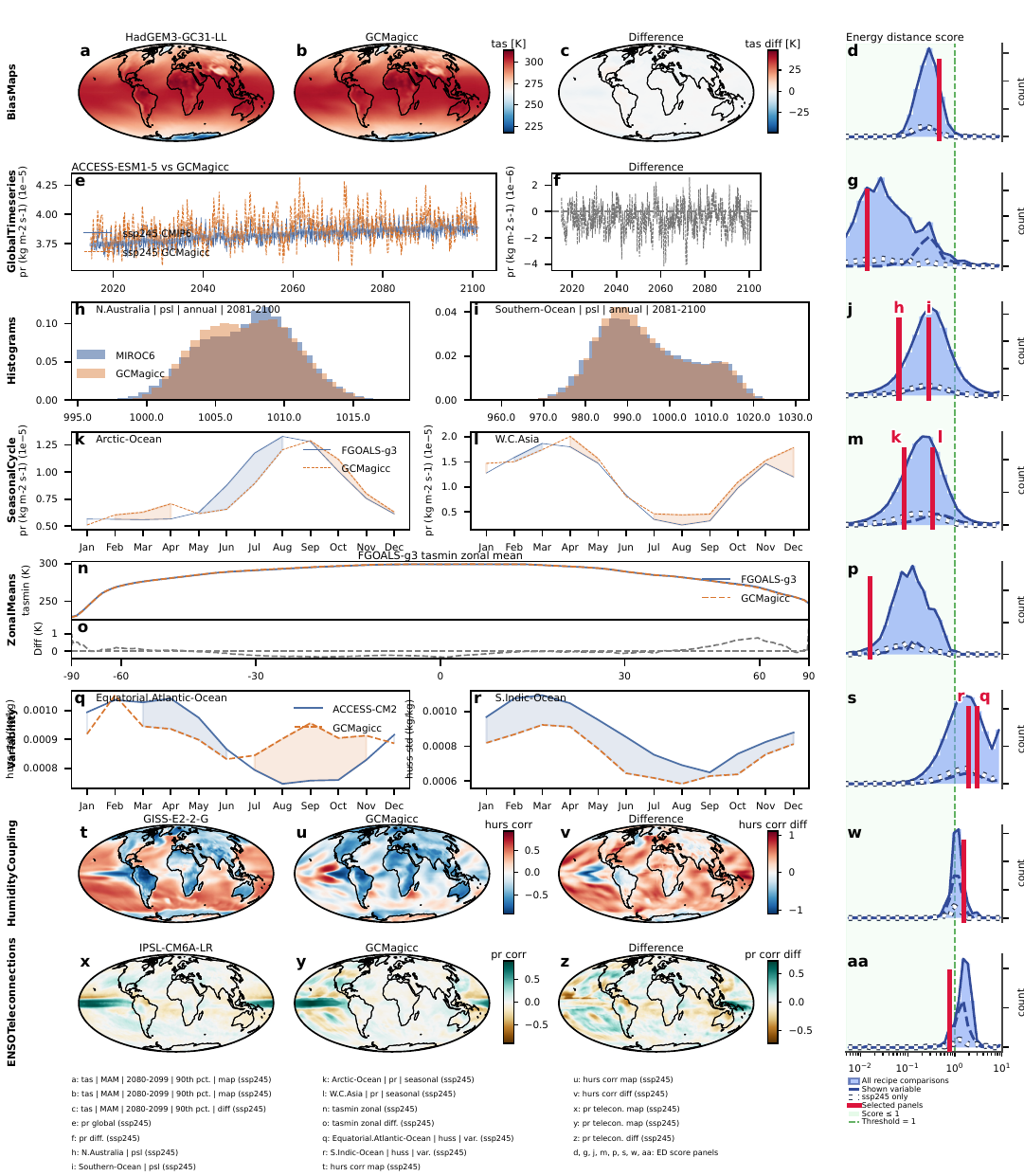}
  \caption{\textbf{Fidelity evaluation of GCMagicc vs CMIP6.} \emph{As Figure~\ref{fig:validation-diagnostics} in the manuscript, but random alternate draws of metric diagnostics}}
  \label{fig:supp-validation-alt-009}
\end{figure}

\clearpage
\begin{figure}[H]
  \centering
  \includegraphics[width=\linewidth,height=0.85\textheight,keepaspectratio]{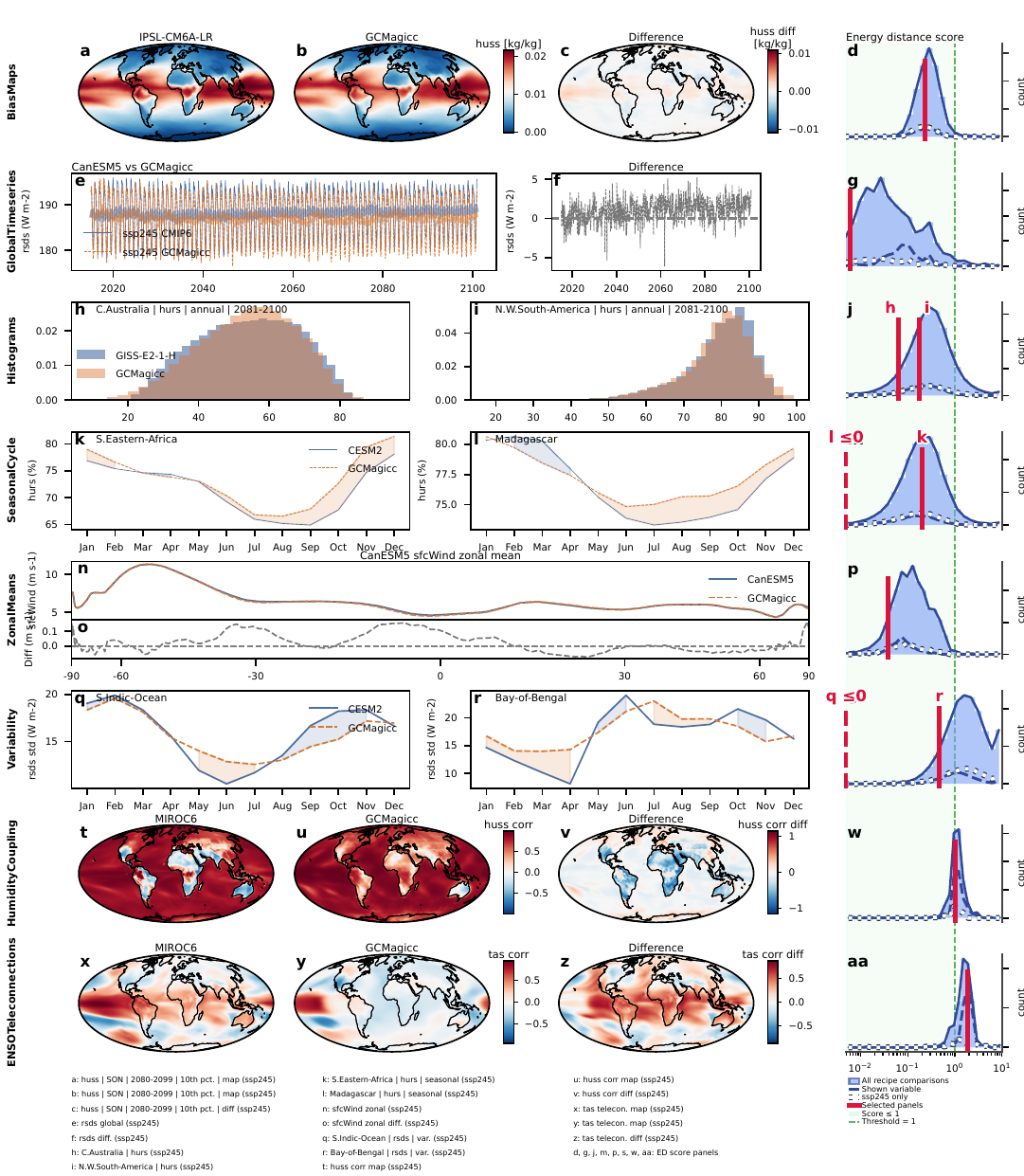}
  \caption{\textbf{Fidelity evaluation of GCMagicc vs CMIP6.} \emph{As Figure~\ref{fig:validation-diagnostics} in the manuscript, but random alternate draws of metric diagnostics}}
  \label{fig:supp-validation-alt-010}
\end{figure}

\clearpage
\begin{figure}[H]
\centering
\includegraphics[width=\textwidth,height=0.68\textheight,keepaspectratio]{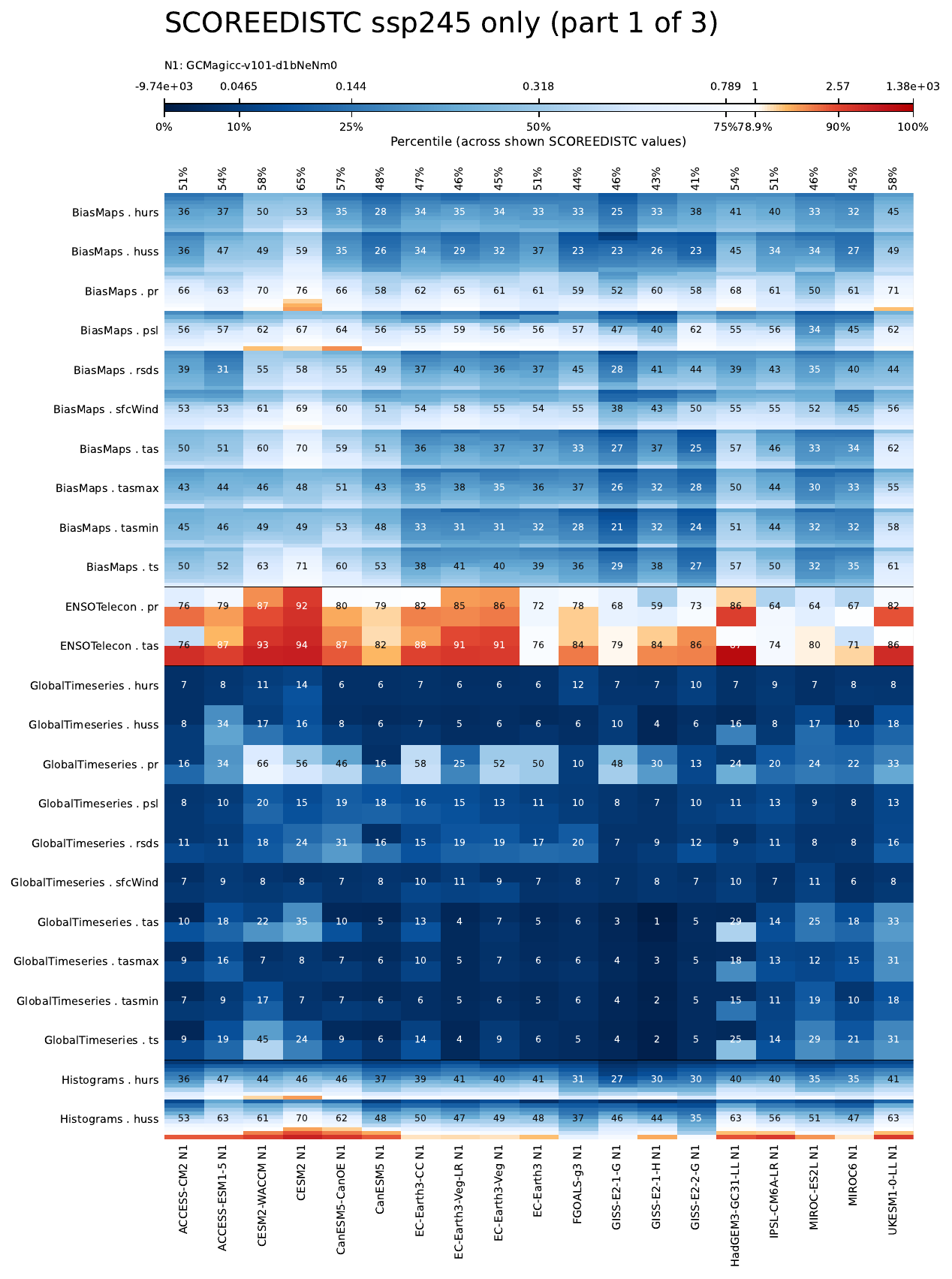}
\caption{\textbf{Part 1 Energy distance scores for the SSP2-4.5 hold-out scenario.} Fidelity checks for the diagnostic-variable combinations comparing the calibrated emulator with individual CMIP6 reference models. Stripe colours within each cell encode the percentile ranks of the underlying energy-distance-score records (SCOREEDISTC) across all values shown; ties receive the mid-rank percentile. The integer in a cell is the mean of its stripe percentiles, and the percentage above each model column is the corresponding column mean. White is centred on the raw score $S=1$ (the 78.9th percentile in this displayed sample), rather than on the median. The colourbar carries two annotations of the same axis: the upper tick labels give the raw energy distance score value and the lower tick labels the rank percentile across the displayed values, so the two scales are not independent axes but a value-to-rank mapping.}
\label{fig:scoreedistc_ssp245_v101_part1of3}
\end{figure}

\clearpage
\begin{figure}[H]
\centering
\includegraphics[width=\textwidth,height=0.86\textheight,keepaspectratio]{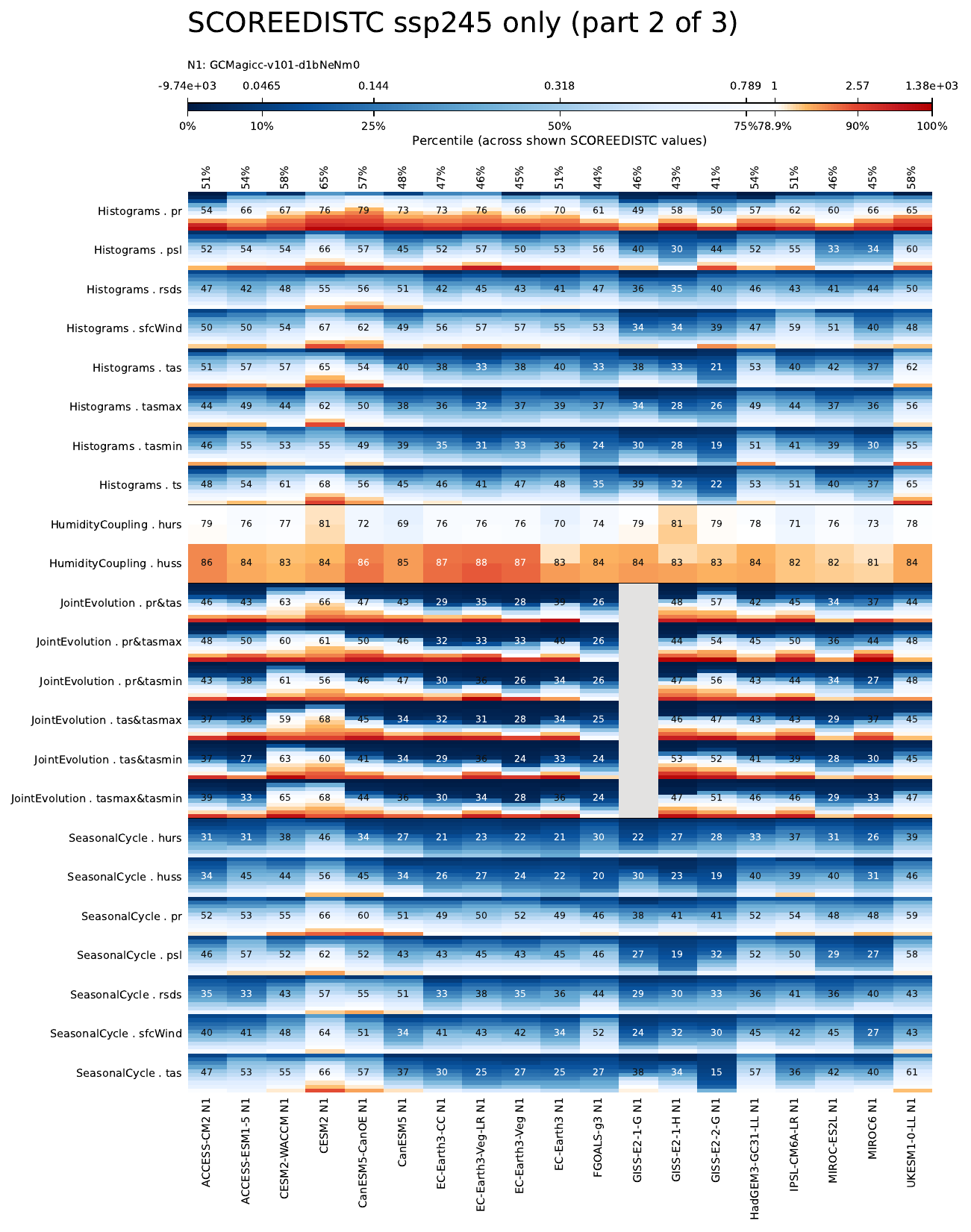}
\caption{\textbf{Part 2 Energy distance scores for the SSP2-4.5 hold-out scenario.} As Figure \ref{fig:scoreedistc_ssp245_v101_part1of3}, continued.}
\label{fig:scoreedistc_ssp245_v101_part2of3}
\end{figure}

\clearpage
\begin{figure}[H]
\centering
\includegraphics[width=\textwidth,height=0.86\textheight,keepaspectratio]{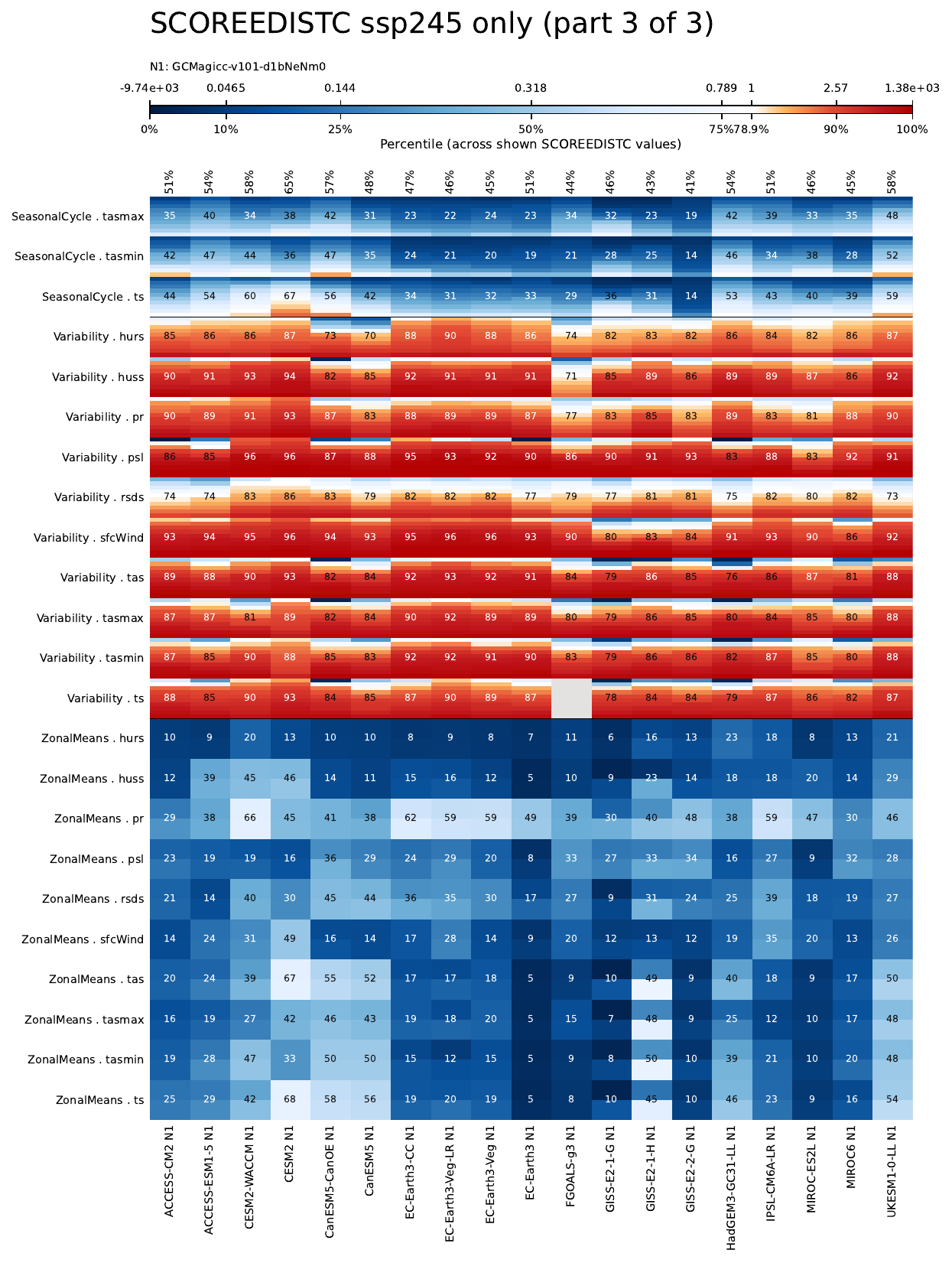}
\caption{\textbf{Part 3 Energy distance scores for the SSP2-4.5 hold-out scenario.} As Figure \ref{fig:scoreedistc_ssp245_v101_part1of3}, continued.}
\label{fig:scoreedistc_ssp245_v101_part3of3}
\end{figure}

\section{Emergent constraints from leave-one-family-out cross-validation}\label{sec:supp-emergent-constraint}

\subsection{Reduced emulator for global-mean warming}

This section uses the GCMagicc-PM architecture (Methods, Table~\ref{tab:gcmagicc-variants}) operated at HEALPix resolution $\textit{nside}=1$ only (12 equal-area pixels), giving effectively a global-mean representation. It retains the bias model $\textit{b}_1$ and the stochastic generator $\textit{g}_1$ described in the main-text Methods, but omits the common-effect model $\textit{e}_1$. That is, the prediction at $\textit{nside}=1$ reduces to
\begin{equation}
\widehat{\textit{y}}^{(1)}_t \;=\; \textit{b}_1(\textit{x}^{\textit{model}}_t,\textit{x}^{\textit{season}}_t)
\;+\;\textit{g}_1(\textit{x}_t,\textit{z}_t),
\label{eq:supp-reduced}
\end{equation}
where $\textit{z}_t$ includes autoregressive lags with $\mathcal{L}_1=\{1,\dots,6\}$.

The forcing vector $\textit{x}_t$ retains all ERF components and the model index, month, and time covariates, but excludes the smoothed global-mean indicators \textit{tas\_smoothed} and \textit{rtmt\_smoothed}. These two covariates encode externally prescribed global temperature and radiative-imbalance trajectories from MAGICC; removing them forces the emulator to predict global-mean warming from the raw ERF forcings alone, without anchoring to a pre-specified warming pathway. This makes the emulator's warming prediction genuinely out-of-sample with respect to the global temperature trajectory.

\subsection{Leave-one-family-out procedure}

The 32 CMIP6 source identifiers are partitioned into 13 model families defined by the first four characters of each model name (Table~\ref{tab:supp-emergent-families}). For each family $\mathcal{F}_k$, the reduced emulator is retrained on data from all families $\mathcal{F}_{j\neq k}$ plus ERA5, together with the historical runs of family $\mathcal{F}_k$ itself, from which its model-index embedding is fitted; no scenario run of $\mathcal{F}_k$ enters the training set. A separate emulator is also trained on ERA5 data alone (the ERA5 model).

We note explicitly that ERA5 is retained in the training data of every fold, and that the constrained predictions are produced by an emulator fitted to that same observational record. The cross-validation therefore measures how well the emulator generalises to a withheld \emph{model} family, not how well it would predict a withheld real world: the residuals $r_i$ quantify within-sample spread with respect to the observational record rather than out-of-sample observational error. This section is accordingly presented as a limitation rather than as a constrained projection. Its purpose is to test whether the reduced emulator, once its own cross-validated uncertainty is accounted for, could deliver end-of-century ranges narrower than those assessed by IPCC AR6. It cannot.

\begin{table}[ht]
\centering
\begingroup
\renewcommand{\arraystretch}{1}
\footnotesize
\caption{Model families used in leave-one-out cross-validation, defined by the first four characters of the CMIP6 source identifier.}\label{tab:supp-emergent-families}
\begin{tabular}{ll}
\tophline
Family prefix & Members \\
\middlehline
ACCE & ACCESS-CM2, ACCESS-ESM1-5 \\
CanE & CanESM5, CanESM5-1, CanESM5-CanOE \\
CESM & CESM2, CESM2-WACCM \\
EC-E & EC-Earth3, EC-Earth3-AerChem, EC-Earth3-CC, EC-Earth3-Veg, EC-Earth3-Veg-LR \\
FGOA & FGOALS-g3 \\
GFDL & GFDL-CM4, GFDL-ESM4 \\
GISS & GISS-E2-1-G, GISS-E2-1-G-CC, GISS-E2-1-H, GISS-E2-2-G, GISS-E2-2-H \\
HadG & HadGEM3-GC31-LL, HadGEM3-GC31-MM \\
INM- & INM-CM4-8, INM-CM5-0 \\
IPSL & IPSL-CM6A-LR, IPSL-CM6A-LR-INCA, IPSL-CM6A-MR1 \\
MIRO & MIROC-ES2H, MIROC-ES2L, MIROC6 \\
SAM0 & SAM0-UNICON \\
UKES & UKESM1-0-LL \\
\bottomhline
\end{tabular}
\endgroup
\end{table}

To produce ERA5 warming estimates for each scenario, the ERA5-trained emulator is run on the same concatenated historical-scenario forcing sequences used above, but with the model index set to $c_t=0$ (ERA5). This gives the emulator's best guess for how the real climate system would warm under each scenario's forcing pathway.

\subsection{Warming metric}\label{supp-sec:emergent-warming}

Global-mean warming $\Delta T$ is defined following the IPCC AR6 convention in Table SPM.1 of IPCC AR6 Summary for Policy Makers \citep{IPCC_AR6_WG1_SPM_2021}. Let $\bar{T}(a,b)$ denote the temporal mean of global-mean \textit{tas} over the period $[a,b]$. Then
\begin{equation}
\Delta T \;=\; \bar{T}(2081,2100) \;-\; \bar{T}(1995,2014) \;+\; 0.85\,^{\circ}\mathrm{C},
\label{eq:supp-warming}
\end{equation}
where $0.85\,^{\circ}\mathrm{C}$ is the IPCC AR6 central estimate of observed warming from the 1850--1900 pre-industrial baseline to the 1995--2014 reference period \citep{IPCC_AR6_WG1_SPM_2021}. This shifts the anomaly so that $\Delta T$ is measured relative to 1850--1900. The global mean at $\textit{nside}=1$ is computed as the unweighted average over the 12 equal-area HEALPix pixels.

Four quantities are recorded for each model $c$, scenario, and stochastic replicate:
\begin{itemize}\setlength{\itemsep}{0pt}
\item $\Delta T^{\text{true}}_c$: warming from the CMIP6 ground-truth \textit{tas} of run~1,
\item $\Delta T^{\text{true,2nd}}_c$: warming from a second independent run of the same model and scenario (if available; NaN otherwise),
\item $\Delta T^{\text{hat}}_c$: warming from the leave-one-out emulator using the model's forcing data and its own model index, whose embedding was fitted on that model's historical run only,
\item $\Delta T^{\text{ERA}}_c$: warming from the ERA5-trained emulator using the model's forcing data and ERA5 model index.
\end{itemize}
Each model-scenario pair is sampled with multiple stochastic replicates (3 per combination) to capture generator variability.

\subsection{Cross-validation scatter plots}

\begin{figure}[H]
\centering
\includegraphics[width=\textwidth,height=0.58\textheight,keepaspectratio]{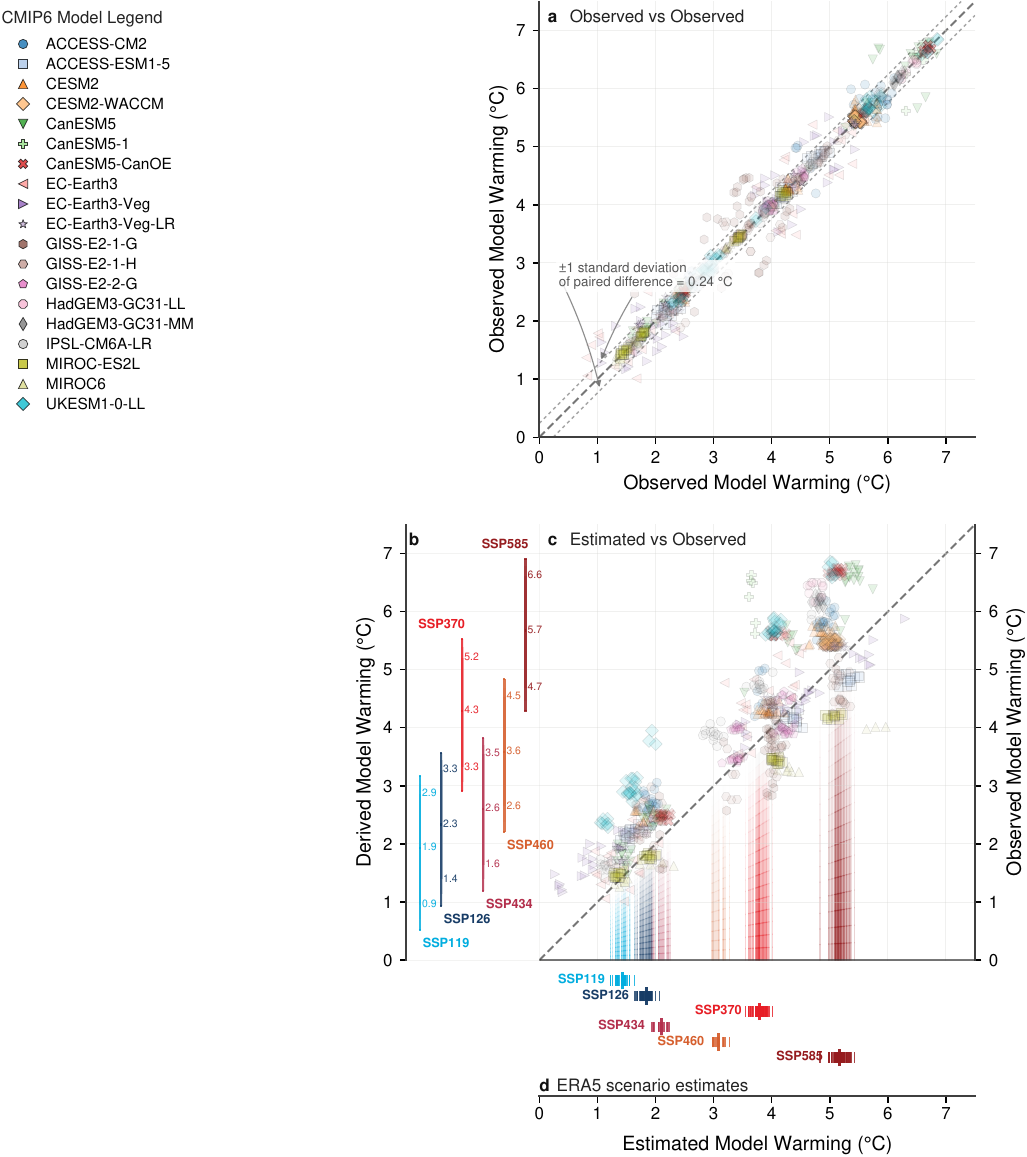}
\caption{\textbf{Leave-one-family-out evaluation and illustrative estimates of end-of-century global warming using GCMagicc-PM.} \textbf{(a)} Agreement between warming simulated by two independent CMIP6 realisations for 799 paired cases from 19 models; colours and symbols identify CMIP6 models, the dashed line denotes the one-to-one relationship, and the dotted lines show $\pm1$ sample standard deviation of the paired difference ($0.24\,^{\circ}$C). \textbf{(b)} Scenario-specific illustrative warming distributions for SSP1-1.9, SSP1-2.6, SSP3-7.0, SSP4-3.4, SSP4-6.0, and SSP5-8.5, obtained by adding resampled leave-one-family-out residuals to ERA5-conditioned GCMagicc-PM estimates. Nested intervals show the 2.5--97.5\,\%, 5--95\,\%, and 10--90\,\% ranges; annotations give the 10th, 50th, and 90th percentiles. \textbf{(c)} Leave-one-family-out GCMagicc-PM estimates versus the corresponding CMIP6-simulated warming; the dashed line denotes the one-to-one relationship and the scenario-coloured vertical plumes show ERA5-conditioned estimates. \textbf{(d)} Individual ERA5-conditioned GCMagicc-PM warming estimates for each scenario. Warming is expressed relative to 1850--1900 and constructed as the 2081--2100 mean minus the 1995--2014 mean plus 0.85\,$^{\circ}$C, the IPCC AR6 central estimate of warming in 1995--2014 relative to 1850--1900 \citep{IPCC_AR6_WG1_SPM_2021}. The illustrative distributions use 2,000 independent draws with replacement of an ERA5-conditioned estimate and a pooled residual.}
\label{fig:supp-emergent-lofo}
\end{figure}

Four panels present the cross-validation and its illustrative uncertainty consequence (Fig.~\ref{fig:supp-emergent-lofo}).

\textbf{Panel (a): observed versus observed.} Each point plots $\Delta T^{\text{true,2nd}}_c$ (x-axis) against $\Delta T^{\text{true}}_c$ (y-axis). Both axes show actual CMIP6 warming from two independent runs of the same model for the end of the 21st century (2081-2100). Points that lie close to the identity line indicate that the two runs agree well; scatter away from the diagonal reflects internal variability and ensemble spread. This panel serves as a baseline: it shows the irreducible inter-run variability that no emulator can improve upon.

\textbf{Panel (c): leave-one-family-out estimate versus observed.} Each point plots $\Delta T^{\text{hat}}_c$ (x-axis) against $\Delta T^{\text{true}}_c$ (y-axis). The x-axis now shows the emulator's leave-one-out prediction rather than a second CMIP6 run. Points near the identity line indicate that the emulator, despite never having seen any member of the held-out family, recovers that family's warming well. Short vertical rug ticks along the x-axis of panel c and in panel d mark the ERA5-trained emulator's warming predictions $\Delta T^{\text{ERA}}_c$ for the same scenario, summarising the range of emulated ERA5 warming across the available forcing sequences.

All 799 paired cases are shown in the scatter panels with low point opacity so that overlapping cases appear as density while individual model colours and markers remain visible. In panel (a), the dotted lines show one sample standard deviation of the paired difference $\Delta T^{\text{true}}-\Delta T^{\text{true,2nd}}$ on either side of the identity line ($0.24\,^{\circ}$C; $n=799$). The model legend uses higher-opacity marker examples for legibility.

\textbf{Panel (b): scenario-specific illustrative distributions.} Panel (b) shows illustrative warming distributions for SSP1-1.9, SSP1-2.6, SSP3-7.0, SSP4-3.4, SSP4-6.0 and SSP5-8.5, obtained by adding resampled leave-one-family-out residuals to the ERA5-conditioned GCMagicc-PM estimates by the procedure of Section~\ref{sec:supp-emergent-constraint-intervals}. Nested intervals show the 2.5--97.5\,\%, 5--95\,\% and 10--90\,\% ranges, and the annotations give the 10th, 50th and 90th percentiles. These distributions carry the section's central negative result: their width is approximately 2.3\,$^{\circ}$C and exceeds the corresponding IPCC WG1 AR6 assessed range for every scenario for which one is assessed except SSP5-8.5, where it is marginally narrower (2.32 against 2.40\,$^{\circ}$C); SSP4-3.4 and SSP4-6.0 have no assessed range. The constraint therefore delivers no reduction in uncertainty except at the top of the scenario range, where the near-invariant width happens to fall just inside the widest assessed range.

\textbf{Panel (d): individual ERA5-conditioned estimates.} Panel (d) shows the individual ERA5-conditioned GCMagicc-PM warming estimates for each scenario, that is the set $\{e_j\}$ that enters the sampling in Eq.~\ref{eq:supp-emergent-sampling} before the residual pool is added.

\clearpage

\subsection{Width implied by the GCMagicc vs CMIP6 cross-validation residuals}\label{sec:supp-emergent-constraint-intervals}

The scatter plots establish that the leave-one-out emulator captures inter-model warming differences. To express this as the width the emulator implies for real-world warming (here proxied by ERA5), we adopt the emergent-constraint sampling procedure described below.

Let $r_i = \Delta T^{\text{true}}_i - \Delta T^{\text{hat}}_i$ denote the cross-validated residual for observation $i$, pooled across all models, scenarios, and replicates in the SSP trends data. Let $\{e_j\}$ denote the set of ERA5-trained emulator warming predictions $\Delta T^{\text{ERA}}_j$ for a given target scenario. The residuals capture the typical discrepancy between the emulator's warming prediction and the true GCM warming, while the ERA5 predictions give the emulator's best guess for observed warming.

To construct an illustrative resampling distribution for warming under the target scenario, we draw $N_{\text{sim}}=2{,}000$ samples
\begin{equation}
\widetilde{\Delta T}_n \;=\; e_{j_n} + r_{i_n}, \qquad j_n\sim\text{Uniform}\{e_j\},\quad i_n\sim\text{Uniform}\{r_i\},
\label{eq:supp-emergent-sampling}
\end{equation}
where $j_n$ and $i_n$ are drawn independently with replacement. Each sample $\widetilde{\Delta T}_n$ combines a random ERA5 prediction with a random residual, so that the distribution of $\widetilde{\Delta T}$ reflects both the emulator's uncertainty about ERA5 warming and the systematic prediction error observed across the model ensemble. Because $r_i$ is computed from single generator draws and $e_{j}$ is itself a single draw, the generator's sampling variance enters the width twice; the generator's sampling variance therefore inflates the width relative to the emulator-implied uncertainty. A second limitation is that the procedure does not capture the warm bias of residuals for low scenarios and resulting illustrative uncertainty ranges for the low scenarios are spuriously low.

From the $N_{\text{sim}}$ samples, we extract the quantiles listed in Table~\ref{tab:supp-emergent-quantiles}. These define symmetric resampling ranges: the 5\%--95\% range spans 90\% of the samples, the 2.5\%--97.5\% range 95\%, and so on.

\begin{table}[ht]
\centering
\begingroup
\renewcommand{\arraystretch}{1}
\footnotesize
\caption{Emergent-constraint prediction quantiles for global-mean warming $\Delta T$ ($^{\circ}$C above 1850--1900) under six SSP scenarios, period 2081--2100. Obtained from $N_{\text{sim}}=2{,}000$ Monte Carlo samples combining ERA5 emulator predictions with cross-validated residuals. These quantiles demonstrate that the emulator's own cross-validated uncertainty is not smaller than the AR6 assessed ranges; they are not offered as calibrated prediction intervals for real-world warming. The residual pool is not stratified by scenario, so the 5--95\,\% width is near-invariant across scenarios (2.30--2.32\,$^{\circ}$C here) by construction, whereas the assessed AR6 widths widen with forcing. Read the table as an illustrative bound on achievable uncertainty rather than as a scenario-resolved projection, noting that the double-counted generator variance biases it up. Two consequences follow: the intervals are too wide for the low scenarios --- the SSP1-1.9 5\,\% bound of 0.71\,$^{\circ}$C lies below already-observed warming --- and the outermost 2.5\,\% and 97.5\,\% quantiles are reported at a precision finer than $N_{\text{sim}}=2{,}000$ samples resolve.}\label{tab:supp-emergent-quantiles}
\begin{tabular}{lrrrrrr}
\tophline
Quantile & SSP1-1.9 & SSP1-2.6 & SSP3-7.0 & SSP4-3.4 & SSP4-6.0 & SSP5-8.5 \\
\middlehline
2.5\%   & 0.52 & 0.94 & 2.92 & 1.19 & 2.21 & 4.29 \\
5\%     & 0.71 & 1.14 & 3.07 & 1.39 & 2.38 & 4.46 \\
10\%    & 0.95 & 1.39 & 3.32 & 1.63 & 2.65 & 4.69 \\
90\%    & 2.86 & 3.28 & 5.21 & 3.54 & 4.54 & 6.61 \\
95\%    & 3.03 & 3.44 & 5.39 & 3.70 & 4.69 & 6.78 \\
97.5\%  & 3.16 & 3.57 & 5.53 & 3.83 & 4.83 & 6.90 \\
\bottomhline
\end{tabular}
\endgroup
\end{table}

For instance, under SSP5-8.5 the 5--95\% range of the resampled distribution spans $[4.46,\,6.78]\,^{\circ}\mathrm{C}$ and under SSP1-2.6 $[1.14,\,3.44]\,^{\circ}\mathrm{C}$; these are emulator-implied widths, not calibrated prediction intervals for real-world warming, and the SSP5-8.5 lower bound lies above both the AR6 assessed median and this study's own SSP5-8.5 ensemble median, which shows how coarsely the unstratified residual pool locates the distribution.

\subsection{Summary}

The leave-one-family-out cross-validation demonstrates that the reduced $\textit{nside}=1$ emulator, operating without the common-effect model and without externally prescribed global temperature trajectories, captures the spread in end-of-century warming across CMIP6 model families from ERF forcings alone. Combining the same emulator's ERA5-forced estimates with the cross-validated residual distribution yields illustrative intervals that are as wide as, or wider than, the corresponding IPCC AR6 assessed ranges for every scenario with an assessed range except SSP5-8.5, where the near-invariant width is marginally narrower. The purpose of this exercise is precisely that negative result: GCMagicc-PM does not deliver a reduction in end-of-century warming uncertainty relative to the assessed ranges, and the projections in this paper should not be read as offering one.
Note that our main application GCMagicc workflow has a different setup. It uses the MAGICC simple climate model to impose IPCC AR6 calibrated ensembles for GCMagicc, hence - by construction - the IPCC AR6 assessed uncertainty distributions for \textit{tas} are honoured.

\section{Observationally and IPCC AR6 aligned projections}

\subsection{Data table for Figure~\ref{fig:observational-alignment}}

Table~\ref{tab:observational-alignment-panel-b-values} tabulates the values plotted in panel b, and Table~\ref{tab:observational-alignment-cmip6-model-index} decodes the CMIP6 line numbers of panels c1 and c2 (24 of the 32 CMIP6 models; the eight not shown are those without the ensemble coverage the panel requires).

\subsection{New scenarios: Estimating NDC and LT-LEDS future emissions}\label{supp-NDCs-LTLEDs}

\textbf{Country-level NDC and LT-LEDS emission quantification. }Country-level greenhouse gas (GHG) emission trajectories were constructed in a bottom-up fashion for all 196 countries using officially submitted inventory data, NDCs, and LT-LEDS. Historical emission data were drawn from the PRIMAP-hist dataset \citep{gutschow2025primaphist, gutschow2016primaphist} to anchor national time series, while NDC targets, expressed as absolute emissions levels, reductions relative to historical or BAU levels, emissions-intensity targets, per-capita targets, or multi-year budgets, were implemented while taking into account the gas-by-gas and sectoral coverage of those targets. Emissions from sectors and gases not covered by a country's NDC were assumed to follow NGFS GCAM Current Policies growth rates \citep{ngfs_phase5_2024}, with LULUCF projections treated separately where reported data were available. For countries with long-term targets, including net-zero CO$_2$ or GHG targets, emissions were linearly interpolated between the 2030 or 2035 NDC level (whichever the latest NDC target year by a country) and the stated long-term target year.

\textbf{Infilling, harmonisation, and gas-by-gas extension to 2100.} To construct complete, gas-by-gas global emission pathways extending to 2100, the country-aggregated GHG totals were complemented, harmonised and infilled. International aviation and shipping emissions were added to aggregated country emissions. LULUCF emissions were adjusted to approximate directly induced anthropogenic sinks (as countries report a mix of directly anthropogenic and non-anthropogenic LULUCF sinks), in line with IPCC AR6 assessed GHG emission scenarios, by subtracting indirectly induced LULUCF CO$_2$ emissions derived from the NGFS downscaled GCAM pathways\citep{ngfs_phase5_2024}. The resulting global GHG emissions were harmonised via the Aneris python package \citep{Gidden2018} and infilling/extension algorithms (via Silicone) \citep{lamboll2020silicone}. Beyond 2060, an adapted equal-quantile-walk approach was applied using the IPCC AR6 scenario database. Also, in order to split up the GHG emissions into separate individual gases (CO$_2$, CH$_4$, N$_2$O, and others) in a manner consistent with the characteristics of that IPCC AR6 scenario ensemble, we used an adapted equal quantile walk approach. This yielded complete, multi-gas emission pathways from the present day to 2100 for each scenario variant. For more details on the NDC and LT-LEDs quantification methods, see \citet{meinshausen_realization_2022}.

\begin{table}[tp]
\centering
\footnotesize
\setlength{\tabcolsep}{3pt}
\resizebox{0.80\linewidth}{!}{%
\begin{tabular}{lrrrrrrrrrrrrrrr}
\tophline
Scenario & \multicolumn{3}{c}{IPCC AR6} & \multicolumn{3}{c}{CMIP6} & \multicolumn{3}{c}{GCMagicc n20} & \multicolumn{3}{c}{n100 AR6} & \multicolumn{3}{c}{n100 CMIP7-ScenarioMIP} \\
  & p5 & med & p95 & p5 & med & p95 & p5 & med & p95 & p5 & med & p95 & p5 & med & p95 \\
\middlehline
\textcolor[HTML]{00addf}{ssp119} & 1.00 & 1.40 & 1.80 & 0.99 & 1.63 & 2.35 & 1.21 & 1.44 & 1.90 & 1.21 & 1.44 & 1.85 & 1.22 & 1.45 & 1.94 \\
\textcolor[HTML]{173c66}{ssp126} & 1.30 & 1.80 & 2.40 & 1.46 & 2.29 & 2.96 & 1.45 & 1.82 & 2.42 & 1.48 & 1.81 & 2.40 & 1.50 & 1.88 & 2.52 \\
\textcolor[HTML]{f79420}{ssp245} & 2.10 & 2.70 & 3.50 & 2.25 & 3.09 & 4.07 & 2.25 & 2.71 & 3.66 & 2.21 & 2.71 & 3.60 & 2.23 & 2.79 & 3.80 \\
\textcolor[HTML]{e71d25}{ssp370} & 2.80 & 3.60 & 4.60 & 3.21 & 4.30 & 5.66 & 3.17 & 3.78 & 4.75 & 3.07 & 3.71 & 4.77 & 3.11 & 3.76 & 4.95 \\
\textcolor[HTML]{b32f4c}{ssp434} & -- & -- & -- & 1.63 & 2.53 & 3.77 & 1.89 & 2.19 & 2.71 & 1.88 & 2.19 & 2.84 & 1.89 & 2.24 & 2.88 \\
\textcolor[HTML]{d65f2e}{ssp460} & -- & -- & -- & 2.17 & 3.25 & 4.68 & 2.58 & 3.07 & 4.16 & 2.57 & 3.09 & 4.10 & 2.57 & 3.18 & 4.12 \\
\textcolor[HTML]{7c2d6f}{ssp534-over} & -- & -- & -- & 1.38 & 2.45 & 3.67 & 1.81 & 2.16 & 2.95 & 1.78 & 2.19 & 2.99 & 1.75 & 2.26 & 3.12 \\
\textcolor[HTML]{951b1e}{ssp585} & 3.30 & 4.40 & 5.70 & 3.67 & 5.13 & 6.68 & 3.76 & 4.51 & 6.08 & 3.56 & 4.44 & 5.94 & 3.65 & 4.55 & 6.17 \\
\middlehline
\textbf{SSP scenarios} & \textbf{1.00} & \textbf{2.78} & \textbf{5.70} & \textbf{0.99} & \textbf{3.09} & \textbf{6.68} & \textbf{1.21} & \textbf{2.71} & \textbf{6.08} & \textbf{1.21} & \textbf{2.70} & \textbf{5.94} & \textbf{1.22} & \textbf{2.76} & \textbf{6.17} \\
\middlehline
\textcolor[HTML]{2aa198}{PA-all-l} & -- & -- & -- & -- & -- & -- & 1.45 & 1.66 & 2.19 & 1.44 & 1.66 & 2.08 & 1.44 & 1.71 & 2.21 \\
\textcolor[HTML]{006d77}{PA-all-u} & -- & -- & -- & -- & -- & -- & 1.94 & 2.32 & 3.06 & 1.92 & 2.29 & 3.01 & 1.94 & 2.35 & 3.19 \\
\textcolor[HTML]{8ab17d}{PA-woUS-l} & -- & -- & -- & -- & -- & -- & 1.55 & 1.77 & 2.13 & 1.49 & 1.78 & 2.35 & 1.51 & 1.81 & 2.52 \\
\textcolor[HTML]{52796f}{PA-woUS-u} & -- & -- & -- & -- & -- & -- & 2.05 & 2.42 & 3.28 & 2.01 & 2.39 & 3.19 & 2.05 & 2.50 & 3.25 \\
\textcolor[HTML]{c87828}{SSP2-com} & -- & -- & -- & -- & -- & -- & 1.67 & 1.99 & 2.82 & 1.65 & 2.04 & 2.78 & 1.66 & 2.07 & 2.80 \\
\textcolor[HTML]{8a4b2a}{CP\_GCAM} & -- & -- & -- & -- & -- & -- & 2.42 & 2.87 & 3.76 & 2.40 & 2.83 & 3.67 & 2.38 & 2.92 & 3.79 \\
\textcolor[HTML]{b6652d}{CP\_MSG} & -- & -- & -- & -- & -- & -- & 2.25 & 2.76 & 3.53 & 2.22 & 2.73 & 3.53 & 2.25 & 2.80 & 3.67 \\
\textcolor[HTML]{d4883a}{CP\_RMD} & -- & -- & -- & -- & -- & -- & 2.40 & 2.94 & 3.98 & 2.36 & 2.91 & 3.90 & 2.38 & 2.99 & 4.06 \\
\middlehline
\textbf{NDCs + others} & \textbf{--} & \textbf{--} & \textbf{--} & \textbf{--} & \textbf{--} & \textbf{--} & \textbf{1.45} & \textbf{2.34} & \textbf{3.98} & \textbf{1.44} & \textbf{2.33} & \textbf{3.90} & \textbf{1.44} & \textbf{2.39} & \textbf{4.06} \\
\middlehline
\textcolor[HTML]{bddcf4}{VL} & -- & -- & -- & -- & -- & -- & 1.17 & 1.56 & 2.07 & 1.22 & 1.54 & 2.07 & 1.21 & 1.59 & 2.21 \\
\textcolor[HTML]{a9cce9}{LN} & -- & -- & -- & -- & -- & -- & 1.38 & 1.64 & 2.20 & 1.41 & 1.66 & 2.17 & 1.44 & 1.72 & 2.23 \\
\textcolor[HTML]{93bbe0}{L} & -- & -- & -- & -- & -- & -- & 1.40 & 1.75 & 2.17 & 1.47 & 1.76 & 2.33 & 1.50 & 1.80 & 2.45 \\
\textcolor[HTML]{7ba5d4}{M} & -- & -- & -- & -- & -- & -- & 2.38 & 2.80 & 3.63 & 2.35 & 2.76 & 3.56 & 2.36 & 2.82 & 3.65 \\
\textcolor[HTML]{4c6fb8}{ML} & -- & -- & -- & -- & -- & -- & 2.00 & 2.34 & 2.96 & 1.97 & 2.33 & 3.05 & 1.97 & 2.40 & 3.17 \\
\textcolor[HTML]{6282be}{HL} & -- & -- & -- & -- & -- & -- & 2.34 & 2.89 & 3.77 & 2.29 & 2.87 & 3.90 & 2.39 & 2.90 & 4.04 \\
\textcolor[HTML]{3b0f70}{H} & -- & -- & -- & -- & -- & -- & 2.77 & 3.30 & 4.22 & 2.71 & 3.24 & 4.23 & 2.77 & 3.31 & 4.37 \\
\middlehline
\textbf{CMIP7 scenarios} & \textbf{--} & \textbf{--} & \textbf{--} & \textbf{--} & \textbf{--} & \textbf{--} & \textbf{1.17} & \textbf{2.33} & \textbf{4.22} & \textbf{1.22} & \textbf{2.31} & \textbf{4.23} & \textbf{1.21} & \textbf{2.36} & \textbf{4.37} \\
\bottomhline
\end{tabular}%
}
\caption{Values from Figure~\ref{fig:observational-alignment}, panel b, relative to 1850--1900 in $^{\circ}$C. Each bold category row reports the minimum p5, mean median, and maximum p95 over available scenarios; dataset coverage may differ. Values are changes from 1995--2014 to 2081--2100 plus the IPCC AR6 estimate of 0.85\,$^{\circ}$C warming from 1850--1900 to 1995--2014 \citep{IPCC_AR6_WG1_SPM_2021}.}
\label{tab:observational-alignment-panel-b-values}
\end{table}

\begin{table}[tp]
\centering
\footnotesize
\begin{tabular}{rl}
\tophline
No. & CMIP6 model \\
\middlehline
1 & ACCESS-CM2 \\
2 & ACCESS-ESM1-5 \\
3 & CESM2 \\
4 & CESM2-WACCM \\
5 & CanESM5 \\
6 & CanESM5-1 \\
7 & CanESM5-CanOE \\
8 & EC-Earth3 \\
9 & EC-Earth3-CC \\
10 & EC-Earth3-Veg \\
11 & EC-Earth3-Veg-LR \\
12 & FGOALS-g3 \\
13 & GFDL-CM4 \\
14 & GFDL-ESM4 \\
15 & GISS-E2-1-G \\
16 & GISS-E2-1-H \\
17 & GISS-E2-2-G \\
18 & HadGEM3-GC31-LL \\
19 & INM-CM4-8 \\
20 & INM-CM5-0 \\
21 & IPSL-CM6A-LR \\
22 & MIROC-ES2L \\
23 & MIROC6 \\
24 & UKESM1-0-LL \\
\bottomhline
\end{tabular}
\caption{Line-number mapping for panels c1 and c2 of Fig.~\ref{fig:observational-alignment}: the 24 CMIP6 models plotted there, numbered as in the panel. The eight remaining models of the 32 used elsewhere in this study are not plotted in these panels.}
\label{tab:observational-alignment-cmip6-model-index}
\end{table}

\subsection{Extended data figure for observational matches of GCMagicc and CMIP6 models}

\begin{figure}[H]
  \centering
  \includegraphics[width=\textwidth,height=0.66\textheight,keepaspectratio]{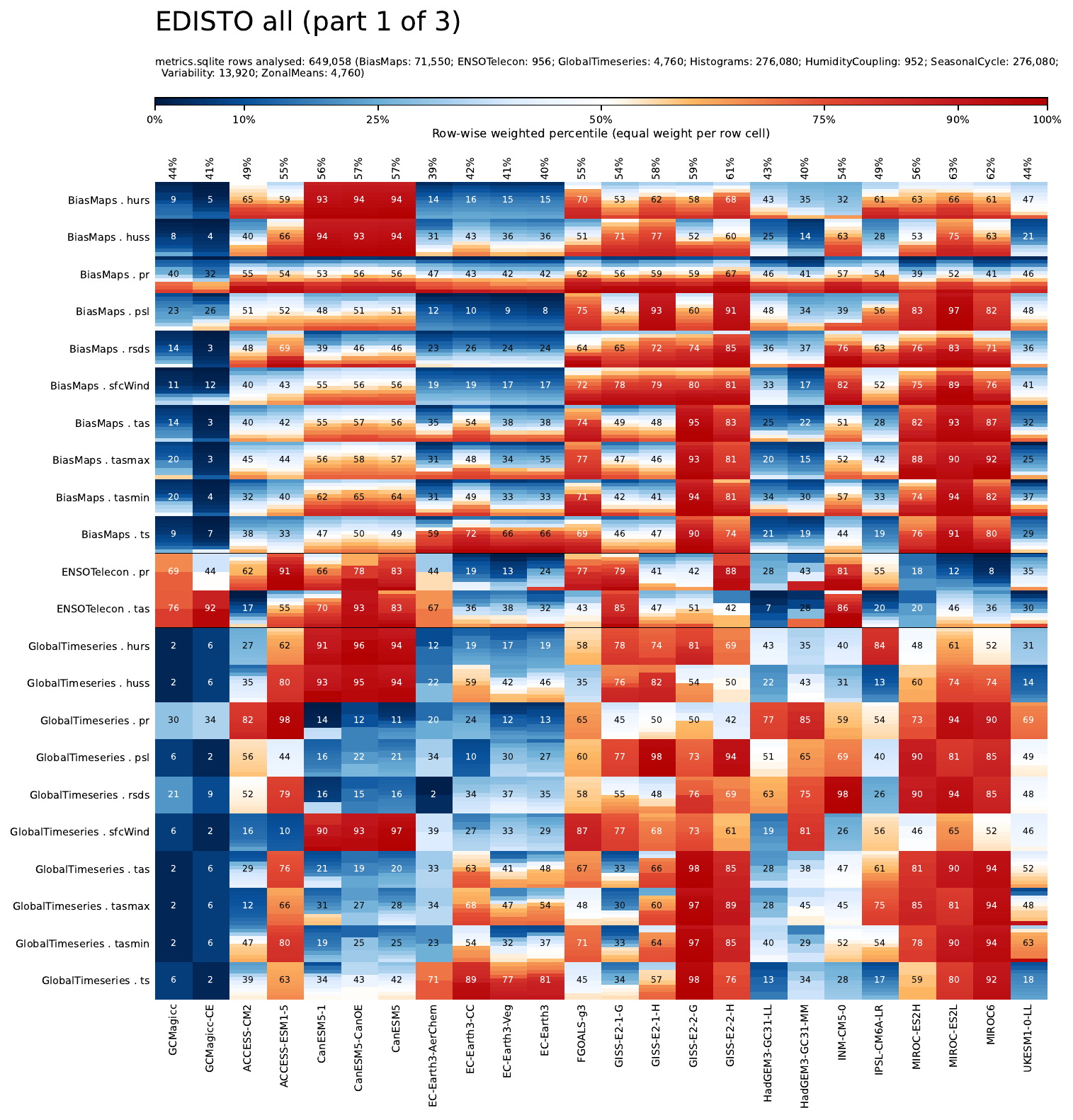}
  \caption{\textbf{Observation-referenced energy-distance performance across diagnostics and variables (Part 1 of 3).} Columns show two GCMagicc configurations - GCMagicc and GCMagicc-CE - followed by 22 CMIP6 models (the models with complete diagnostic coverage in the frozen evaluation database for all displayed diagnostic-variable rows). Rows represent combinations of diagnostic and climate variable. Part 1 contains BiasMaps, ENSOTelecon and GlobalTimeseries; Part 2 contains Histograms, HumidityCoupling and SeasonalCycle except surface temperature (\textit{ts}); Part 3 contains SeasonalCycle--\textit{ts}, Variability and ZonalMeans. Together, the panels summarise 649{,}058 evaluations in 64 diagnostic--variable rows and 24 columns in total (two GCMagicc configurations plus the 22 CMIP6 models). For the CMIP6 columns, the observation-referenced energy-distance measure is $\mathrm{EDISTO}_C=2\,\mathrm{GOF}_{OC}-\overline{\mathrm{GOF}_{CC}}$; for GCMagicc it is $\mathrm{EDISTO}_N=2\,\mathrm{GOF}_{ON}-\overline{\mathrm{GOF}_{NN}}$. Here, $O$, $C$ and $N$ denote the observational reference (ERA5), CMIP6 and GCMagicc, respectively, and the within-model terms correct for internal ensemble spread. Lower energy-distance values indicate a smaller discrepancy to observations.}
  \label{fig:supp-edisto-observational}
\end{figure}

\begin{figure}[H]
  \centering
  \includegraphics[width=\textwidth,height=0.66\textheight,keepaspectratio]{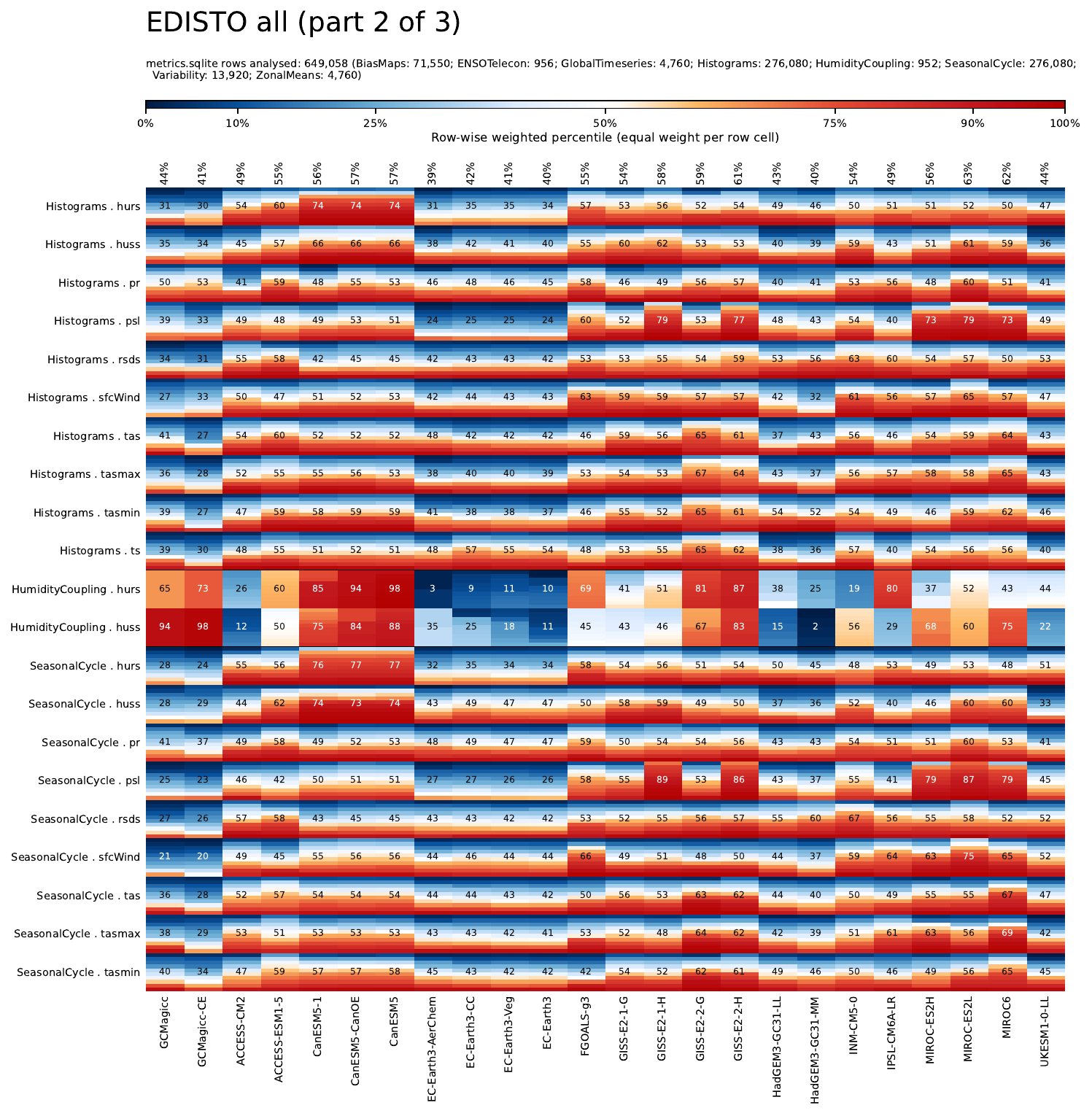}
  \caption[]{\textbf{Observation-referenced energy-distance performance (continued; Part 2 of 3).} Because the underlying diagnostics have different numerical scales, energy-distance values are converted separately within every diagnostic--variable row to weighted percentiles. Each model cell is assigned equal total weight, irrespective of how many underlying records it contains. Blue therefore denotes a relatively low energy-distance percentile and closer agreement with ERA5, white denotes approximately the row median, and red denotes a relatively high energy-distance percentile and larger discrepancy. Colours should consequently be compared between models within a row rather than interpreted as absolute differences between different diagnostics. Each cell is subdivided vertically into multiple horizontal colour bands. These bands represent the distribution of individual evaluations contributing to that diagnostic--variable--model combination, which can differ by diagnostic subtype, region, season or time window, source member and comparison member. Records are ordered from lower percentiles at the top to higher percentiles at the bottom in each cell. When ten or fewer records contribute, each record is displayed; for larger samples, ten representative mid-bin order statistics are shown.}
  \label{fig:supp-edisto-observational-part2}
\end{figure}

\begin{figure}[H]
  \centering
  \includegraphics[width=\textwidth,height=0.66\textheight,keepaspectratio]{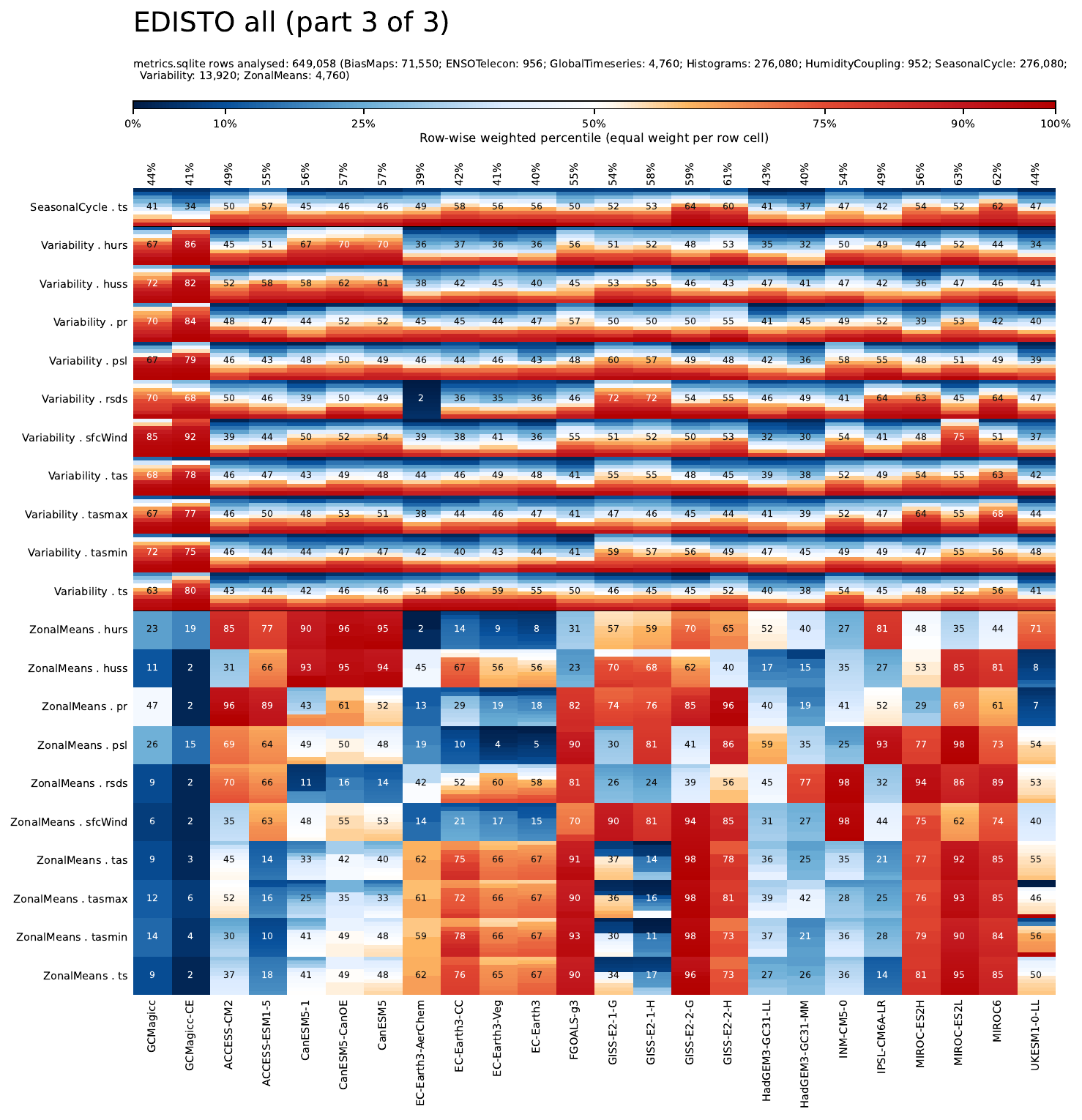}
  \caption[]{\textbf{Observation-referenced energy-distance performance (continued; Part 3 of 3).} The bands therefore expose within-cell heterogeneity that would be hidden by a single mean colour. The integer printed in each cell is the rounded mean percentile calculated from all contributing records, not the number of records. Percentages above the columns give the corresponding mean percentile across all displayed records for that model; lower percentages indicate better overall relative performance across the included rows. Thin horizontal separators delineate diagnostic groups. These files exclude the explicit oscillation-index variables NAO, SOI, Ni\~no-3, Ni\~no-3.4, Ni\~no-4 and PDO, and therefore omit the ENSOCharacteristics and IndicatorFreq row groups; the precipitation- and temperature-based ENSOTelecon diagnostics remain included. The three parts differ only in which rows are displayed and share the same normalisation and colour scale. Because the visualisation is percentile-based, it describes relative model ordering and within-cell spread, not the absolute magnitude or statistical significance of energy distances or their differences.}
  \label{fig:supp-edisto-observational-part3}
\end{figure}

The CMIP6 subset shown differs between evaluation displays (Fig.~\ref{fig:observational-alignment}d and Figs.~\ref{fig:supp-edisto-observational}, \ref{fig:supp-edisto-observational-part2} and \ref{fig:supp-edisto-observational-part3}) because a model is included in a display only where the frozen evaluation database provides every diagnostic--variable row that the display shows; the remaining training models lack at least one row, sometimes due to lack of raw data for the investigated scenarios.
 The specific models entering each display can be read from that display's own axis labels and legend.

\clearpage

\clearpage

\clearpage

\section{Suitability of existing climate emulators and ESM workflows for multivariate drought attribution}\label{sec:supp-emulator-comparison}

The drought-attribution and projection analyses presented in the main manuscript impose seven concurrent requirements on any candidate emulator or ESM-based workflow:
\begin{enumerate}\setlength\itemsep{2pt}\setlength\topsep{2pt}
\item[(i)] joint emulation of at least the seven variables required to evaluate Penman--Monteith potential evapotranspiration (\textit{tas}, \textit{tasmin}, \textit{tasmax}, \textit{hurs} or \textit{huss}, \textit{sfcWind}, \textit{rsds}, \textit{pr});
\item[(ii)] free-running operation without prescribed sea-surface temperatures or ocean state, so that all-forcing and counterfactual natural-only forcing climates remain distinguishable;
\item[(iii)] decomposed effective radiative forcing inputs (GHG, aerosol, solar, volcanic, ozone), so that natural-only (and potentially aerosol-perturbed) counterfactuals can be constructed rather than only a CO$_2$ or total forcing implementation;
\item[(iv)] a stochastic, generative formulation supporting $\mathcal{O}(100)$ independent realisations per scenario at low marginal cost, as needed for tail-event probability estimation;
\item[(v)] temporally coherent multi-decadal output with stable low-frequency variability sufficient for SPEI-48, which accumulates 48 months of climatic water balance;
\item[(vi)] present-day observational alignment so that inferred event probabilities are anchored to the realised climate;
\item[(vii)] applicability to new scenarios without re-training, since the most decision-relevant scenario space (CMIP7, NDC-aligned and current-policy pathways) lags both ESM simulation and any per-scenario ML retraining cycle.
\end{enumerate}

To our knowledge, no published emulator or off-the-shelf workflow satisfies all seven requirements simultaneously (Table~\ref{tab:emulator-comparison-main}); this includes multi-resolution compound-hazard emulators such as MERCURY~\citep{nath_2025_mercury}, which is not tabulated here.

To our knowledge, the best toolset at present is the conventional ESM-driven pipeline using CanESM5, MIROC6, and GISS-E2-1-G historical, natural-only, and SSP2-4.5 ensembles. Our inventory contains 65 / 50 / 50 historical / SSP2-4.5 / historical-natural CanESM5 members, 50 / 50 / 50 MIROC6 members, and 46 / 22 / 20 GISS-E2-1-G members after exclusion of one incomplete historical file. The quantitative comparison uses the monthly Penman--Monteith protocol in Section~\ref{supp-sec:monthly-drought-diagnostic}; a single-model large ensemble samples internal variability conditional on that model but does not estimate model-structural uncertainty (see Section~\ref{drought-attrib}).

In the following, we briefly discuss the illustrative entries in our comparison table of different approaches, as shown in the main manuscript Table~\ref{tab:emulator-comparison-main}. The table is not exhaustive, but it illustrates the diversity of approaches and their relative strengths and weaknesses with respect to the seven requirements. It is far from a comprehensive review of these approaches, and we do not attempt to provide a complete list of all existing climate emulators.

\subsection{CMIP6 + statistical bias adjustment.} The conventional ESM-driven attribution workflow combines CMIP6 ensembles with statistical bias adjustment. ESMs provide the Penman--Monteith variables and multi-decadal simulations, but complete single-forcing archives, natural-only extensions, and large ensembles are available for relatively few models. Bias adjustment can also modify multivariate dependence among PM drivers. Applicability to a new scenario remains conditional on an upstream ESM simulation, motivating the partial column~'New scen.' rating in Table~\ref{tab:emulator-comparison-main}.

\subsection{MESMER family.} MESMER~\citep{beusch_mesmer_2020}, MESMER-M~\citep{nath_mesmer_m_2022}, and MESMER-X~\citep{quilcaille_mesmer_x_2022} are stochastic pattern-scaling (and, for MESMER-X, distribution-scaling) emulators trained per-ESM. MESMER and MESMER-X are conditioned on a global $tas$ trajectory; MESMER-M is conditioned instead on local yearly temperatures. They produce stochastic ensembles via AR(1) processes with spatially correlated innovations (MESMER, MESMER-M) or draws from conditional generalized extreme value (GEV) distributions (MESMER-X), and are stable on multi-decadal timescales. Recent extensions add precipitation (MESMER-M-TP), fire-weather indices and soil moisture (extended MESMER-X~\citep{quilcaille_esd-14-1333-2023}), but the family does not produce the full Penman--Monteith driver set ($tasmin$/$tasmax$, $hurs$/$huss$, $sfcWind$, $rsds$ are absent). Volcanic forcing enters via a stratospheric-AOD response correction~\citep{beusch_mesmer_2020}, which is a single-variable adjustment; the partial (\ensuremath{\circ}) column~'Dec.ERF' rating reflects this limited form of forcing handling. Scenario transferability without retraining is most clearly established for MESMER-X (GMT-conditioned scaling); MESMER and MESMER-M flag full inter-scenario transfer as work-in-progress, but no retraining step is technically required to apply trained parameters to a new GMT trajectory.

\subsection{MESMER-MAGICC~\citep{beusch_magicc_mesmer_2022}.} Couples MAGICC to MESMER, so the global response is by construction tied to the IPCC AR6 calibration of MAGICC. The published chain is emission-driven and passes only the forced global-mean temperature change to MESMER - with explorations of adding ocean heat uptake as predictor \citep{beusch_magicc_mesmer_2022}; the decomposed effective radiative forcings handled internally by MAGICC do not reach the regional emulator, hence the partial (\ensuremath{\circ}) rating on column~'Dec.ERF'. Output remains restricted to MESMER's variable set (annual-mean $tas$ in the published v0.8.3), and the spatial patterns inherit from the CMIP6 ESMs that MESMER was calibrated against. Recent additions include a combination of MAGICC and MESMER-M-TP to monthly heat and drought indicators, the latter being based on the Thornthwaite PET formulation \citep{Thornthwaite1948} that depends on temperature and precipitation \citep{schongart2025highincome}.

\subsection{STITCHES~\citep{tebaldi_stitches_2022}.} The STITCHES approach 're-stitches' existing CMIP6 simulation segments (default nine-year windows) under user-specified GSAT trajectories. It inherits all archived variables and source-ESM bias. The nine-year window disrupts low-frequency variability modes such as NAO, PDO, and ENSO, and the method cannot extend to GSAT trajectories outside the envelope available in the archive. No retraining is required to construct new scenarios within that envelope, although scenarios with different aerosol loadings would be considered the same as long as GSAT is identical, hence the partial (\ensuremath{\circ}) column~'New scen.' rating; the limitation is reach and forcing specific responses, not retraining overhead, given that there is no training step per se.

\subsection{ClimaX~\citep{nguyen_climax_2023}.} A foundation model with weather and climate configurations, pretrained on five CMIP6 sources and fine-tuned on ERA5 for weather tasks. In its documented ClimateBench configuration, ClimaX uses the preceding ten years of CO$_2$, SO$_2$, BC, and CH$_4$ forcing fields and predicts annual-mean surface temperature, diurnal temperature range, precipitation, and 90th-percentile precipitation \citep{watsonparris_climatebench_2022,nguyen_climax_2023}. Surface shortwave radiation, surface wind, and humidity are not outputs of this configuration. The published ClimateBench implementation is deterministic, rather than emphasizing the distributional match.

\subsection{Aurora~\citep{bodnar_2025_aurora}.} A foundation model pre-trained on a six-dataset mixture (ERA5, CMCC, IFS-HR, HRES Forecasts, GFS Analysis, GFS Forecasts) and demonstrated for medium-range weather forecasts up to roughly ten days, plus fine-tuned variants for air quality, ocean waves and tropical-cyclone tracks. Aurora's standard surface output set comprises 2~m temperature, 10~m wind components and mean sea-level pressure, omitting humidity, surface shortwave radiation, precipitation, and daily temperature extremes, which is why we assign the partial (\ensuremath{\circ}) rating in the column~'Penman-M vars'. The published paper does not document climate-scale free-running stability or a mechanism for accepting time-varying greenhouse-gas, aerosol, or solar forcings. The model is deterministic in the strong sense -- one forward pass yields a single forecast with no stochastic generator -- and the authors flag probabilistic ensembles as an open problem to be addressed either by a probabilistic fine-tuning or by ensembles of separately trained deterministic models; it therefore receives \ensuremath{\times}\ on column~'Stoch'.

\subsection{cBottle~\citep{brenowitz_cbottle_2025}.} A generative diffusion-based foundation model trained on hourly ERA5 and ICON kilometre-scale simulations, conditioned on prescribed sea-surface temperature, solar position, and a dataset label. It samples atmospheric states and outputs temperature, winds, pressure, precipitation, radiative/cloud variables, and humidity on atmospheric levels, but the primary release does not document a dedicated 2-m relative- or specific-humidity output. It does not accept anthropogenic forcing inputs or run as a continuous time-stepping simulator; Table~\ref{tab:emulator-comparison-main} therefore treats its Penman--Monteith variable coverage as partial.

\subsection{NeuralGCM~\citep{kochkov_neural_2024}.} A hybrid neural--physical atmospheric model trained on 40 years of ERA5. In climate mode, it is run AMIP-style with prescribed sea-surface temperatures; multi-decadal stability is partially demonstrated, with 22 of 37 initial 40-year runs completing stably. It supports stochastic ensembles via learned space-time-correlated Gaussian random fields. Prognostic variables are 3D temperature, humidity components, winds, and surface pressure; the published standard output set does not include surface downwelling shortwave radiation.

\subsection{ACE2~\citep{wattmeyer_2025}.} An atmosphere-only emulator trained on ERA5, with a 1000-year stable AMIP-style rollout demonstrated. ACE2 requires a prescribed sea-surface-temperature field; the only free-running configuration shown is a short-horizon ($\sim$10-day) persisted-SST forecast mode, motivating the partial (\ensuremath{\circ}) column~'Free-run' rating in this isolated form rather than a flat \ensuremath{\times}. ACE2 accepts time-varying SST and a global-mean CO$_2$ scalar but no other separable anthropogenic forcing components. It is deterministic and outputs (at 6-hourly resolution) 2-m temperature, specific humidity, 10-m winds, surface pressure, surface temperature, surface radiative fluxes, and precipitation.

\subsection{SamudrACE~\citep{Duncan_2025}.} Couples the ACE2 atmosphere--land emulator to the Samudra ocean emulator, providing a free-running coupled ML climate model. It was trained on a GFDL CM4 pre-industrial control simulation with constant forcing. Atmospheric output inherits the ACE2 variable set. Related coupled designs include ACE2-SOM, which couples ACE2 to a slab ocean and learns equilibrium responses across prescribed CO$_2$ levels \citep{clark_ace2som_2025}, and DLESyM, which couples learned atmosphere and ocean components for stable simulations of the observed climate \citep{cresswellclay_dlesym_2025}; neither is scored as a separate row in Table~\ref{tab:emulator-comparison-main}.

\subsection{LUCIE-3D~\citep{guan_lucie3d}.} A free-running three-dimensional atmospheric emulator trained on 30 years of ERA5, with optional prescribed-SST AMIP mode. It accepts CO$_2$ as a forcing variable and demonstrates 40-year free-running rollouts. The LUCIE-3D forward model is deterministic -- no stochastic perturbation is injected during the autoregressive rollout -- but the original LUCIE~\citep{guan_lucie2d} is explicitly designed for $O(1000)$-member ensembles, generated by stochastic weight averaging (an ensemble of parameter snapshots saved during extended stochastic-gradient-descent training) rather than by a stochastic internal-variability generator; in a chaotic rollout these parameter-space differences diverge into ensemble spread. Because this delivers large ensembles suitable for tail-risk assessment, albeit through parameter-space spread and not demonstrated in the cited three-dimensional forced-response configuration, we assign the partial (\ensuremath{\circ}) column~'Stoch' rating rather than \ensuremath{\times}. Its output set comprises temperature, specific humidity, and horizontal wind components on log surface pressure levels, and total precipitation.

\subsection{DiffESM~\citep{bassetti_diffesm_2024}.} A 3D diffusion model that downscales monthly mean ESM output for temperature and precipitation only to daily values. DiffESM is therefore not a self-contained climate emulator: it requires an existing ESM simulation as input, and inherits its forcing dependence, stability, and scenario coverage from the ESM output it is fed. Observational alignment is not directly evaluated in the DiffESM paper (which compares only to held-out ESM members), so any present-day alignment is inherited rather than validated; the partial (\ensuremath{\circ}) column~'Obs' rating reflects this inherited-but-unvalidated status.

\subsection{Score-based ESM emulator~\citep{bouabid2026score}.} This score-based diffusion emulator operates on an equal-area HEALPix mesh as in GCMagicc and learns the joint monthly distribution of near-surface air temperature, precipitation, relative humidity, and near-surface wind speed. It is trained separately for each of MPI-ESM1-2-LR, MIROC6, and ACCESS-ESM1-5 using large initial-condition ensembles, rather than being trained jointly across ESMs. Its four demonstrated outputs cover part of the Penman--Monteith driver set, but omit daily minimum and maximum temperature and surface downwelling shortwave radiation, motivating the partial (\ensuremath{\circ}) column~'Penman-M vars' rating. The diffusion formulation draws independent samples conditioned on global-mean surface-temperature anomaly and calendar month, efficiently supporting large stochastic ensembles and reproducing cross-variable statistics and distribution tails; this supports \ensuremath{\checkmark} on column~'Stoch'. However, it neither dynamically advances a climate state nor uses a prescribed or prognostic ocean state, and the authors explicitly note that its instantaneous formulation cannot generate temporally consistent samples. It therefore receives \ensuremath{\times} for both 'Free-run' and 'Multi-dec.' and cannot directly supply temporally coherent 48-month sequences for SPEI-48. The conditioning does not separate GHG, aerosol, solar, volcanic, or ozone forcing contributions, giving \ensuremath{\times} for 'Dec.ERF'. Training and validation are against each parent CMIP6 ESM rather than observations, while bias-aware training or transfer learning is identified as future work, giving \ensuremath{\times} for 'Obs'. The model successfully generates the held-out SSP2-4.5 and SSP3-7.0 pathways without retraining, supporting \ensuremath{\checkmark} for 'New scen.'; this demonstrated transfer is interpolation within the SSP1-2.6 to SSP5-8.5 training range, requires an externally supplied global-temperature trajectory, and is not claimed to extrapolate beyond that range. These task-specific ratings do not diminish the method's demonstrated strengths in multivariate distributional and tail fidelity, and they do not imply that GCMagicc has perfect temporal fidelity: GCMagicc's remaining low-frequency and spectral limitations are discussed in the main manuscript Limitations section.

\subsection{ArchesClimate-SSP~\citep{clyne_2026_archesclimate}.}
A flow-matching generative emulator of the IPSL-CM6A-LR model: a deterministic mean-state prediction is followed by a stochastic residual, and the climate state is advanced autoregressively as a coupled atmosphere--ocean system (ocean temperature \textit{thetao} is predicted rather than prescribed, motivating the free-running \ensuremath{\checkmark}\ rating). Conditioning forcings comprise CO\textsubscript{2}, CH\textsubscript{4} and N\textsubscript{2}O, multi-species sulphate, nitrate and black-carbon aerosols, solar irradiance in six wavelength bands and a ten-level ozone profile, but no separable volcanic channel (assumed implicit in the aerosol fields); this is the most complete forcing decomposition among the machine-learning emulators considered here, yet the missing volcanic component motivates the partial (\ensuremath{\circ}) column~'Dec.ERF' rating rather than a full \ensuremath{\checkmark}. Surface output includes \textit{tas}, \textit{pr}, \textit{evspsbl} and \textit{uas}/\textit{vas} (so \textit{sfcWind} is derivable), with near-surface humidity inferable from the lowest level of the three-dimensional \textit{hus} field; \textit{tasmin}/\textit{tasmax} and surface shortwave radiation (\textit{rsds}) are absent, giving the partial (\ensuremath{\circ}) column~'Penman-M vars' rating on par with LUCIE-3D. The model is generative, but only five- to ten-member ensembles are demonstrated and the stochastic component is a residual around a deterministic mean, so the partial (\ensuremath{\circ}) column~'Stoch' rating is assigned rather than the large-ensemble \ensuremath{\checkmark}. Free-running rollouts of 85 years (2015--2100) and a 100-year abrupt-4$\times$CO\textsubscript{2} control are numerically stable, supporting the multi-decadal (\ensuremath{\checkmark}) column~'Multi-dec.' rating; we note, however, that the out-of-distribution SSP5-8.5 scenario exhibits a systematic warm drift of order 2$^\circ$C by 2100 and that an ozone-ablated variant destabilises, so this stability is best established within, rather than beyond, the training forcing envelope. Training and validation use the single ESM IPSL-CM6A-LR and the statistical emulator MESMER-M rather than observations or reanalysis, hence the \ensuremath{\times}\ on column~'Obs'. Scenario transfer is the paper's headline result: trained on \textit{historical}, \textit{piControl}, SSP1-1.9, SSP1-2.6, SSP2-4.5, SSP3-7.0 and SSP4-6.0, it generates the held-out SSP4-3.4, SSP5-3.4 and SSP5-8.5 scenarios without retraining, supporting the \ensuremath{\checkmark}\ on column~'New scen.'.

\subsection{GCMagicc (this work).} The partial (\ensuremath{\circ}) column~'Free-run' rating reflects that the free-running claim holds for the coupled MAGICC+GCMagicc package: GCMagicc and GCMagicc-CE are conditioned on MAGICC global temperature and heat uptake, making them less free-running than GCMagicc-PM and GCMagicc-XS. The partial (\ensuremath{\circ}) column~'Multi-dec.' rating reflects that the amplitude and frequency spectra of the low-frequency modes are not yet at parity with the range across CMIP6 ESMs (main manuscript Limitations section). The \ensuremath{\checkmark^{\dagger}} on column~'Obs.' records that alignment is imposed for global-mean \textit{tas} by construction, while the regional and multivariate alignment is learned and evaluated separately (main manuscript Section~\ref{sec:Obs-aligned}).

\section{Drought attribution and drought projections}\label{supp-Drought-Methods}\label{Drought-Methods}

\subsection{Penman--Monteith SPEI-48}

We quantify multi-year meteorological drought with the 48-month Standardised Precipitation Evapotranspiration Index (SPEI-48) \citep{vicenteserrano2010spei,Begueria2014SPEIrevisited}. Monthly climatic water balance is $WB_m=P_m-PET_m$ and its trailing accumulation is
\begin{equation}
WB^{(48)}_m=\sum_{j=0}^{47}WB_{m-j}.
\end{equation}
Potential evapotranspiration is calculated only with the FAO-56 Penman--Monteith reference-evapotranspiration formulation \citep{Allen1998FAO56,Penman1948,Monteith1965}:
\begin{equation}
ET_0 =
\frac{
0.408\,\Delta\,(R_n-G)+\gamma\,\frac{900}{T+273}\,u_2\,(e_s-e_a)
}{
\Delta+\gamma(1+0.34\,u_2)
}.
\end{equation}
Here $\Delta$ is the slope of the saturation-vapour-pressure curve, $R_n$ is net radiation, $G$ is soil heat flux, $\gamma$ is the psychrometric constant, $T$ is air temperature, $u_2$ is wind speed, and $e_s-e_a$ is vapour-pressure deficit. Daily values are converted to monthly totals. The required monthly drivers are precipitation, mean/minimum/maximum air temperature, relative humidity, surface downwelling shortwave radiation, surface wind, and pressure.

For every product, SPEI is standardised separately for each calendar month and grid cell with a 1991--2010 baseline and a generalised logistic (GLO) distribution fitted by unbiased probability-weighted moments \citep{Begueria2014SPEIrevisited,Hosking1990LMoments}. The GCMagicc family fit is pooled across all ensemble members and is shared by its all-forcing and natural-only ensembles. Each SMILE uses a fit pooled across all eligible historical members of that model and applies the same fitted parameters to every all-forcing and natural-only member. Surface downwelling shortwave radiation is used without an ERA5 offset. Grid-cell SPEI is aggregated over Iran with cosine-latitude weights and finite-point renormalisation:
\begin{equation}
\overline{SPEI}_R(t)=
\frac{\sum_{i\in R}SPEI_i(t)\cos(\varphi_i)}
{\sum_{i\in R}\cos(\varphi_i)}.
\end{equation}

\subsection{Monthly below-threshold probability}\label{supp-sec:monthly-drought-diagnostic}

For each ERA5 year $y$ and calendar month $m$ in 2021--2025, probability is estimated from the ensemble-member cross-section:
\begin{equation}
\widehat p_{y,m}=\frac{1}{M}\sum_{i=1}^{M}
\mathbb I\!\left[
\overline{SPEI}^{(i)}_{y,m}\le
\overline{SPEI}^{ERA5}_{y,m}
\right].
\end{equation}
Conditional on the model and forcing state, the $M$ members are the primary independent sampling units. The analysis therefore yields 60 same-year/month probabilities for 2021--2025. For the 2041--2060 diagnostic, each of the same 60 ERA5 thresholds is compared with the distribution for that calendar month pooled across the 20 future years and all members. In addition, the pooled comparison estimates the fraction of all modeled values in each window at or below the latest ERA5 threshold from December 2025. A zero count is reported as an estimated probability of 0\%; the empirical resolution of an estimate is $1/n$, and zero is not evidence that the event is physically impossible.

Table~\ref{tab:drought-attribution-irn-v100-ensemble-comparison} reports the pooled latest-threshold probabilities and the medians and descriptive 5--95\% ranges over the 60 month-matched probabilities. These ranges represent seasonal and interannual variation and are not confidence intervals. SPEI-48 autocorrelation does not invalidate a probability calculated across independent members at a fixed year and month. It does mean, however, that neither the 60 diagnostics nor the nominal $60M$ recent values (or $20M$ values per future calendar month) can be treated as independent observations. Any inferential resampling must therefore resample members as complete trajectories, preserving all within-member autocorrelation.

\input{Figures/main/iran_drought_attribution/iran_drought_attribution_gcmagicc_v100_supplementary_ensemble_comparison.tex}

The recent comparison is structurally model-dependent. GCMagicc and CanESM5 show higher all-forcing than natural-only probabilities, whereas MIROC6 and GISS-E2-1-G show little or no all-forcing--natural-only separation for current times, while presenting that separation for the future. All three SMILEs show higher all-forcing drought probabilities in 2041--2060 than in 2021--2025, although the magnitude differs substantially among models.

\subsection{SMILE inventory and repeated natural-only tails}

The live inventory contains 65 / 50 / 50 historical / SSP2-4.5 / historical-natural CanESM5 files, 50 / 50 / 50 MIROC6 files, and 46 / 22 / 20 GISS-E2-1-G files. All-forcing diagnostics use the 50, 50, and 22 SSP2-4.5 members, respectively. Natural-only CanESM5 and MIROC6 output ends in 2020 and GISS-E2-1-G output ends in 2014. To display and diagnose later periods, each member's final five complete years is repeated member-wise. Natural-only comparisons after the observed endpoints are explicitly flagged as synthetic tails and are interpreted as stationary diagnostic extensions, not future natural-forcing simulations.

\subsection{Predictor sensitivity and SMILE extension}

\begin{figure*}[p]
  \centering
  \includegraphics[width=\linewidth,height=0.85\textheight,keepaspectratio]{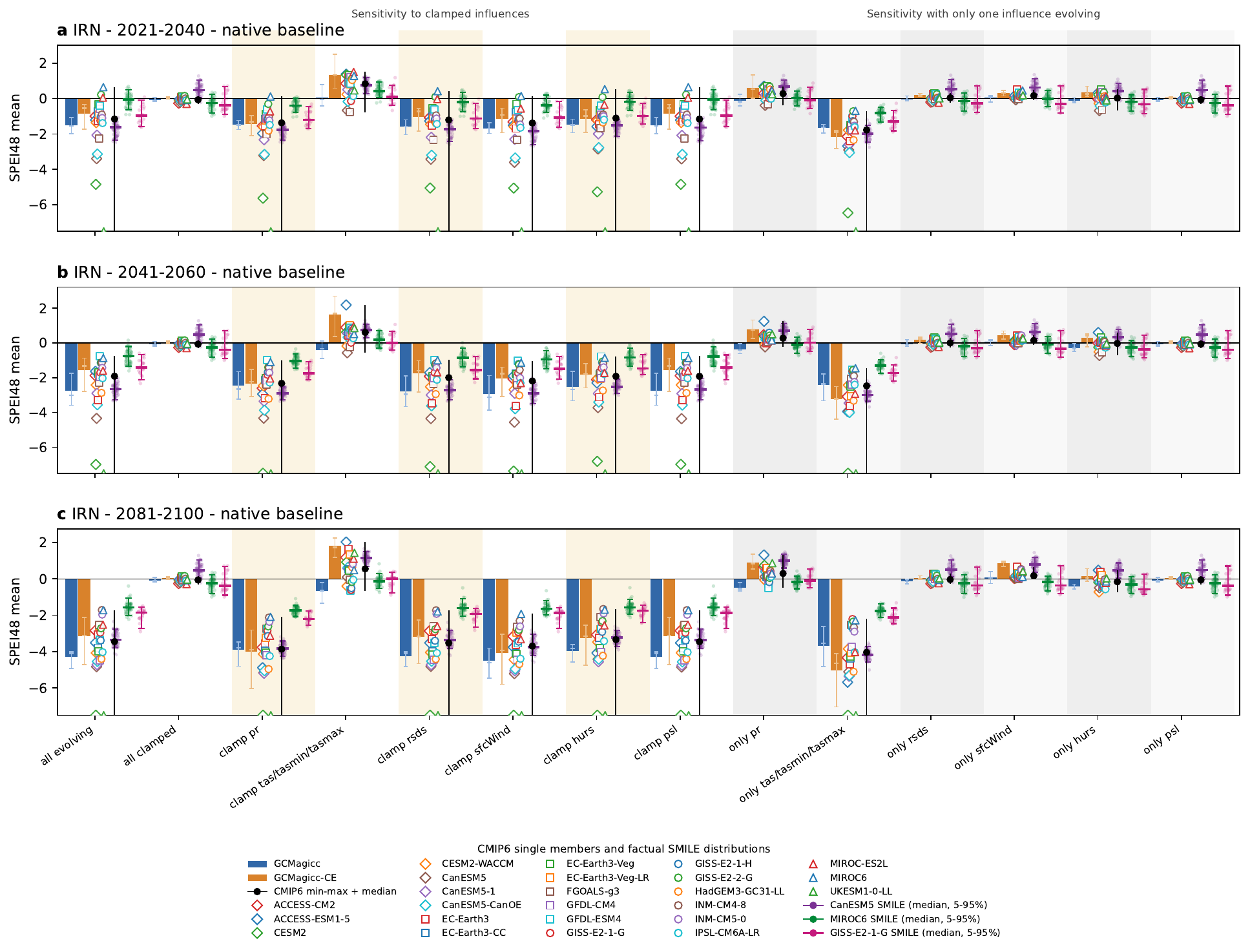}
  \caption{\textbf{Penman--Monteith SPEI-48 sensitivity to evolving meteorological drivers.} Iranian projections are summarised for 2021--2040, 2041--2060, and 2081--2100. Cases on the left clamp one driver group after 2001; cases on the right retain evolution in only one group. Existing GCMagicc summaries and single-member CMIP6 dots retain their native member-specific baseline treatment. Values beyond the plotted range are clipped to the axis floor at $-7.5$. Clipping affects two models: UKESM1-0-LL in 21 of its 42 period--case cells, spanning outliers of $-23.5$ to $-28.8$ SPEI-48, and CESM2 in 9 cells, spanning $-7.5$ to $-9.0$, possibly because of variability or limited robustness in the single-member baseline estimate. Each CMIP6 model is identified by a unique combination of outline colour and marker shape. CanESM5, MIROC6, and GISS-E2-1-G factual SMILE columns show faint member points, the ensemble median, and 5--95\% whiskers, using each model's shared pooled 1991--2010 historical GLO fit.}
  \label{fig:iran-drought-attribution-sensitivity}
\end{figure*}

The predictor-sensitivity experiment is shown at the bottom of Fig.~\ref{fig:iran-drought-attribution-sensitivity}. For each member, one case leaves all drivers evolving, one clamps all drivers after 2001, six clamp one driver group, and six allow only one group to evolve. For each single-member CMIP6 ESM, we compute a member-specific climatology and SPEI baseline over 1975--2000. The three added SMILE columns instead apply each model's shared pooled 1991--2010 historical GLO fit, raw \textit{rsds}, and cosine-latitude regional weighting to every member and sensitivity case.

\subsection{Assumptions and limitations}

SPEI-48 diagnoses multi-year atmospheric supply--demand anomalies rather than plant-available soil water. Results remain conditional on the 1991--2010 GLO fit, Penman--Monteith assumptions, reanalysis uncertainty, cosine-latitude aggregation, forcing construction, and each model's representation of internal variability and long-term change. The monthly cross-member estimator substantially limits the direct effect of strong SPEI-48 autocorrelation on each probability estimate, because members rather than adjacent months are the independent units.

%
%

\codedataavailability{The standalone GCMagicc v1 model and evaluation release is archived at Zenodo at \url{https://doi.org/10.5281/zenodo.22139683} \citep{meinshausen_etal_2026_gcmagicc_dataset} and at \url{https://github.com/maltemeinshausen/GCMagicc_v1_manuscript} under Apache License 2.0. The MAGICC simple climate model, used here as the global-mean predictor generator, is described by \citet{meinshausen_magicc_2011}; all predictors used in this paper were generated with MAGICC~v7.5.3 driven by the AR6 probabilistic parameter set, as configured by the \texttt{AR6} workflow default. That parameter set is the AR6 drawn set \texttt{magicc-ar6-0fd0f62-f023edb} (sub-sample set 0fd0f62 of derived-metrics set f023edb, drawn 9 February 2021), licensed CC~BY-NC~4.0. MAGICC binaries are obtained by registration at \url{https://live.magicc.org}; For MAGICC releases, see \url{https://gitlab.com/magicc/magicc/-/tree/v7.6.0a3}. MAGICC~v7.6.0a3 is the CMIP7 ScenarioMIP fast-track configuration and with methane feedbacks switched off is functionally the same as MAGICC~v7.5.3.

\medskip\noindent The input data underlying this work are CMIP6 model output distributed through the Earth System Grid Federation and the ERA5 reanalysis distributed through the Copernicus Climate Data Store \citep{era5_dataset}. The 76 CMIP6 data references covering the 32 source models and the four MIP activities used here (CMIP, ScenarioMIP, DAMIP and CFMIP) are tabulated by DOI in Table~\ref{tab:cmip6-dois} in the Supplement, with their corresponding full dataset citations in its reference list. The model--experiment--member combinations drawn from CMIP6 are listed in Table~\ref{tab:runs} in the Supplement. GCMagicc data are released under CC~BY~4.0 in four packages:
(1) output replicating CMIP6 model--scenario combinations;
(2) ERA5-aligned scenario projections;
(3) CMIP6-format output for the 20-member ERA5-spliced ensembles; and
(4) analysis data underlying the manuscript figures.
}

\authorcontribution{N.M. conceived, built, developed, trained, and implemented the GCMagicc model. Methodological work also involved X.S. M.M. distilled the CMIP6 training data, built the evaluation and figure suite, analysed GCMagicc output, and created the gcmagicc.org portal. Z.N. and M.M. adapted and ran the MAGICC workflows for the predictors. A.S., K.S., J.L. and M.M. analysed NDCs and LT-LEDS for the NDC-related scenarios. S.S. provided Figure~\ref{fig:gcmagicc-model-architecture} and contributed to the writing of the manuscript. E.V. contributed to the evaluation methodology and to the writing of the manuscript. All authors contributed to interpretation and writing.
}

\competinginterests{All GCMagicc code and data presented here are released under Apache-2.0 and CC~BY~4.0 and are free to use, including commercially; the MAGICC probabilistic parameter distribution used to generate the predictors is licensed CC~BY-NC~4.0 and is excluded from that permission. M.M., Z.N., J.L., K.S. and A.S. are employees of Climate Resource, which also engages in geophysical climate hazard consultancy. The remaining authors declare no competing interests.
}

\begin{acknowledgements}
We acknowledge the World Climate Research Programme, which, through its Working Group on Coupled Modelling, coordinated and promoted CMIP6. We thank the climate modelling groups for producing and making available their model output, the Earth System Grid Federation for archiving the data and providing access, and the agencies supporting CMIP6 and ESGF. We thank Urs Beyerle for providing access to a CMIP6 archive at ETH Zurich and Navid Constantinou, Andrew King and Mandy Freund for helpful discussions on the manuscript in general and on the attribution and El Ni\~no related sections in particular. 

Anthropic Claude, OpenAI Codex and Google Gemini were used as assistive tools for code generation, high-performance-computing operations, development of the gcmagicc.org portal, and analysis support, and for copy-editing and readability assistance on the manuscript. No generative AI tool produced the manuscript text or the scientific interpretations presented here. The authors inspected, modified and verified (or threw out) the resulting code, calculations, citations and text, and retain full responsibility for the work. 
\end{acknowledgements}

\financialsupport{M.M. and Z.N. appreciate the involvement in and support from the National Environmental Science Program (NESP) Climate Systems Hub [grant number C036648]), project 'uncharted futures'.}

\begingroup
\sloppy
\hbadness=10000
\bibliographystyle{copernicus}
\bibliography{GMDreferences_v6}
\endgroup

\end{document}

%% file: Figures/main/iran_drought_attribution/iran_drought_attribution_gcmagicc_v100_supplementary_ensemble_comparison.tex
\begin{table*}[t]
\centering
\caption{Estimated probabilities of modeled SPEI48 being at or below ERA5 thresholds for Iran, Islamic Republic of. The pooled comparison uses the latest ERA5 value (2025-12; SPEI48=-1.871). Each estimate is the fraction of modeled values at or below the corresponding ERA5 threshold. The month-matched comparison reports the median [5th--95th percentile] across the 60 ERA5 months in 2021--2025.}
\label{tab:drought-attribution-irn-v100-ensemble-comparison}
\footnotesize
\setlength{\tabcolsep}{3.2pt}
\renewcommand{\arraystretch}{1.05}
\resizebox{\textwidth}{!}{%
\begin{tabular}{@{}lllrll@{}}
\toprule
Panel (window) & Product & Forcing state & Members & \shortstack{Probability at/below latest ERA5, \%\\(count/$N$)} & \shortstack{Month-matched probability, \%\\median [P$_5$--P$_{95}$] ($n$)} \\ 
\midrule
d (2021-2025) & GCMagicc & Factual & 100 & 11.25\% (675/6{,}000) & 29.00\% [13.00--57.05] (n=100) \\
d (2021-2025) & GCMagicc & Natural-only & 100 & 0.35\% (21/6{,}000) & 0.00\% [0.00--7.05] (n=100) \\
\addlinespace[2pt]
d (2021-2025) & CanESM5 & Factual & 50 & 28.10\% (843/3{,}000) & 44.00\% [26.00--76.10] (n=50) \\
d (2021-2025) & CanESM5 & Natural-only & 50 & 0.00\% (0/3{,}000) & 0.00\% [0.00--0.00] (n=50) \\
\addlinespace[2pt]
d (2021-2025) & MIROC6 & Factual & 50 & 0.37\% (11/3{,}000) & 2.00\% [0.00--18.00] (n=50) \\
d (2021-2025) & MIROC6 & Natural-only & 50 & 0.63\% (19/3{,}000) & 2.00\% [0.00--30.00] (n=50) \\
\addlinespace[2pt]
d (2021-2025) & GISS-E2-1-G & Factual & 22 & 6.21\% (82/1{,}320) & 13.64\% [8.86--45.45] (n=22) \\
d (2021-2025) & GISS-E2-1-G & Natural-only & 20 & 7.08\% (85/1{,}200) & 15.00\% [5.00--55.25] (n=20) \\
\midrule
e (2041-2060) & GCMagicc & Factual & 100 & 58.11\% (13{,}947/24{,}000) & 75.10\% [58.17--91.41] (n=2{,}000) \\
e (2041-2060) & GCMagicc & Natural-only & 100 & 0.30\% (73/24{,}000) & 1.40\% [0.25--6.39] (n=2{,}000) \\
\addlinespace[2pt]
e (2041-2060) & CanESM5 & Factual & 50 & 82.49\% (9{,}899/12{,}000) & 91.20\% [82.75--96.91] (n=1{,}000) \\
e (2041-2060) & CanESM5 & Natural-only & 50 & 0.00\% (0/12{,}000) & 0.00\% [0.00--0.00] (n=1{,}000) \\
\addlinespace[2pt]
e (2041-2060) & MIROC6 & Factual & 50 & 9.69\% (1{,}163/12{,}000) & 24.55\% [10.32--50.83] (n=1{,}000) \\
e (2041-2060) & MIROC6 & Natural-only & 50 & 0.63\% (76/12{,}000) & 2.40\% [0.80--18.88] (n=1{,}000) \\
\addlinespace[2pt]
e (2041-2060) & GISS-E2-1-G & Factual & 22 & 31.44\% (1{,}660/5{,}280) & 49.55\% [31.07--76.16] (n=440) \\
e (2041-2060) & GISS-E2-1-G & Natural-only & 20 & 7.08\% (340/4{,}800) & 16.00\% [7.00--49.20] (n=400) \\
\bottomrule
\end{tabular}%
}
\begin{minipage}{\textwidth}
\vspace{2pt}\footnotesize\textit{Notes.} Probabilities are empirical fractions of modeled SPEI values less than or equal to the corresponding ERA5 threshold. For panel d, each monthly probability compares ERA5 with the ensemble-member cross-section at the same year and calendar month; for panel e, it compares ERA5 with all member-years in 2041--2060 for the same calendar month. A zero count is reported as an estimated probability of 0\%; the empirical resolution of an estimate is $1/n$. The 5th--95th percentile range describes month-to-month variation, not a confidence interval. Factual denotes SSP2-4.5 for the GCMagicc product and historical plus SSP2-4.5 for the SMILEs; natural-only denotes SSP2-4.5-nat for the GCMagicc product and hist-nat for the SMILEs. For the SMILE natural-only rows, the final five complete hist-nat years are repeated member-wise through 2100; all panel d/e values are therefore synthetic repetitions and add no temporal degrees of freedom. Nominal member-month counts are not independent sample sizes because values share ensemble trajectories and SPEI48 is autocorrelated.
\end{minipage}
\end{table*}